%% file: DPhilThesis.tex
\documentclass[12pt]{ociamthesis}  

\usepackage{amssymb}
\usepackage{cite}

\usepackage[utf8]{inputenc}
\usepackage{amsmath}
\usepackage{amsfonts}
\usepackage{latexsym}
\usepackage{mathrsfs}
\usepackage{graphicx}
\usepackage{color}
\usepackage{slashed}
\usepackage{twistor}
\usepackage[all]{xy}

\newcommand{\sa}{\mathsf{a}}

\newcommand{\vepsilon}{\varepsilon}
\newcommand{\sA}{\mathsf{A}}
\newcommand{\sK}{\mathsf{K}}
\newcommand{\tepsilon}{\tilde{\epsilon}}
\newcommand{\tvepsilon}{{\tilde\varepsilon}}

\usepackage{braket}		
\usepackage{xcolor}

\newcommand{\sT}{\mathsf{T}}
\newcommand{\sG}{\mathsf{G}}
\newcommand{\sH}{\mathsf{H}}

\newcommand{\sd}{\mathsf{d}}
\newcommand{\se}{\mathsf{e}}

\renewcommand{\d}{\mathrm{d}}
\renewcommand{\sb}{\mathsf{b}}
\renewcommand{\sc}{\mathsf{c}}

\usepackage{hyperref}
\renewcommand{\d}{\mathrm{d}}

\newtheorem{theorem}{Theorem}

\title{Ambitwistor strings and \\[1ex]     
        amplitudes in curved backgrounds}   

\author{Stefan Nekovar}             
\college{Jesus College}  

\degree{Doctor of Philosophy}     
\degreedate{Trinity 2018}         

\begin{document}

\baselineskip=24pt plus1pt

\setcounter{secnumdepth}{3}
\setcounter{tocdepth}{3}

\maketitle                  
\include{dedication}        
\include{abstract}          
\include{originality}  
\include{acknowledgements}  

\begin{romanpages}          
\tableofcontents            
\listoffigures              
\end{romanpages}            

\include{chapter1}

\include{chapter2}

\include{chapter3}

\include{chapter4}

\include{conclusions}

\appendix
\include{appendix1}

\include{appendix2}

\include{appendix3}

\include{appendix4}

\addcontentsline{toc}{chapter}{Bibliography}
\bibliography{bibliography}        	
\bibliographystyle{JHEP} 		

\end{document}

%% file: dedication.tex
\begin{dedicationNV}
To Anton, Birgit and Carola 
\\
for their unconditional love and support
\end{dedicationNV}

%% file: abstract.tex
\begin{abstract}

Feynman diagrams have been superseded as the tool of choice for calculating scattering amplitudes. Various other methods are not only more efficient but also explicitly exhibit beautiful structures obscured by Feynman diagrams. This thesis aims to lay some groundwork on how two of these methods, ambitwistor strings and the double copy, can be generalised to scattering in curved backgrounds.

In the first part of this thesis, a heterotic ambitwistor string model coupled to a non-abelian background gauge field is constructed. It is shown that after decoupling gravity this model is anomaly free if and only if the background field is a solution to the Yang-Mills equations. A fixed gluon vertex operator for the aforementioned heterotic model as well as a vertex operator encoding graviton, B-field and dilaton for type II ambitwistor strings in a curved background are presented. It is shown that they are BRST closed if and only if they correspond to physical on-shell states.

In the second part, sandwich plane waves are considered. It is shown that scattering of gluons and gravitons is well defined on these backgrounds. 3-point amplitudes are calculated using quantum field theory techniques and a double copy relation between gluons on a gauge theory plane wave and gravitons on a gravitational plane wave is proposed. Using the results from the first part of this thesis, it is then shown that curved background heterotic and type II ambitwistor string models correctly reproduce these 3-point amplitudes on sandwich plane waves.

\end{abstract}

%% file: originality.tex
\begin{originalitylong}
This thesis contains results obtained from work done in collaboration with Tim Adamo, Eduardo Casali and Lionel Mason. 
\\
Chapter~\ref{C3.Heterotic} is mostly based on 
\begin{quote}
T. Adamo, E. Casali and S. Nekovar, \textit{Yang-Mills theory from the worldsheet}, \textit{Phys. Rev.} \textbf{D98} (2018) 086022, \href{https://arxiv.org/abs/1807.09171}{arXiv:1807.09171 [hep-th]}. \cite{Adamo:2018hzd}
\end{quote}
Chapter~\ref{C3.VertexOps} is mostly based on 
\begin{quote}
T. Adamo, E. Casali and S. Nekovar, \textit{Ambitwistor string vertex operators on curved backgrounds}, \href{https://arxiv.org/abs/1809.04489}{arXiv:1809.04489 [hep-th]}. \cite{Adamo:2018ege} 
\end{quote}
Chapter~\ref{C4.Spacetime} is mostly based on 
\begin{quote}
T. Adamo, E. Casali, L. Mason and S. Nekovar, \textit{Scattering on plane waves and the double copy, Class. Quant. Grav.} \textbf{35} (2018) 015004, \href{https://arxiv.org/abs/1706.08925}{arXiv:1706.08925 [hep-th]}. \cite{Adamo:2017nia}
\end{quote}
Chapter~\ref{C4.Worldsheet} is mostly based on 
\begin{quote}
T. Adamo, E. Casali, L. Mason and S. Nekovar, \textit{Amplitudes on plane waves from ambitwistor strings, JHEP} \textbf{11} (2017) 160, \href{https://arxiv.org/abs/1708.09249}{arXiv:1708.09249 [hep-th]}. \cite{Adamo:2017sze}
\end{quote}
Chapter~\ref{chapter2} is expanded from the relevant review sections of the aforementioned papers.

\end{originalitylong}

%% file: acknowledgements.tex
\begin{acknowledgementslong}
\pagestyle{empty}	
First and foremost, I would like to thank my supervisor, Lionel Mason, for his guidance and support throughout my DPhil. He has always been generous with his time and very patient when explaining some new piece of mathematics or physics to me. I very much enjoyed our joint work, but am also grateful that I was given the freedom to pursue my own interests during my DPhil.

Working with Tim Adamo and Eduardo Casali has not only taught me a great deal about physics but also been an incredibly enjoyable experience, for which I am very grateful. Another thanks is due to Eduardo for carefully proofreading this thesis and the many useful comments resulting from that.

During my first two years in Oxford, Yvonne Geyer has been a constant source of explanations, support and encouragement, thank you for that.

Thanks to the entire mathematical physics group in Oxford for being so welcoming and approachable. I enjoyed our many entertaining and instructive discussions, academic and otherwise, in particular with my officemate Robert.

This research has been funded by EPSRC grant EP/M50659X/1 and a Studien\-stiftung des deutschen Volkes scholarship.
\bigskip

Over the past few years, my housemates Thomas, Jack and Alison shared my joys and had to put up with my miseries on a daily basis. Their friendship has provided a point of sanity throughout my DPhil. 

Out of the many friends I made in college, without whom my time in Oxford would not have been this enjoyable, Thomas and Karan deserve a special mention. Working together as social secretaries was a great experience and whenever I was getting too busy, they took over my part of the job without much complaining.

I would like to thank my teammates Fabio, Krzysztof, Andy, Gytis, Sanders, Christos, Adam, David, Jonas, Kuba, Sven, Nick and Tilo of the 2015/16 blues volleyball team for much needed distraction and encouragement during a difficult year and them and all other friends I made at OUVC for the great times we spent together.
\bigskip

Thanks are due to my family, in particular to my parents and my sister, who have supported and encouraged me throughout my education.

Last but not least, I would like to thank Medha for her unwavering love and support in the past years.

\end{acknowledgementslong}

%% file: chapter1.tex
\chapter{Introduction}\label{chapter1}

Physics has seen two major revolutions in the twentieth century: General relativity, linking gravity directly to the inseparable notion of spacetime and describing the universe on its largest scales, and quantum theory, fundamentally changing our understanding of the nature of elementary particles that make up matter on its smallest scales.

From the mathematical point of view, the proper tool to formulate general relativity is differential geometry. It provides a very clear and precise understanding of the theory. This geometric formulation of gravity is widely considered to be the most elegant physical theory known today. In more practical terms, general relativity has been incredibly successful, both explaining and predicting physical effects that are incompatible with Newton's theory of gravity. This ranges from the early tests using the perihelion advance of mercury and bending of light rays by the sun to the recent experimental discovery of gravitational waves. Its results even need to be taken into account in the ubiquitous satellite navigation systems like GPS.
\medskip

Quantum theory, in particular in the guise of (perturbative) quantum field theory, can boast of similar successes when it comes to predicting the behaviour of physical systems. The anomalous magnetic dipole moment of the electron can be predicted by QED and measured very precisely. This yields the most accurate agreement between theoretical prediction and experimental results known in physics. In fact, the entire standard model is a quantum field theory. Furthermore quantum field theory has spread beyond elementary particle physics and is also used in fields like condensed matter theory or topology. While many of the initial problems of quantum field theory have been resolved, most notably via the systematic treatment of infinities by renormalisation, it still lacks a precise mathematical formulation. This clearly shows that there is a considerable amount of work to be done to reach a similar level of conceptual clarity as in general relativity.

Even from the perspective of physics, there have been recent developments in quantum field theory that are entirely unexpected from the standard perturbative point of view and require us to search for a more thorough understanding of the underlying theory. These include supersymmetric localisation, the existence of integrable quantum field theories, quantum field theories without Lagrangian formulation and dualities like the famous AdS/CFT correspondence.
\medskip

Another development indicating that our understanding of quantum field theory is still rather incomplete has taken place in the theory of \emph{scattering amplitudes}, objects firmly in the realm of perturbative QFT. These are the main physical observables considered in quantum field theory. Particle physics experiments like the LHC essentially measure these quantities and compare them to predictions from theoretical models.

The standard method to calculate the S-matrix is to evaluate Feynman diagrams order by order in perturbation theory. This yields results in good agreement with experiments, however calculations tend to get out of hand quickly when increasing the number of loops or external particles. Furthermore, the final answers are often much simpler and more elegant than one would expect from looking at the individual contribution of each Feynman diagram. This is illustrated nicely by the famous \emph{Parke-Taylor formula} for the tree level, colour ordered, maximally helicity violating $n$-gluon amplitude. Using the spinor helicity formalism\footnote{This formalism makes use of the fact that the momentum of a massless particle can be decomposed into two-component spinors and makes manifest that amplitudes of massless particles transform under little group representations determined by the external states. An introduction to this formalism can be found in~\cite{Elvang:2015rqa}.} and stripping off the momentum conservation delta functions, the amplitude is simply
\begin{equation}
\label{ParkeTaylor}
A[1^-,2^-,3^+,...,n^+] = \frac{\langle 12 \rangle^4}{\langle 12 \rangle ... \langle n1 \rangle}
\end{equation}
and this equation holds for all $n \ge 3 $. From the point of view of Feynman diagrams, the existence of such a formula is a miracle: The number of diagrams contributing to the tree level $n$-gluon amplitude grows fast~\cite{Elvang:2015rqa} and already exceeds $100$ for $n=7$. The Parke-Taylor amplitude was initially conjectured in~\cite{Parke:1986gb} and its proof two years later used recursion relations instead of Feynman diagrams~\cite{Berends:1987me}. Amplitudes with unexpectedly simple structure like equation~\eqref{ParkeTaylor} led physicists to believe that there ought to be more direct ways to arrive at these results than the standard diagram expansions, like the off-shell recursions used to prove the Parke-Taylor formula. This resulted in the development of various new methods to calculate scattering amplitudes over the past years, a recent review of (some of) these can be found in~\cite{Elvang:2015rqa}.
\medskip

One of these methods is known as the \emph{double copy}, often heuristically written as $\text{gravity} = (\text{gauge theory})^2$, which essentially means that a gravity amplitude can be obtained from the product of two gauge theory amplitudes (all stripped of their respective momentum conserving delta functions and coupling constants). The first relation of this kind was discovered with the help of string theory, where it was found that closed string tree amplitudes can be obtained from summing over certain products of two open string tree amplitudes and kinematic coefficients~\cite{Kawai:1985xq}. In the $\alpha' \rightarrow 0$ limit, where these tree level string amplitudes with massless external states turn into regular tree amplitudes for massless particles, this turns into an analogous statement about graviton and colour ordered gluon amplitudes. This is known as \emph{KLT relations}. While they follow the general theme of the double copy, they are not honest squaring relations, as the KLT kernel for a large number of external particles is highly non-trivial. 

Another incarnation of the double copy makes use of the \emph{BCJ relations} or \emph{colour kinematics duality}~\cite{Bern:2008qj,Bern:2010ue,Bern:2010yg}. Remember that a general tree level gauge theory amplitude can be written as 
\begin{equation}
\label{GluonTree}
A = \sum_i \frac{c_i n_i}{P_i},
\end{equation}
where $c_i$ are the colour factors, $n_i$ the kinematic numerators and $P_i$ the propagator factors contributing to the $i$th diagram\footnote{We blow up each 4-point vertex by inserting the appropriate propagator in the numerator and denominator at the $s,t$ or $u$-channel colour contribution of said vertex, so that we only need to consider trivalent graphs.}. Recall that the colour factors are built from structure constants and hence satisfy certain algebraic properties that follow from the antisymmetry of these structure constants as well as their Jacobi identities. The kinematic numerators are not unique. Using so called generalised gauge transformations, which leave the amplitude invariant, they can be brought into a form $\tilde{n}_i$, that satisfies the same algebraic identities as the colour factors $c_i$. Then the gravity amplitude can be obtained by replacing the colour factors in~\eqref{GluonTree} by these new kinematic numerators $\tilde{n}_i$:
\begin{equation}
\label{GravitonTree}
M = \sum_i \frac{\tilde{n}_i n_i}{P_i} 
\end{equation}

Various proofs for this relation exist at tree level~\cite{BjerrumBohr:2009rd,Stieberger:2009hq,BjerrumBohr:2010zs,Feng:2010my,Tye:2010dd}. While the BCJ version of the double copy is equivalent to the KLT relations at tree level, it makes the squaring manifest directly and more importantly is conjectured to also hold at loop level~\cite{Bern:2010ue}. This conjecture has been applied to calculations at increasingly high loop orders, which have been considered computationally inaccessible prior to the discovery of the colour kinematics duality~\cite{Bern:2009kd,Bern:2012cd,Bern:2012gh,Bern:2013uka,Bern:2014sna}. These calculations have shown that the onset of UV divergences in supergravity has to happen at much higher loop order than expected from standard techniques~\cite{Bern:2015xsa,Bern:2017lpv,Bern:2017tuc}. There have been papers suggesting that four-dimensional $\cN=8$ supergravity could even be perturbatively finite in the UV~\cite{Bern:2006kd,Bern:2011qn}, however recent work does not support this conjecture~\cite{Bern:2018jmv}.

Note that the double copy simplifies considerably for the 3-point tree level amplitudes, where the colour factor is simply the structure constant itself. It is antisymmetric under the interchange of external particles. However, as gluons are bosons, the amplitude needs to be symmetric. Hence the kinematic ``numerator'' has to be antisymmetric under the interchange of external particles as well and therefore automatically satisfies the same algebraic property as the colour factor. There obviously is no propagator in this case either. Then the gravity amplitude literally is the square of the colour stripped gluon amplitude.

There also exist examples of a classical version of the double copy, relating non-linear solutions in gauge theory and gravity~\cite{Anastasiou:2013hba,Monteiro:2014cda,Luna:2015paa,Ridgway:2015fdl,Borsten:2015pla,Luna:2016due,Goldberger:2016iau,Cardoso:2016amd,Luna:2016hge,Goldberger:2017frp,Berman:2018hwd}. Choosing an appropriate gauge, certain spacetimes, for example the Schwarzschild black hole, can be related to classical solutions of gauge field theories, in this case the Coulomb solution of electromagnetism, via a squaring relation. These results heavily rely on properties of the algebraically special solutions studied and there seems to be no clear general formalism analogous to the amplitudes one yet.  
\medskip

Another direction of progress in amplitudes arose from \emph{twistor theory}. Twistor space is non-locally related to complexified spacetime via the incidence relations: a point in spacetime corresponds to a line in twistor space, while points in twistor space correspond to certain two dimensional null hyperplanes, usually referred to as $\alpha$ planes, in complexified spacetime. It was noted that the Parke-Taylor formula~\eqref{ParkeTaylor} adapted to $\mathcal{N}=4$ supersymmetric Yang-Mills theory can be expressed elegantly in twistor space~\cite{Nair:1988bq}. This was subsequently extended to a full formulation of $\mathcal{N}=4$ SYM as a string theory with supersymmetric twistor space as target space~\cite{Witten:2003nn,Berkovits:2004hg,Berkovits:2004jj}. The resulting formula for $n$ particle scattering amplitudes is a remarkably simple integral over the moduli space of rational curves from the worldsheet\footnote{Strictly speaking, this is only known to be true for genus zero corresponding to tree level amplitudes.} with $n$ marked points to supersymmetric twistor space, where the MHV degree is fixed by the degree of the rational curve. To simplify our life even further, the integration over the moduli space localises to solutions of the so called \emph{scattering equations}. The result is known as RSV formula~\cite{Roiban:2004yf} and it is completely unexpected from the Feynman diagram perspective, the non-local nature of twistor space essentially allowed us to get around all the usual complexity of graph combinatorics. There also is an analogous twistor string model for $\mathcal{N}=8$ supergravity~\cite{Skinner:2013xp}. 

This localisation to the scattering equations is a persistent feature of all twistor string theories, however they remained somewhat unsatisfactory in the sense that they were confined to four spacetime dimensions and required maximal supersymmetry. This was remedied when the so called \emph{CHY formulae} for scattering amplitudes were discovered~\cite{Cachazo:2013gna,Cachazo:2013hca,Cachazo:2013iea}. They express $n$ particle tree level amplitudes of a multitude of theories, both with and without supersymmetry, as integrals over the moduli space of the Riemann sphere with $n$ punctures. These integrals again localise to solutions of the scattering equations and the formulae are valid in any dimension. The precise integrand is determined by the choice of theory for which one wants to compute the amplitudes.
\medskip

The form of these amplitudes as integrals over the moduli space of $n$-punctured Riemann spheres is highly suggestive of genus zero string theory amplitudes with $n$ vertex operator insertions. This was confirmed by the discovery of \emph{ambitwistor strings}~\cite{Mason:2013sva}, which reproduced the CHY amplitudes from a chiral worldsheet model with finite massless spectrum and no free parameters on the worldsheet. The target space of this model is the space of complex null geodesics, which is a close relative of twistor space\footnote{In the case of four dimensional flat space, the corresponding projective ambitwistor space is a quadric in the product of projective twistor space and its dual. This also explains its name, ambi is a Latin prefix meaning ``both''.} known as \emph{ambitwistor space}. The localisation of the integral to solutions of the scattering equations is automatically built into the model by the ambitwistor Penrose transform of momentum eigenstates. 

A variety of ambitwistor string models have been constructed to reproduce the tree level scattering amplitudes for a wide array of field theories~\cite{Ohmori:2015sha,Casali:2015vta,Azevedo:2017lkz}. Loop amplitudes have been obtained from these models by considering worldsheet correlators at higher genus~\cite{Adamo:2013tsa,Casali:2014hfa,Adamo:2015hoa} or on the nodal Riemann sphere~\cite{Geyer:2015bja,Geyer:2015jch,Geyer:2016wjx,Geyer:2017ela,Geyer:2018xwu}. Ambitwistor strings can also be viewed as null strings quantised in an unusual way~\cite{Casali:2016atr,Casali:2017zkz}. Other aspects that have been studied include four dimensional models~\cite{Geyer:2014fka,Bandos:2014lja}, pure spinor versions~\cite{Berkovits:2013xba,Chandia:2015sfa,Jusinskas:2016qjd,Azevedo:2017yjy}, ambitwistor string field theory~\cite{Reid-Edwards:2015stz,Reid-Edwards:2017goq}, a proposed ambitwistor string propagator~\cite{Roehrig:2017gbt} and the spectrum in both GSO projections~\cite{Berkovits:2018jvm}. Ambitwistor string methods have also been applied to the study of asymptotic symmetries and soft theorems~\cite{Adamo:2014yya,Geyer:2014lca,Lipstein:2015rxa,Adamo:2015fwa} and even spacetime conformal invariance~\cite{Adamo:2017zkm}.

A remarkable feature of type II ambitwistor strings is that they remain a free worldsheet CFT in a curved supergravity background. Quantum consistency of this model is equivalent to the background fields obeying the NS-NS supergravity (gravity coupled to a B-field and dilaton) equations of motion~\cite{Adamo:2014wea}. This yields an exact description of a non-linear field theory as free two dimensional CFT.
\bigskip

A major goal of modern theoretical physics is to obtain a theory that incorporates both gravity and quantum theory. Various attempts at such a theory have been made with varying degrees of success or lack thereof, including loop quantum gravity as well as M-theory and most prominently string theory, which can be viewed as a limiting case of M-theory. 

One of the first attempts to marry quantum theory to gravity was to simply put a perturbative quantum field theory on a classical curved background spacetime. The best known triumph of this program is the discovery of Hawking radiation emitted by black holes~\cite{Hawking:1975iha}, allowing their treatment as thermodynamic objects with finite temperature. Despite the successes of quantum field theory in curved spacetime, curved background amplitude\footnote{Or other physical observables like correlation functions, as the usual notion of scattering amplitudes is not well defined in generic curved backgrounds.} calculations have not been able to keep pace with the recent progress in flat space. While standard perturbative QFT computations tend to get out of hand even faster in curved spacetimes, none of the modern tools mentioned above are available there. In this thesis, we aim to make some initial steps towards the exploitation of two of these modern methods in curved backgrounds: 
\begin{itemize}
\item Like regular string theory, \emph{ambitwistor strings} can be placed in curved backgrounds. This raises the usual questions about quantum consistency of these theories and if sensible correlators can still be calculated. While it is known that ambitwistor strings yield elegant formulas for amplitudes in flat space, the corresponding statement in curved backgrounds is not clear (even in those backgrounds where scattering amplitudes are still well defined).
\item Recall that there are two incarnations of the \emph{double copy}, one for scattering amplitudes in flat space and another one for classical solutions of the equations of motion. This gives rise to hope that a suitably adapted combination of the two might still work in curved backgrounds. Similar to the heuristic double copy formula, this can be written as $\text{Gravity + gravity} = (\text{Gauge theory + gauge theory})^2$.
\end{itemize}
While these questions cannot be answered definitively at the moment, we are able to report some progress supporting the above ideas in this thesis.


\newpage
\section{Outline}

This thesis is divided in two main parts, one mostly concerned with ambitwistor strings in curved backgrounds, while the other focuses on scattering on plane waves. Before we present our research, we review relevant previous work in chapter~\ref{chapter2}. In particular, we describe ambitwistor string models in chapter~\ref{WSM} and (sandwich) plane wave spacetimes and gauge fields in chapter~\ref{Background}.

Chapter~\ref{chapter3} is concerned with results about ambitwistor string models in generic curved backgrounds. In chapter~\ref{C3.Heterotic}, we construct a heterotic ambitwistor string in a non-abelian background gauge field and show that the anomaly cancellation conditions impose the Yang-Mills equations on this background field \emph{exactly}, if one decouples the gravitational degrees of freedom. In chapter~\ref{C3.VertexOps}, vertex operators for the curved background heterotic and type II ambitwistor string are proposed and we show that BRST closure imposes appropriate gauge conditions and equations of motion for these operators to correspond to linear perturbations of the background fields by gluon as well as graviton, B-field and dilaton insertions. 

Chapter~\ref{chapter4} is concerned with scattering on sandwich plane wave backgrounds. In chapter~\ref{C4.Spacetime}, we generalise results about the well definedness of the scattering problem in these backgrounds from scalars to gluons and gravitons. This allows us to calculate 3-point amplitudes for gluons on a gauge field plane wave and gravitons on a plane wave spacetime. We then propose a double copy construction relating these two objects. In chapter~\ref{C4.Worldsheet}, we calculate the 3-point correlators of gluon vertex operators in the plane wave background heterotic ambitwistor string and of graviton vertex operators in the plane wave background type II ambitwistor string. These correlators are shown to agree with the amplitudes obtained in the previous chapter.

In chapter~\ref{conclusions}, we summarise and discuss our findings and propose some directions for future research.

%% file: chapter2.tex
\chapter{Review}\label{chapter2}


\section{Ambitwistor strings}
\label{WSM}

Ambitwistor string theories are worldsheet models whose spectra contain only massless degrees of freedom. Our focus will be on those models whose spectra include ordinary, massless supergravity and gauge theory; these are known as the type II and heterotic ambitwistor strings, respectively. After a brief review of the CHY formulae and ambitwistor space, we describe these models on flat backgrounds and how the type II model can be coupled to curved background fields.

The material in this chapter is relevant to chapter~\ref{chapter3} and chapter~\ref{C4.Worldsheet} and can be skipped, if the reader is familiar with these topics.


\subsection{The scattering equations and CHY formulae}

While we do not make explicit use of them in this thesis\footnote{We calculate some curved background 3-point amplitudes using ambitwistor strings in chapter~\ref{C4.Worldsheet}. However, it will soon become clear that the scattering equations cannot appear below 4 points.}, the scattering equations are an essential feature of ambitwistor strings and a brief review is in order. These equations have appeared in the context of dual models~\cite{Fairlie:1972zz,Roberts:thesis,Fairlie:2008dg} and string theory~\cite{Gross:1987kza,Gross:1987ar} much before the search for alternatives to Feynman diagrams began. We, however, will focus on their role in ``modern'' amplitudes as presented in~\cite{Cachazo:2013gna,Cachazo:2013hca,Cachazo:2013iea}. Given a set of null momenta $k_{i\, \mu}$ with $i\in \{1,...,n\}$ satisfying momentum conservation $\sum_{i=1}^n k_{i\, \mu} = 0$, let us consider the maps $P_\mu : \mathbb{CP}^1 \rightarrow \mathbb{C}$ defined as
\be
\label{P}
P(z) = \sum_{i=1}^n \frac{k_i}{z-z_i} .
\ee
Its residues associate an external momentum $k_i$ to each pole $z_i$ on the Riemann sphere. In chapter~\ref{ASflat} it will become clear that from the perspective of ambitwistor strings, this is best viewed as a meromorphic section of the canonical bundle of the (genus zero) worldsheet, $P_\mu \d z \in\Omega^0(\Sigma,K_\Sigma)$. The meromorphic quadratic differential $P^2$ has only simple poles, as our external momenta $k_{i\, \mu}$ are null:
\be
\label{SquaredP}
P^2(z) = \sum_{i=1}^n \sum_{j \ne i} \frac{k_i \cdot k_j}{\left(z-z_i\right)\left(z-z_j\right)} .
\ee
Since non-vanishing meromorphic quadratic differentials on the Riemann sphere cannot be free of poles, one can set $P^2 = 0$ by setting the residues of~\eqref{SquaredP} to zero. This yields the scattering equations:
\be
\label{ScatteringEquations}
\text{Res}_{z_i} P^2(z) = \sum_{j \ne i} \frac{k_i \cdot k_j}{ z_i-z_j } = k_i \cdot P(z_i)
\ee
The minimum number of poles for meromorphic quadratic differentials on $\mathbb{CP}^1$ is actually four. It should therefore be sufficient to kill the residues of all but three poles to set $P^2$ to zero globally. This is also reflected in the scattering equations: Using momentum conservation, one can show that they are invariant under the $\text{SL}(2,\mathbb{C})$ symmetry of the moduli space of the $n$-punctured Riemann sphere. Hence only $n-3$ of them are independent and these have $(n-3)!$ different solutions~\cite{Dolan:2014ega}. 

An important feature of the scattering equations is that they relate the factorisation channels of amplitudes to those boundary points of the moduli space of the $n$-punctured Riemann sphere, where some of the punctures approach each other. These boundary points are included in the Deligne-Mumford compactification of the moduli space, which has been studied in the context of string perturbation theory~\cite{Witten:2012bh}. Another remarkable feature is that they lead to a particularly simple version of the KLT relations by effectively diagonalising the KLT kernel. 
\medskip

The scattering equations make their star appearance in the CHY formulae for scattering amplitudes~\cite{Cachazo:2013hca,Cachazo:2013iea}. Tree level amplitudes are written as 
\be
\label{CHY}
A_n^{(0)} = \int_{\left(\mathbb{CP}^1\right)^n} \frac{\prod_{i=1}^n \d z_i}{\text{vol}\,\text{SL}(2,\mathbb{C})} \prod_j{}' \bar{\delta} \left( k_j \cdot P(z_j) \right) \, I_n\left(z_l, k_l, \epsilon_l \right),
\ee
where $I_n\left(z_l, k_l, \epsilon_l \right)$ is a theory dependent integrand encoding the external kinematics. The primed product of delta functions is defined as 
\be
\prod_j{}' \bar{\delta} \left( k_j \cdot P(z_j) \right) := z_{kl} z_{lm} z_{mk} \prod_{j\ne k,l,m} \bar{\delta} \left( k_j \cdot P(z_j) \right)
\ee
where $z_{kl} = z_{k} - z_{l}$. This does not depend on the choice of $k,l,m$ and imposes the scattering equations, remember that only $n-3$ of them are independent. Notice that we wrote the measure in its permutation invariant form and had to therefore divide it by the (infinite) volume of $\text{SL}(2,\mathbb{C})$. By using the $\text{SL}(2,\mathbb{C})$ invariance to fix three points $z_k,z_l,z_m$ on the Riemann sphere we can remove this redundancy. This introduces a Jacobian $\frac{z_{kl} z_{lm} z_{mk}}{ \d z_k \d z_l \d z_m}$, so that the measure becomes
\be
\frac{\prod_{i=1}^n \d z_i}{\text{vol}\,\text{SL}(2,\mathbb{C})} \rightarrow \prod_{i\ne k,l,m} \d z_i \; z_{kl} z_{lm} z_{mk},
\ee
which leaves us with $n-3$ integrations which are exactly fixed by the delta functions imposing the scattering equations.
\medskip

As mentioned above, the structure of~\eqref{CHY} is identical for all theories and the integrand $I_n\left(z_l, k_l, \epsilon_l \right)$ determines, which amplitudes we are actually calculating. Integrands for a whole zoo of theories are known, see e.g.~\cite{Cachazo:2014nsa,Cachazo:2014xea}, however we will focus on the original integrands for Yang-Mills theory, Einstein gravity coupled to a B-field and dilaton as well as a bi-adjoint scalar theory to illustrate these formulae. There are only two basic ingredients necessary to construct the integrands for these three theories. First one needs the Parke-Taylor factors encoding the colour structure
\be
\mathcal{C}_{1...n} = \sum_{\sigma \in S_n / \mathbb{Z}_n} \frac{\text{Tr} \left( T^{\sa_{\sigma(1)}} \, ... \, T^{\sa_{\sigma(n)}} \right)}{ z_{\sigma(1)\sigma(2)} z_{\sigma(2)\sigma(3)} \, ... \,  z_{\sigma(n-1)\sigma(n)} z_{\sigma(n)\sigma(1)} }.
\ee
The second ingredient is obtained from the matrix
\be
\Psi = \begin{pmatrix}
A & -C^{T}
\\
C & B
\end{pmatrix}
\ee
encoding the kinematic data via
\begin{align}
A_{ij} &= \frac{k_i \cdot k_j}{z_{ij}} &
B_{ij} &= \frac{\epsilon_i \cdot \epsilon_j}{z_{ij}} &
C_{ij} &= \frac{\epsilon_i \cdot k_j}{z_{ij}} 
\\
A_{ii} &= 0 &
B_{ii} &= 0 &
C_{ii} &= -\sum_{j \ne i} \frac{\epsilon_i \cdot k_j}{z_{ij}} .
\end{align}
Using the scattering equations, it is easy to see that the vector $(1,...,1,0,...,0)$ is in the kernel of $\Psi$. The second vector in the kernel of $\Psi$ is $(z_1,...,z_n,0,...,0)$, showing this explicitly requires momentum conservation and $k_i^2 = k_i \cdot \epsilon_i =0$ in addition to the scattering equations. These two vectors actually span the two dimensional kernel of $\Psi$. Therefore, the Pfaffian (square root of the determinant) of $\Psi$ vanishes. However the quantity 
\be
\text{Pf}' (\Psi) = \frac{(-1)^{i+j}}{z_{ij}} \text{Pf} \left(\Psi^{ij} \right),
\ee
where $\Psi^{ij}$ is the same matrix with columns and rows $i$ and $j$ removed, does not depend on the choice of $i,j$ and will yield a non-vanishing answer. This so called reduced Pfaffian is the second ingredient we were looking for.

We are now ready to write down the integrands $I_n\left(z_l, k_l, \epsilon_l \right)$. Yang-Mills theory requires both kinematic and colour data, so the integrand is
\be
I^{\text{YM}}_n\left(z_l, k_l, \epsilon_l \right) = \mathcal{C}_{1...n} \cdot \text{Pf}' (\Psi) .
\ee
In a stunning manifestation of the double copy, the gravity integrand is obtained from this by replacing the colour factor by another kinematic Pfaffian
\be
I^{\text{gr}}_n\left(z_l, k_l, \epsilon_l \right) = \text{Pf}' \left(\Psi\right) \cdot \text{Pf}' \left(\tilde{\Psi}\right) ,
\ee
note that the tilde on the second kinematic matrix corresponds to $\Psi$ with a tilded polarisation vector and the same external momenta. The graviton and B-field polarisations are the traceless symmetric and skew part of $\epsilon_\mu \tilde{\epsilon}_\nu$, while the dilaton corresponds to the trace. The bi-adjoint scalar will simply have two colour factors and no reduced Pfaffian, which is sometimes referred to as zeroth copy.
\medskip

Notice that the structure of equation~\eqref{CHY} resembles a genus zero string theory amplitude. It is natural to expect that it actually originates from some worldsheet theory. This is indeed the case, ambitwistor strings yield exactly this type of amplitude. We will turn to a review of these models after a brief introduction to ambitwistor space.


\subsection{Ambitwistor space}
The space of all complex null geodesics in a complex spacetime manifold $(M,g)$ is called projective ambitwistor space $\mathbb{PA}$. It has been used to study physical fields, which can be encoded in its holomorphic structure~\cite{Isenberg:1978kk,Witten:1978xx,Lebrun:1983pa}. As in the case of twistor space, these fields are automatically defined up to gauge transformations on $\mathbb{PA}$, however unlike in twistor theory, they are not forced to obey any field equations\footnote{This can be seen as a disadvantage, as we want these fields to obey their respective equations of motion. On the other hand side, twistor theory imposes equations that are too restrictive to be physical, reducing the fields to their (anti-)self dual sectors. Moreover, ambitwistor strings seem to partially remedy this problem by imposing the correct equations of motion via BRST cohomology.}. We will simply outline some of the most important properties of ambitwistor space here, a recent review in the context of ambitwistor strings\footnote{Amongst other things, the authors discuss the details of the supersymmetric version of ambitwistor space. While this is the version relevant for the ambitwistor string, the concepts are similar to those of the standard ambitwistor space presented here.} can be found in~\cite{Mason:2013sva}.

The construction of projective ambitwistor space works as follows: Consider the holomorphic cotangent bundle $T^*M$ and restrict covectors $p$ to be null with respect to the metric $p^2:= g^{-1}(p,p)=0$ to obtain the null cotangent bundle
\be
T^*_N M = \left\{(x,p) \in T^*M | p^2 = 0 \right\} \,.
\ee
Remember that a rescaling of null momenta simply corresponds to a reparametrisation of null geodesics. Hence we quotient $T^*_N M$ by the vector field $p_\mu \frac{\partial}{\partial p_\mu}$ to find its projective version $\mathbb{P} T^*_N M$. To obtain the space of null geodesics from this scale free null cotangent bundle, we have to now quotient by the generator of null geodesics
\be
p\cdot\nabla = p_\mu \left(\frac{\partial}{\partial x_\mu} + \Gamma^\rho_{\mu\nu} p_\rho \frac{\partial}{\partial p_\nu} \right) \,,
\ee
which yields: 
\be
\mathbb{PA} \;=\; ^{\textstyle \mathbb{P} T^*_N M} \Big/_{\textstyle p\cdot\nabla } \,
\ee
The double fibration property of this construction is nicely illustrated by
\begin{align}
\begin{split}
\mathbb{P} T^*_N M \qquad \,
\\
\pi_\mathbb{PA} \swarrow \qquad \searrow \pi_M
\\
\mathbb{PA} \qquad\qquad M \;\,
\end{split}\,,
\end{align}
where a fibre of $\pi_M$ is the lightcone of the base point and a fibre of $\pi_\mathbb{PA}$ is the null geodesic (along with its unscaled momentum at every point) corresponding to the base point. This also allows us to understand the relation between spacetime and ambitwistor space: A point in ambitwistor space corresponds to a null geodesic in spacetime by construction, while a point in spacetime corresponds to the (quadric) surface of all null geodesics through said point in $\mathbb{PA}$.

Any cotangent bundle carries a natural symplectic structure with symplectic potential (or tautological one form) $\theta = p_\mu dx^\mu$. This symplectic structure obviously cannot survive on the $2d-3$ dimensional space $\mathbb{PA}$, however the symplectic potential descends to projective ambitwistor space in the shape of a non-degenerate, line bundle valued one form $\theta \in \Omega^1 \left(\mathbb{PA},\mathcal{O}(1)\right)$, which is called a \emph{contact structure}.

Using this contact structure in combination with Kodaira theory, the construction of $\mathbb{PA}$ from spacetime can be reversed: Given $\mathbb{PA}$ with its contact structure, one can reconstruct the original spacetime manifold up to conformal transformations~\cite{Lebrun:1983pa}. Going one step further, a small deformation of the complex structure of $\mathbb{PA}$ that preserves the contact structure $\theta$ yields a small deformation of the conformal structure of spacetime. 
\medskip

Like in twistor theory, the Penrose transform relates fields in spacetime to cohomology classes on ambitwistor space. Its precise statement is
\begin{theorem}[Ambitwistor Penrose transform]
\label{APTrafo}
Trace free symmetric fields on spacetime modulo gauge transformations are mapped to cohomology classes on ambitwistor space as:
\begin{equation*}
H^1\left(\mathbb{PA}, \mathcal{O}(n)\right) = 
\begin{cases}
0 &n<-1
\\
&
\\
^{\textstyle H^0\left(\mathbb{P} T^*_N M, \mathcal{O}(n+1)\right)} \Big/_{\textstyle p\cdot\nabla H^0\left(\mathbb{P} T^*_N M, \mathcal{O}(n)\right)}  &n\ge -1
\end{cases}
\end{equation*}
\end{theorem}
A (trace free symmetric) field on spacetime corresponds to a cohomology class in $H^0\left(\mathbb{P} T^*_N M, \mathcal{O}(n+1)\right)$ by contracting all its free indices with $p_\mu$. Consider $n=1$ for example, this yields a graviton in the form $h^{\mu\nu}p_\mu p_\nu$. Similarly, spin $s$ fields are obtained from $n=s-1$. A detailed proof can be found in the literature~\cite{Baston:1987av,Mason:2013sva}, we will simply sketch the main idea here. Consider the following short exact sequence:
\be
0 \rightarrow \mathcal{O}_\mathbb{PA}(n) \rightarrow \mathcal{O}_{\mathbb{P} T^*_N M}(n)  \stackrel{p\cdot \nabla }{\rightarrow} \mathcal{O}_{\mathbb{P} T^*_N M}(n+1) \rightarrow 0
\ee
The Penrose transform can then be obtained directly from the corresponding long exact sequence in cohomology. The relevant subsequence is ($\delta$ is the usual connecting homomorphism)
\be
0 \rightarrow H^0\left(\mathbb{P} T^*_N M, \mathcal{O}(n)\right)  \stackrel{p\cdot \nabla }{\rightarrow} H^0\left(\mathbb{P} T^*_N M, \mathcal{O}(n+1)\right) \stackrel{\delta }{\rightarrow} H^1\left(\mathbb{PA}, \mathcal{O}(n)\right) \rightarrow 0,
\ee
which implies theorem~\ref{APTrafo}.

From the point of view of ambitwistor strings, a key feature of the Penrose transform is the way it acts on momentum $k$ eigenstates in flat space. Fields proportional to $e^{ikx}$ in spacetime turn into Dolbeault cohomology classes with representatives proportional to $\bar{\delta}(k \cdot p)$ on $\mathbb{PA}$. This will ultimately impose the scattering equations~\eqref{ScatteringEquations} in ambitwistor string models~\cite{Mason:2013sva}.


\subsection{Ambitwistor strings in flat space}\label{ASflat}

For flat backgrounds, ambitwistor strings are given by constrained chiral CFTs in two dimensions, governing holomorphic maps from a Riemann surface $\Sigma$ to ambitwistor space. As mentioned in the introduction, there is a small zoo of these ambitwistor strings~\cite{Ohmori:2015sha,Casali:2015vta,Azevedo:2017lkz}. We will be interested in two particular models: the \emph{type II} and \emph{heterotic} ambitwistor strings, which were introduced in~\cite{Mason:2013sva}.


\subsubsection{The type II ambitwistor string}
The type II ambitwistor string is described by the worldsheet action (in conformal gauge):
\begin{align}
\label{IIAction}
S=\frac{1}{2\pi}\int_\Sigma P_\mu\, \bar{\partial}X^\mu + \bar{\psi}_\mu\, \bar{\partial}\psi^\mu - \frac{e}{2}\,\eta ^{\mu\nu} P_\mu P_\nu +\bar\chi\, \psi^\mu P_\mu + \chi\, \eta^{\mu\nu} \bar{\psi}_\mu P_\nu\,,
\end{align}
with the worldsheet matter fields $\{P_{\mu}, X^{\mu}, \bar{\psi}_{\mu}, \psi^{\mu}\}$ having holomorphic conformal weight $\{1,0,\frac{1}{2}, \frac{1}{2}\}$, respectively.\footnote{This form of the type II ambitwistor string, given in~\cite{Adamo:2014wea}, combines the two real worldsheet Majorana fermion systems of the original formulation~\cite{Mason:2013sva} into a single complex fermion system.} The $PX$-system has bosonic statistics, while the $\bar{\psi}\psi$-system is fermionic. In other words,
\be
P_\mu \in\Omega^0(\Sigma,K_\Sigma)\,, \qquad X^{\mu}\in\Omega^0(\Sigma)\,, \qquad \bar{\psi}_{\mu},\psi^{\nu}\in\Pi\Omega^0(\Sigma, K^{1/2}_\Sigma)\,.
\ee
The gauge fields $e$, $\bar{\chi}$, $\chi$ act as Lagrange multipliers, enforcing the constraints $P^2=0$, $\psi\cdot P=0=\bar{\psi}\cdot P$, and carry non-trivial conformal weights:
\be
e\in\Omega^{0,1}(\Sigma, T_{\Sigma})\,, \qquad \bar\chi,\chi \in\Pi\Omega^{0,1}(\Sigma,T^{1/2}_\Sigma)\,.
\ee
The constraints imposed by these Lagrange multipliers are conjugate to the gauge transformations
\begin{align}
 \delta X^{\mu}=\alpha\,\eta^{\mu\nu}P_{\nu} -\epsilon\,\eta^{\mu\nu}\bar{\psi}_{\nu}-\bar{\epsilon}\,\psi^{\mu}\,, \qquad \delta P_{\mu}=0\,, \nonumber \\
 \delta \psi^{\mu}=\epsilon\,\eta^{\mu\nu}P_{\nu}\,, \qquad \delta\bar{\psi}_{\mu}=\bar{\epsilon}\,P_{\mu}\,, \nonumber \\
 \delta e= \dbar\alpha +2(\chi\,\bar{\epsilon}+\bar{\chi}\epsilon)\,, \qquad \delta\chi=\dbar\epsilon\,, \qquad \delta\bar{\chi}=\dbar\bar{\epsilon}\,, \nonumber
\end{align}
where $\alpha$ is a bosonic gauge parameter of holomorphic conformal weight $-1$ and $\epsilon,\bar{\epsilon}$ are fermionic gauge parameters of holomorphic conformal weight $-\frac{1}{2}$. These gauge symmetries effectively reduce the target space to (super) ambitwistor space.

The gauge freedoms can be used to set $e=\chi=\bar{\chi}=0$ via the standard Fadeev-Popov procedure. The resulting gauge-fixed action is manifestly free:
\be\label{gfIIAct}
S=\frac{1}{2\pi}\int_{\Sigma} P_\mu\, \bar{\partial}X^\mu + \bar{\psi}_\mu\, \bar{\partial}\psi^\mu + b\,\dbar c+\bar{b}\,\dbar\bar{c}+\beta\,\dbar\gamma +\bar{\beta}\,\dbar\bar{\gamma}\,.
\ee
The $c$-ghost, a fermionic field of conformal weight $(-1,0)$, is associated with holomorphic reparametrisation invariance on the worldsheet, and $\bar{c}$ (with the same quantum numbers as $c$) is associated with the gauge transformations generated by the $P^2=0$ constraint. The bosonic ghosts $\gamma,\bar{\gamma}$ are both left-moving, with conformal weight $(-\frac{1}{2},0)$, and are associated with the gauge transformations generated by the constraints $\psi\cdot P=0=\bar{\psi}\cdot P$. 

The BRST-charge resulting from this gauge fixing procedure is
\be\label{IIBRST}
Q=\oint c\,T+bc\,\partial c +\frac{\bar{c}}{2}\,\eta^{\mu\nu}P_{\mu}P_{\nu} + \bar{\gamma}\,\psi^{\mu} P_{\mu}+\gamma\,\eta^{\mu\nu}\bar{\psi}_{\mu} P_{\nu} -2\gamma\bar{\gamma}\bar{b}\,,
\ee
where $T$ is the holomorphic stress tensor for all fields except the $(b,c)$ ghost system. Using the free OPEs associated with \eqref{gfIIAct}
\be\label{flatOPEs}
X^{\mu}(z)\,P_{\nu}(w)\sim \frac{\delta^{\mu}_{\nu}}{z-w}\,, \qquad \psi^{\mu}(z)\,\bar{\psi}_{\nu}(w)\sim \frac{\delta^{\mu}_{\nu}}{z-w}\,,
\ee
and likewise for the ghost fields, it is straightforward to calculate any possible anomalies. Indeed, one finds
\be\label{IIQ2}
Q^2=(d-10)\,\frac{c\,\partial^{3} c}{4}\,,
\ee
so the only anomaly is the central charge, which is eliminated in the critical target dimension $d=10$. As long as the worldsheet is genus zero, $\Sigma\cong\CP^1$, this conformal anomaly will not affect the computation of worldsheet correlation functions, except for an overall numerical ambiguity we can ignore. So from the point of view of scattering amplitudes, the type II ambitwistor string is well-defined on Minkowski space of any dimension at genus zero (A caveat is in order here: Even at genus zero, ambitwistor strings only exhibit the correct factorisation properties in the critical dimension $d=10$).

Using the BRST operator \eqref{IIBRST}, one can investigate the spectrum of the model, which is in one-to-one correspondence with that of type II supergravity~\cite{Mason:2013sva,Adamo:2013tsa,Berkovits:2018jvm}. For instance, it is easy to see that in the NS-NS sector, fixed vertex operators of the form
\be\label{IIfixed}
c\,\bar{c}\,\delta(\gamma)\,\delta(\bar{\gamma})\,\bar{\epsilon}^{\mu}\epsilon_{\nu}\,\bar{\psi}_{\mu}\,\psi^{\nu}\,\e^{\im\, k\cdot X}\,,
\ee
are $Q$-closed provided $k^2=k\cdot\epsilon=k\cdot\bar{\epsilon}=0$. The trace-free symmetric part of $\bar{\epsilon}^{\mu}\epsilon_{\nu}$ encodes the massless graviton of type II supergravity. As we are using a complex fermion system, B-field and dilaton vertex operators are more subtle than in~\cite{Mason:2013sva} and can be obtained from flat space limits of the vertex operators presented in chapter~\ref{C3.VertexOps}. A key feature of the ambitwistor string is that the $n$-point sphere correlation functions of these vertex operators, along with their picture number zero descendants, are equal to the CHY formulae for the tree level scattering amplitudes of supergravity. 

So to summarise: the type II ambitwistor string on a Minkowski background has the spectrum of massless type II supergravity (after GSO projection), is well-defined up to a conformal anomaly (which is under control at genus zero), and produces the tree level S-matrix of supergravity perturbatively around Minkowski space in terms of worldsheet correlation functions.


\subsubsection{The heterotic  ambitwistor string}

The heterotic ambitwistor string, as its name suggests, is obtained by replacing the complex fermion system of the type II model with a single real fermion system while simultaneously adding a holomorphic worldsheet current algebra. In Minkowski space, the worldsheet action in conformal gauge is given by
\be\label{HetAct}
S=\frac{1}{2\pi}\int_{\Sigma}P_\mu\, \bar{\partial}X^\mu + \Psi_\mu\, \bar{\partial}\Psi^\mu -\frac{e}{2} \eta^{\mu\nu}\,P_{\mu}P_{\nu}+\chi\,\Psi\cdot P + S_C\,,
\ee
where $\Psi^{\mu}$ are fermionic with holomorphic conformal weight $\frac{1}{2}$, and $S_C$ is the action of a holomorphic worldsheet current algebra for some gauge group $G$ (assumed to be simple and compact). The current, $j^{\sa}$, associated to $S_C$ has conformal weight $(1,0)$ and its OPE on $\Sigma$ takes the form:
\be\label{calg}
j^{\sa}(z)\,j^{\mathsf{b}}(w)\sim \frac{k\,\delta^{\mathsf{ab}}}{(z-w)^2} + \frac{f^{\mathsf{abc}}\,j^{\mathsf{c}}(w)}{z-w}\,,
\ee
where the sans-serif Roman indices $\sa, \sb, \ldots =1,\ldots, \dim \mathfrak{g}$ run over the adjoint representation of $G$, $k$ is the level of the worldsheet current algebra, and $f^{\mathsf{abc}}$ are the structure constants of $\mathfrak{g}$.

As before, holomorphic reparametrisation invariance and the gauge freedoms associated with the constraints $P^2=0$ and $\Psi\cdot P=0$ can be used to set $e=\chi=0$. This results in a gauge fixed action
\be\label{gfhAct}
S=\frac{1}{2\pi}\int_{\Sigma}P_\mu\, \bar{\partial}X^\mu + \Psi_\mu\, \bar{\partial}\Psi^\mu + b\,\dbar c+\bar{b}\,\dbar\bar{c}+\beta\,\dbar\gamma + S_{C},
\ee
and BRST charge
\be\label{HetBRST}
Q=\oint c\,T+bc\,\partial c +\frac{\bar{c}}{2}\,\eta^{\mu\nu}P_{\mu}P_{\nu} + \gamma\,\Psi^{\mu} P_{\mu}-\frac{\bar{b}}{2}\,\gamma\,\gamma\,,
\ee
where the ghost systems have the same statistics and quantum numbers as before. 

The only obstruction to $Q^2=0$ for the heterotic model is again given by the central charge, which is $\frac{5}{2}d-41 +\mathfrak{c}$, with $\mathfrak{c}$ the central charge of the worldsheet current algebra. So for any fixed $d\leq 16$, this anomaly can be eliminated by choosing the worldsheet current algebra appropriately. However, at genus zero the conformal anomaly is practically irrelevant.

In the gauge theory sector, the spectrum of the heterotic model agrees with that of $\cN=1$ super-Yang-Mills theory. Take the fixed NS sector vertex operators
\be\label{hetfixed}
c\,\bar{c}\,\delta(\gamma)\,\epsilon\cdot\Psi\,j^{\sa}\,\sT^{\sa}\,\e^{\im\,k\cdot X}\,.
\ee
These vertex operators are $Q$-closed provided $k^2=k\cdot\epsilon=0$, and therefore represent gluons. Correlation functions of such vertex operators (and their descendants) at genus zero lead to the CHY expressions~\cite{Cachazo:2013hca} for the tree level scattering amplitudes of Yang-Mills theory in $d$-dimensional Minkowski space. 

The gravitational sector of the heterotic ambitwistor string corresponds to a certain non-unitary $R^2$ supergravity~\cite{Azevedo:2017lkz}. At genus zero, these modes can be projected out consistently by isolating the single trace contributions to the correlator from the worldsheet current algebra. Double (and higher) trace terms -- which contribute with higher powers of the level $k$ -- are mediated by the non-unitary gravitational modes.

%


\subsection{Type II model on a curved background}
\label{CurvedTypeII}

In~\cite{Adamo:2014wea} it was shown how to couple the type II ambitwistor string to curved background fields from the NS-NS supergravity sector which has a metric $g_{\mu\nu}$, B-field $B_{\mu\nu}$ and dilaton $\Phi$ in its spectrum.

The curved space analogue of the gauge-fixed worldsheet action \eqref{gfIIAct}, taking into account certain subtleties associated with worldsheet re\-pa\-ra\-metri\-sa\-tion invariance (see~\cite{Adamo:2014wea} for details), is
\begin{align}
\begin{split}\label{IIcurv1}
S=\frac{1}{2\pi}&\int_{\Sigma} P_\mu\, \bar{\partial}X^\mu + \bar{\psi}_\mu\, \bar{\partial}\psi^\mu +\bar{\psi}_{\mu}\,\psi^{\nu}\,\Gamma^{\mu}_{\nu\rho}\,\dbar X^{\rho} + b\,\dbar c+\bar{b}\,\dbar\bar{c}+\beta\,\dbar\gamma +\bar{\beta}\,\dbar\bar{\gamma} 
\\
&+\frac{1}{8\pi}\,\int_{\Sigma}R_{\Sigma}\log\left(e^{-2 \Phi} \sqrt{g}\right)\,,
\end{split}
\end{align}
where $\Gamma^{\mu}_{\nu\rho}$ are the Christoffel symbols for the Levi-Civita connection of $g_{\mu\nu}$, $R_{\Sigma}$ is the scalar curvature of the worldsheet and $g$ is the (absolute value of the) determinant of the metric. At first, this action may not seem very promising: Even if we ignore the dilaton dependent contribution in the second line, the connection term (required to ensure spacetime covariance of the worldsheet action) couples the fermions to $X^{\mu}$ non-polynomially. However, it was observed in~\cite{Adamo:2014wea} that the field redefinition
\be\label{Pidef}
P_{\mu}\rightarrow \Pi_{\mu}:=P_{\mu}+\bar{\psi}_{\rho}\,\psi^{\nu}\,\Gamma^{\rho}_{\mu\nu}\,,
\ee
leaves a free worldsheet action; the price for this simplification is that the new field $\Pi_{\mu}$ does not transform covariantly under spacetime diffeomorphisms. This is a small price to pay for a manifestly solvable 2d CFT on \emph{any} curved target spacetime, though.

After this field redefinition, the worldsheet action for the type II model on a curved target space metric is:
\be\label{IIcurv2}
S=\frac{1}{2\pi}\int_{\Sigma} \Pi_\mu\, \bar{\partial}X^\mu + \bar{\psi}_\mu\, \bar{\partial}\psi^\mu + b\,\dbar c+\bar{b}\,\dbar\bar{c}+\beta\,\dbar\gamma +\bar{\beta}\,\dbar\bar{\gamma} +\frac{R_{\Sigma}}{4}\,\log\left(e^{-2 \Phi} \sqrt{g}\right)\,,
\ee
This still contains the dilaton dependent term (the final term in the action) we ignored previously. However, $R_{\Sigma}$ can always be taken to vanish locally, so this term does not affect the worldsheet OPEs of the model, which are the same as in flat space with $\Pi_\mu$ playing the role of $P_\mu$:
\begin{align}
 \label{OPE}
 X^{\mu}(z)\,\Pi_{\nu}(w)\sim \frac{\delta^{\mu}_{\nu}}{z-w}\,, \qquad \psi^{\mu}(z)\,\bar\psi_{\nu}(w)\sim \frac{\delta^{\mu}_{\nu}}{z-w}\,
\end{align}

Associated with this gauge-fixed action is a curved version of the BRST charge \eqref{IIBRST}, taking the form:
\be\label{IIcQ}
Q=\oint c\,T+bc\,\partial c +\frac{\bar{c}}{2}\,\cH + \bar{\gamma}\,\cG +\gamma\,\bar{\cG}-2\gamma\bar{\gamma}\bar{b}\,,
\ee
where the currents $\cG$, $\bar{\cG}$ and $\cH$ generalise $\psi\cdot P$, $\bar{\psi}\cdot P$ and $P^2$ to curved space, respectively. The fermionic spin $\frac{3}{2}$ currents are given by 
\begin{align}
\label{GeneralG}
 \mathcal{G}=&\psi^\mu\Pi_\mu + \partial(\psi^\mu\Gamma^\kappa{}_{\mu\kappa})-2\partial(\psi^\mu\partial_\mu\Phi)+\frac{1}{3!}\psi^\mu\psi^\nu\psi^\kappa H_{\mu\nu\kappa}\,,\\
\label{GeneralGbar}
 \bar{\mathcal{G}}=&g^{\mu\nu}\bar\psi_\nu(\Pi_\mu-\Gamma^\kappa{}_{\mu\lambda}\bar\psi_\kappa\psi^\lambda) - g^{\mu\nu}\partial(\bar\psi_\kappa\Gamma^\kappa{}_{\mu\nu})-2\partial(g^{\mu\nu}\bar\psi_\mu\partial_\nu\Phi)+\frac{1}{3!}\bar\psi_\mu\bar\psi_\sigma\bar\psi_\lambda H^{\mu\sigma\lambda}\,,
\end{align}
where $H_{\mu\nu\sigma}$ is the background three-form. The third current is bosonic of spin $2$, given by\footnote{This expression for $\mathcal{H}$ corrects some typos made in~\cite{Adamo:2014wea}. We have checked that these modifications do not alter any of the results in~\cite{Adamo:2017sze}.}
\begin{align}
\begin{split}
\label{GeneralH}
\mathcal{H}=& g^{\mu\nu}\left(\Pi_\mu-\Gamma^\kappa{}_{\mu\lambda}\bar\psi_\kappa\psi^\lambda\right)\left(\Pi_\nu-\Gamma^\kappa{}_{\nu\lambda}\bar\psi_\kappa\psi^\lambda\right) -\frac{1}{2}R^{\kappa\lambda}{}_{\mu\nu}\bar\psi_\kappa\bar\psi_\lambda\psi^\mu\psi^\nu 
 \\ 
& - g^{\mu \nu} \partial\left( \Pi_\rho \Gamma^\rho_{\mu \nu} \right) -\bar\psi_\kappa\partial\psi^\lambda g^{\mu\nu}\partial_\lambda\Gamma^\kappa{}_{\mu\nu} + \psi^\mu \partial_\mu \left( g^{\rho \sigma} \partial (\bar{\psi}_\kappa \Gamma^\kappa_{\rho \sigma} )\right)
\\
&+\frac{1}{2} g^{\mu\nu} H_{\mu \kappa \lambda} \psi^\kappa \psi^\lambda \left(\Pi_\nu - \Gamma^ \rho_{\nu\sigma} \bar{\psi}_\rho \psi^\sigma \right) + \frac{1}{2} \left( \Pi_\mu - \Gamma^\kappa_{\mu\lambda} \bar{\psi}_\kappa \psi^\lambda \right) H_\nu^{\;\; \rho \sigma} \bar{\psi}_\rho \bar{\psi}_\sigma
\\
&+\frac{1}{4} g^{\mu\nu} H_{\mu\kappa\lambda} \psi^\kappa \psi^\lambda H_\nu^{\;\; \rho\sigma} \bar{\psi}_\rho \bar{\psi}_\sigma -\frac{1}{3!} \psi^\mu \bar{\psi}_\nu \bar{\psi}_\kappa \bar{\psi}_\lambda \nabla_\mu H^{\nu\kappa\lambda} - \frac{1}{3!} \bar{\psi}_\mu \psi^\nu \psi^\kappa \psi^\lambda \nabla^\mu H_{\nu\kappa\lambda}
\\
&+ \frac{1}{2} H^{\mu\nu\kappa} \bar{\psi}_\kappa \partial \left( H_{\mu\nu\lambda} \psi^\lambda\right) + \partial \left(H_{\kappa \lambda \nu} \psi^\nu \right) g^{\kappa \sigma} \Gamma^\lambda_{\sigma \rho} \psi^\rho  \,- \partial \left(H_{\kappa \lambda \nu} \psi^\nu  g^{\kappa \sigma} \Gamma^\lambda_{\sigma \rho} \psi^\rho \right)
\\
& - \frac{1}{2} \partial_\sigma H_{\mu\nu\rho} \psi^\nu \psi^\rho \partial g^{\sigma\mu} - \frac{1}{12}  H^{\mu \nu \rho} \partial^2 H_{\mu \nu \rho} + \frac{1}{2} \Gamma^\rho_{\mu\nu} H_{\sigma \lambda \rho} \psi^\sigma \psi^\lambda \partial g^{\mu \nu}
\\
&-2 \partial \left(g^{\mu \nu} \Pi_\mu \partial_\nu \Phi \right) - \partial \left(\bar{\psi}_\kappa \psi^\lambda ( 2 \nabla^\kappa \partial_ \lambda \Phi -2 g^{\mu\nu} \Gamma^\kappa_{\mu \lambda} \partial_\nu \Phi ) \right).
\end{split}
\end{align}
These currents are covariant with respect to target space diffeomorphisms and conformal primaries of the worldsheet CFT. This is despite the fact that they contain various terms which do not appear to be manifestly covariant, due to the requirement of normal-ordering on the worldsheet.

The stress tensor appearing in~\eqref{IIcQ} can be broken into matter and ghost contributions $T=T_{\mathrm{m}}+T_{\mathrm{gh}}$, with
\begin{align}\label{stress_tensor}
 T_\mathrm{m}= -\Pi_\mu\partial X^\mu -\frac{1}{2}(\bar\psi_\mu\partial\psi^\mu+\psi^\mu\partial\bar\psi_\mu)-\frac{1}{2}\partial^2\log(e^{-2\Phi}\sqrt{g})
\end{align}
for the matter fields and
\begin{align}
 T_{\mathrm{gh}} = \bar{c}\partial \bar{b} - 2\bar{b}\partial \bar{c} - \frac{3}{2}\beta\partial\gamma - \frac{1}{2}\gamma\partial\beta - \frac{3}{2}\bar\beta\partial\bar\gamma - \frac{1}{2}\bar\gamma\partial\bar\beta
\end{align}
for the ghost fields, where we exclude the $(b,c)$ ghost system as before.

Using the BRST charge and free OPEs of the worldsheet action, the anomalies of the type II model on a curved background can be computed exactly. As in flat space, there is a conformal anomaly; remarkably, this anomaly is unaffected by the background fields. In particular, it vanishes in $d=10$ spacetime dimensions and can be ignored at genus zero for the purposes of calculating scattering amplitudes. 

However, $Q^2=0$ is also obstructed by anomalies related to the gauged currents $\{\cG,\bar\cG,\cH\}$. These anomalies vanish if the algebra of currents is quantum mechanically closed:
\begin{align}\label{curalg1}
\cG(z)\,\cG(w)\sim 0\,, \qquad &\bar{\cG}(z)\,\bar{\cG}(w)\sim 0\,, 
\\ 
\cG(z)\,\bar{\cG}(w)&\sim \frac{\cH}{z-w}\,,\label{curalg2}
\end{align}
and these conditions do impose constraints on the background fields. The requirement that the $\cG(z)\cG(w)$ and $\bar{\cG}(z)\bar{\cG}(w)$ OPEs be non-singular imposes
\be\label{bianchi1}
\partial_{[\mu}H_{\nu\rho\sigma]}=0\,, \quad R_{\mu[\nu\rho\sigma]}=0\,, \quad R_{(\mu\nu)\rho\sigma}=0\,,
\ee
which are the usual Bianchi identities and symmetries of the Riemann tensor of the background metric, along with $\d H=0$. This latter statement indicates that (locally) $H=\d B$; that is, $H$ arises as the field strength of the background B-field. Note that the conventional normalisation for the exterior derivative in this context is slightly unusual, $H_{\mu\nu\sigma}=\partial_\mu B_{\nu\sigma}+\partial_\nu B_{\sigma\mu}+\partial_\sigma B_{\mu\nu}$. 

Dynamical constraints on the background fields emerge from the final closure requirement of \eqref{curalg2}, which imposes
\begin{align}
 R+4\nabla_\mu\nabla^\mu\Phi-4\nabla_\mu\Phi\nabla^\mu\Phi-\frac{1}{12}H^2 & =0\,,\nonumber\\
 R_{\mu\nu}+2\nabla_\mu\nabla_\nu\Phi-\frac{1}{4}H_{\mu\rho\sigma}H_\nu{}^{\rho\sigma} & =0\,,\label{SugraEOM}\\
 \nabla_\kappa H^\kappa{}_{\mu\nu}-2H^\kappa{}_{\mu\nu}\nabla_\kappa\Phi & =0\,.\nonumber
\end{align}
These are precisely the field equations for the NS-NS sector of type II supergravity, so vanishing of BRST anomalies enforces the appropriate equations of motion on the background fields. This is analogous to the statement that ordinary string theory is anomaly free at lowest order in $\alpha'$ if and only if the Einstein equations hold~\cite{Callan:1985ia,Fradkin:1985ys,Banks:1986fu}. But unlike ordinary string theory on a curved background, where the worldsheet action is a complicated interacting 2d CFT necessitating perturbation theory to determine anomalies, the ambitwistor string remains solvable, the anomaly is obtained exactly, and no perturbative expansion is required.



\newpage

\section{Plane wave backgrounds}
\label{Background}

In this chapter, we review plane wave backgrounds in both the gravitational and gauge theoretic contexts. There is a vast amount of literature on plane waves; we will only cover a small area of the subject necessary for this thesis.

The material in this chapter is only relevant to chapter~\ref{chapter4} and can be skipped until the end of chapter~\ref{chapter3}. 


\subsection{Gravitational plane waves}

Non-linear plane waves are among the oldest exact solutions to the field equations of general relativity, and have many fascinating properties (cf. \cite{Baldwin:1926,Ehlers:1962zz,Griffiths:1991zp,Stephani:2003tm,Blau:2011}). These metrics describe spacetimes composed of pure radiation of the gravitational field itself or a Maxwell field, propagating from past to future null infinity along a given constant null direction. Our focus will be on purely gravitational plane wave metrics, which can be interpreted as a coherent superposition of gravitons. There are two standard coordinate systems: the \emph{Einstein-Rosen}~\cite{Einstein:1937qu} and the \emph{Brinkmann}~\cite{Brinkmann:1925fr} coordinates.

In Einstein-Rosen coordinates, the metric is given by:
\be\label{ER1}
\d s^2 = 2\,\d U\,\d V - \gamma_{ij}(U)\,\d y^{i}\,\d y^{j}\,,
\ee
where the indices $i,j,\ldots=1,\ldots,d-2$. The only non-trivial metric components, $\gamma_{ij}$, depend on $U$. As usual, the inverse of $\gamma_{ij}$ is denoted by $\gamma^{ij}$. These coordinates are useful because they manifest many of the symmetries of the spacetime which are `hidden' in the other coordinates. The metric  \eqref{ER1} clearly has Killing vectors $\frac{\partial}{\partial V}$, $\frac{\partial}{\partial y^{i}}$, and the vectors
\be\label{Sym1}
\mathcal{X}^i = y^{i}\frac{\partial}{\partial V} + F^{ij}(U)\, \frac{\partial}{\partial y^j}\,, \quad F^{ij}(U):= \int^{U}\d s\,\gamma^{ij}(s)\,,
\ee
are also Killing. The vectors $\partial_{V}$, $\partial_{i}$ and $\mathcal{X}^{i}$ form a Heisenberg algebra,
\be\label{Sym2}
\left[\cX^{i},\, \cX^{j}\right]=0\,, \qquad \left[\frac{\partial}{\partial y^i},\,\cX^{j}\right]=\delta^{j}_{i}\frac{\partial}{\partial V}\,,
\ee
so plane wave metrics are endowed with an abelian isometry group generated by translations of the constant $U$ planes as well as this (solvable) Heisenberg symmetry. We will also see that massless field equations are most easily solved in these coordinates.

The main drawback of Einstein-Rosen coordinates is that they are essentially never global coordinates: the metric will develop coordinate singularities due to the focusing of the null geodesic congruence tangent to $\p_U$~\cite{Penrose:1965rx,Bondi:1989vm}. Furthermore, the curvature and field equations are given by somewhat complicated expressions in terms of $\gamma_{ij}$. For instance, the Ricci curvature is
\begin{equation*}
R_{UU}=-\frac{\gamma^{ij}}{2}\left(\ddot{\gamma}_{ij}+\frac{1}{2}\dot{\gamma}_{ik}\gamma^{kl}\dot{\gamma}_{lj}\right)\,,
\end{equation*}
where $\dot{f}=\partial_{U}f$ for any function $f(U)$. Thus the vacuum equations impose conditions on $\gamma_{ij}$ in the form of a second-order ODE.

Brinkmann coordinates have the advantage that they are global, and the curvature is easily identified. In the Brinkmann chart, the metric only has one non-trivial component:
\be\label{Br1}
\d s^{2}=2\,\d u\,\d v - H(u,\mathbf{x})\,\d u^2 - \d x_{a}\,\d x^{a}\,,
\ee
with indices $a,b,\ldots=1,\ldots,d-2$. In these coordinates, the $u=\mathrm{const.}$ metric is completely flat. For pp-waves $H(u,x)$ can have general $x$-dependence, but for plane waves it is constrained to be quadratic in $x^{a}$:
\be\label{Br2}
H(u,\mathbf{x})=H_{ab}(u)\,x^{a}\,x^{b}\,.
\ee
The non-vanishing Christoffel symbols in these coordinates are:
\be\label{Christoffel}
\Gamma^{a}_{uu}=-H_{ab}(u)\,x^{b}\,, \quad \Gamma^{v}_{ua}=-H_{ab}(u)\,x^{b}\,, \quad \Gamma^{v}_{uu}=-\frac{\dot{H}(u,\mathbf{x})}{2}\,,
\ee
and the non-vanishing curvature components are directly encoded in the metric via
\be\label{Rcurv}
R^{a}_{\:\:ubu}=-H^{a}_{b}(u)\,,
\ee
so the vacuum equations in Brinkmann coordinates simply impose that $H_{ab}$ be trace-free: $H^{a}_{a}=0$.

\medskip

\begin{figure}[t]
\centering
\includegraphics[scale=.6]{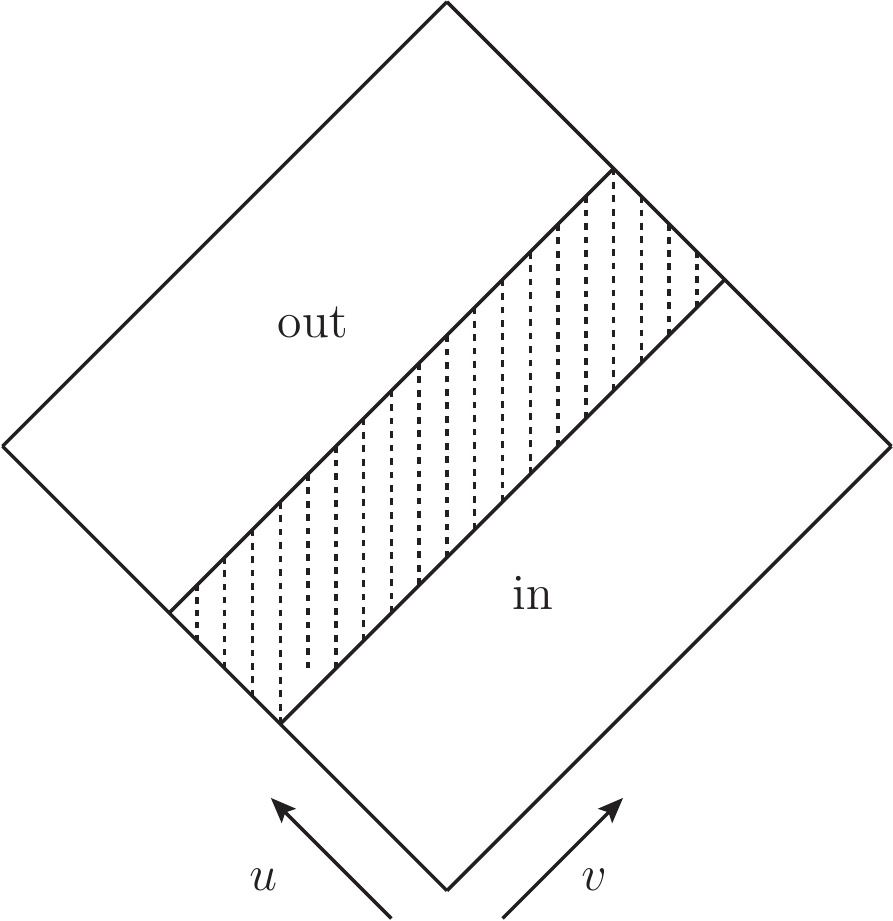}
\caption{The sandwich plane wave with $x^a$-directions suppressed. The function $H_{ab}(u)$ is non-vanishing only in the shaded region; the in- and out-regions are both flat.}\label{SandwichPW}
\end{figure}

The \emph{sandwich} plane wave setup is one for which $H_{ab}(u)$ is compactly supported in $u$~\cite{Bondi:1958aj}. Without loss of generality, we assume that $H_{ab}(u)\neq0$ only for $u_{1}\leq u\leq u_{2}\leq0$; for $u<u_1$ or $u> u_{2}$, the spacetime is a flat. The flat region $u<u_1$ is referred to as the \emph{in-region}, while $u>u_2$ is the \emph{out-region}. See figure~\ref{SandwichPW} for a schematic of this setup.
  
Although we work mostly in Brinkmann coordinates, Einstein-Rosen coordinates are a key tool when solving the linearised Einstein equations on plane wave backgrounds. Hence the relationship between the Brinkmann and Einstein-Rosen coordinate systems will be important. It can be understood in terms of the solutions to the equation:
\be\label{gde}
\ddot{e}_{a}=H_{ab}\,e^{b}\,,
\ee
for some functions $e^{a}(u)$ . Setting $e^{a}(u)=\Delta x^{a}$, \eqref{gde} is the geodesic deviation equation in Brinkmann coordinates; this follows from the fact that the connecting vectors between the geodesics,
\begin{equation*}
e^{a}\frac{\partial}{\partial x^a} - \dot{e}_{a}\,x^{a}\frac{\partial}{\partial v}\,,
\end{equation*}
are  Killing vectors. A set of $(d-2)$ Killing vectors is obtained by choosing a full $(d-2)\times (d-2)$ matrix of solutions to \eqref{gde}, $E^{a}_{i}(u)$ (and its inverse $E^{i}_{a}(u)$), subject to
\be\label{sym}
\dot{E}^{a}_{[i}\,E_{|a|\,j]}=0\,.
\ee
The Killing vectors are then:
\begin{equation*}
\mathcal{D}^{i}=E^{a\,i}\frac{\partial}{\partial x^a} - \dot{E}^{i}_{a}\,x^{a}\frac{\partial}{\partial v}\,.
\end{equation*}
The commutation relations between the $\mathcal{D}^{i}$ and the $\mathcal{X}^i$ (transformed to Brinkmann coordinates) give the Heisenberg algebra which was more manifest in Einstein-Rosen coordinates.

By comparing the line elements \eqref{ER1}, \eqref{Br1}, the diffeomorphism linking Einstein-Rosen and Brinkmann coordinates is identified as:
\begin{subequations}\label{diffeo1}
\begin{eqnarray}
U & = & u\,, \\
V & = & v +\frac{1}{2} \dot{E}^{i}_{a}\,E_{b\,i}\,x^{a}x^{b}\,, \\
y^{i} & = & E^{i}_{a}\,x^{a}\,.
\end{eqnarray}
\end{subequations}
The array $E^{a}_{i}$ and its inverse will be referred to as vielbeins since they give the $d-2$ orthonormal 1-forms $\d x^a=E^a_i\, \d y^i$ in terms of the Einstein-Rosen coordinates. They obey
\be\label{vbrel}
\ddot{E}_{a\,i}=H_{ab}\,E^{b}_{i}\,, \qquad \gamma_{ij}=E^{a}_{(i}\,E_{|a|\,j)}\,.
\ee
As part of the geometry of the Einstein-Rosen waves, the hypersurfaces $V=\mathrm{constant}$ are null and transverse to the geodesic shear-free null congruence $\p_v$ that rules the $u=\mathrm{constant}$ null hypersurfaces.  The $\p_U$ null congruence has a deformation tensor, that often plays a role in the study of perturbative gravity on a plane wave background. In Brinkmann coordinates this tensor is measured by
\begin{equation}
\label{shear} 
\sigma_{ab}=\dot{E}^i_a\,E_{b\,i}\,,
\end{equation}
whose trace is the expansion and trace-free part is the shear.


Note that any other choice of vielbein, say $f^{a}_{i}$, is related to $E^{a}_{i}$ by
\be\label{newvb}
f^{a}_{i}=E^{a}_{j}\left( F^{jk}\,b_{ki}+c^{j}_{i}\right)\,,
\ee 
for constant matrices $b_{ij}$, $c^{i}_{j}$, and $F^{ij}(u)$ defined as:
\be\label{Fij}
F^{ij}(u):=\int^{u} \d s\,\gamma^{ij}(s) = \int^{u} \d s\,E^{a\,(i}(s)\,E^{j)}_{a}(s)\,.
\ee
In particular, given some initial value for the vielbein on the in-region of a sandwich plane wave, \eqref{newvb} encodes how the vielbein changes after passing through the curved interior to the out-region. For the sandwich wave, two natural initial values are given by requiring the vielbein to become trivial in the past or future:
\be\label{vbbc0}
\lim_{u\rightarrow\pm\infty}E^{i\,\pm}_{a}(u) = \delta_{a}^{i}\,.
\ee
Since solutions to \eqref{gde} are simply linear in flat regions, we have
\begin{align}
\begin{split}\label{memory}
E^{a\,-}_i(u)&=b^{a\,+}_i\,u+c^{a\,+}_i \quad \mbox{as} \quad u\rightarrow +\infty\, , 
\\
E^{a\,+}_i(u)&=b^{a\,-}_i\,u+c^{a-}_i \quad \mbox{as} \quad u\rightarrow -\infty\, .
\end{split}
\end{align}
From \eqref{sym} and the conservation of the Wronskian between $E^+$ and $E^-$, it follows that 
\begin{equation}
b_{[i}^{a\,\pm}c_{j]\,a}^{\pm}=0,\qquad b^{a\,+}_i=\delta^{aj}\,\delta_{bi}\,b^{b\,-}_j
\end{equation}
and we can use a rotation of the Brinkmann coordinates to make $b$ symmetric if desired.

Note that it is essentially impossible to have $E$ invertible for all $u$ for non-trivial $b$, so the Einstein-Rosen coordinates are generically singular. This is the inevitable consequence of null geodesic focusing of the $V=\mathrm{constant}$ null hypersurfaces as emphasised by Penrose \cite{Penrose:1965rx}. Both $E^{a\,+}_{i}$ and $E^{a\,-}_{i}$ will describe the \emph{same} flat metric in the asymptotic regions but with different Einstein-Rosen forms. In particular, if the deformation tensor $\sigma_{ab}$ vanishes in one asymptotic region, it will generically be non-trivial in the other, albeit falling off as $1/u$.  
This non-trivial change in $\sigma_{ab}$  is an example of the \emph{memory effect}~\cite{Braginsky:1986ia,Braginsky:1987,Ludvigsen:1989kg}, which has been studied in detail for sandwich plane waves (e.g., \cite{Zhang:2017rno,Zhang:2017geq}).  


\subsection{Gauge theory plane waves}\label{GTPW}

An `Einstein-Rosen' plane wave in gauge theory is a gauge potential which satisfies properties similar to a plane wave metric in Einstein-Rosen coordinates. It is often used to model the electromagnetic fields of lasers (cf. \cite{Reiss:1962,Brown:1964zzb,Ilderton:2012qe}). In particular, we demand that $\sA$ -- \emph{a priori} taking values in the adjoint of some Lie algebra $\mathfrak{g}$ -- manifests the symmetries generated by $\frac{\partial}{\partial v}$ and $\frac{\partial}{\partial x^{a}}$. The most general connection satisfying these conditions has the form:
\be\label{gER1}
\sA=\sA_{0}(u)\,\d v + \sA_{a}(u)\,\d x^{a}\,,
\ee
where we write the potential in the coordinates
\be\label{Mink}
\d s^2 =2\,\d u\,\d v - \d x_{a}\,\d x^{a}\,,
\ee
of Minkowski space. 

We want \eqref{gER1} to be preserved under the same Heisenberg symmetry algebra \eqref{Sym2} that generated the isometries of the plane wave metrics in Einstein-Rosen coordinates. This requires there to be a vector field
\be\label{gSym1}
\cX_{\varphi}^{a}=x^{a}\,\frac{\partial}{\partial v} + u\,\frac{\partial}{\partial x_{a}} +\varphi^{a}\,,
\ee
with $\varphi^{a}$ a Lie algebra-valued function for which
\be\label{gSym2}
\left[\cX^{a}_{\varphi},\, \cX^{b}_{\varphi}\right]=0\,, \qquad \left[\frac{\partial}{\partial x^{a}},\,\cX^{b}_{\varphi}\right]=\delta^{b}_{a}\frac{\partial}{\partial v}\,.
\ee
These conditions imply that $\varphi^{a}=\varphi^{a}(u)$ and $[\varphi^{a},\varphi^{b}]=0$. Furthermore, we require that $\cX^{a}_{\varphi}$ generates a further symmetry of the gauge connection; namely, that $\D=\d +\sA$ is covariantly Lie-dragged along the $\cX^{a}_{\varphi}$. This imposes further constraints on $\sA$:
\be\label{gSymCon}
\sA_{a}=-\dot{\varphi}_{a}\,, \quad  \left[\sA_{0},\,\varphi^{a}\right]=0\,, \quad \left[\sA_{a},\,\varphi^{b}\right]=\delta^{b}_{a}\,\sA_{0}\,.
\ee 
For simplicity, we restrict our attention to the special case where $\varphi^{a}$ is valued in the Cartan subalgebra $\mathfrak{h}\subset\mathfrak{g}$. With this choice, consistency of the symmetry algebra reduces to
\be\label{gSym3}
\sA_{0}=0\,, \qquad \varphi^{a}(u)=-\int^{u}\d s\,\sA^{a}(s)\,,
\ee
and the functional form of $\cX^{a}_{\varphi}$ closely resembles that of its gravitational counterpart \eqref{Sym1}.

To summarise, our definition of an `Einstein-Rosen' plane wave gauge field (valued in the Cartan of the gauge group) results in a gauge potential of the form:
\be\label{gER2}
\sA=-\sA_{a}(u)\,\d x^{a}\,,
\ee
where an overall negative sign has been included for convenience. Just as the Brink\-mann form of a plane wave metric can be obtained by the diffeomorphism \eqref{diffeo1} from Einstein-Rosen form, a gauge transformation of \eqref{gER2} gives the plane wave gauge potential in `Brinkmann' form. In particular, taking $\sA\rightarrow \sA+\d(x^{a}\sA_{a})$ gives 
\be\label{gBr1}
\sA=x^{a}\,\dot{\sA}_{a}\,\d u\,.
\ee
The fact that $\sA$ is a linear polynomial in $x^{a}$, rather than a quadratic function as in the gravitational setting \eqref{Br2}, is a first glimpse of the double copy. It has already been noted that plane wave background geometries (for gauge theory and gravity) exhibit the double copy structure~\cite{Monteiro:2014cda}, although the distinction between linear and quadratic functions does not seem to have been noticed previously. Although we obtained \eqref{gBr1} from the Einstein-Rosen gauge by working in the Cartan subalgebra of the gauge group, general non-abelian plane waves also take this functional form~\cite{Coleman:1977ps}.

The field strength is
\be\label{gFS}
F=\dot{\sA}_{a}\,\d x^{a}\wedge\d u\,.
\ee
As for the Brinkmann metric, the gauge field \eqref{gBr1} directly encodes the field strength; \eqref{gFS} obeys the Maxwell equations, and hence the Yang-Mills equations when valued in the Cartan subalgebra of the gauge group.

The sandwich gauge field plane wave is analogous to that for gravity; the field strength  $F_a=\dot{\sA}_{a}(u)$ is taken to be compactly supported for $u_{1}\leq u\leq u_{2}\leq0$, so that it is flat in the in-region ($u<u_1$) and out-region ($u>u_2$). The memory effect here is associated with the fact that if $\sA$ is taken to vanish in the past, it will be constant and non-zero in the future
\be\label{gmem}
\sA_a|_{\mathrm{out}}-\sA_a|_{\mathrm{in}}=\int_{u_1}^{u_2} F_a\, \d u\,,
\ee
By analogy with the gravitational case, \eqref{gmem} can be viewed as encoding the \emph{electromagnetic memory effect}~\cite{Bieri:2013hqa} for plane wave gauge theory backgrounds. 

%% file: chapter3.tex
\chapter{Ambitwistor strings on non-trivial backgrounds}\label{chapter3}


\section{Heterotic ambitwistor strings on gauge field backgrounds}\label{C3.Heterotic}

The equations of motion for a classical field theory are usually understood as the Euler-Lagrange equations of a corresponding action functional. For certain theories, such as gauge theory and gravity, the equations of motion can also famously be derived from a low-energy expansion of string theory~\cite{Callan:1985ia,Abouelsaood:1986gd,Banks:1986fu}. Coupling the Polyakov action to background gauge or gravitational fields leads to a conformal anomaly. Since the resulting worldsheet action is a complicated interacting 2d conformal field theory (CFT), this anomaly can only be computed perturbatively in a small parameter, taken to be the inverse string tension. To lowest order in this parameter, anomaly cancellation imposes the field equations of gauge theory or gravity on the background fields; the higher-order corrections impose an infinite tower of additional higher-derivative equations. 

Ambitwistor strings are a worldsheet theory with finite massless spectrum and no tunable parameter on the worldsheet. A natural question for this relative of string theory is: can these models be coupled to background fields to give a non-linear description of the underlying field theories? Unlike ordinary string theory, such a description should be exact -- that is, computable without recourse to an infinite perturbative expansion. In the case of the NS-NS supergravity this was answered in the affirmative as described in chapter~\ref{CurvedTypeII}. More recently, it was shown that the abelian Maxwell equations could also be obtained in a similar fashion~\cite{Adamo:2017sze}. However, this success has not been extended to other field theories due to a variety of subtleties associated with coupling to background fields and non-unitary gravitational modes which do not exist in the case of NS-NS supergravity.

In this chapter, which is based on~\cite{Adamo:2018hzd}, we extend the exact worldsheet description of classical field theories to include Yang-Mills theory. This is accomplished by coupling a \emph{heterotic} version of the ambitwistor string to a non-abelian background gauge field\footnote{In~\cite{Azevedo:2016zod} a form of the heterotic ambitwistor string with background fields was also studied, but only classically on the worldsheet.}. The model contains both gauge theoretic and (non-unitary) gravitational degrees of freedom, but the latter can be locally decoupled on the worldsheet. Gauge fixing the worldsheet action leads to potential anomalies; remarkably, the \emph{only} conditions imposed on the background gauge field by anomaly cancellation are the (non-linear) Yang-Mills equations. We also show that re-coupling the gravitational modes leads to gauge anomalies, analogous to but distinct from the well-known anomaly~\cite{Green:1984sg} of the standard heterotic string.


\subsection{The worldsheet model}

As first observed in the context of similar chiral heterotic-like worldsheet models~\cite{Berkovits:2004jj}, the `bad' fourth derivative gravitational modes of the heterotic ambitwistor string can be decoupled at genus zero by taking a limit where $k\rightarrow 0$ while $\frac{k}{g^2_s}$ is held fixed, for $g_s$ the `string' coupling constant which effectively counts the genus of $\Sigma$. In this limit, only single trace contributions to a worldsheet correlation function survive at genus zero. Globality and unitarity of the worldsheet current algebra dictate that $k$ be a positive integer, so the $k\rightarrow 0$ limit must be viewed as a purely formal one which effectively removes the second order pole from the OPE \eqref{calg}. Since our primary concern here will be anomalies, which can be computed \emph{locally} on $\Sigma$, the formality of the limit will not be a problem. 

Fixing $k\rightarrow 0$, introduce a background gauge potential $\sA^{\sa}_{\mu}(X)$ valued in the adjoint of $G$, which couples to the worldsheet current algebra by
\be\label{back1}
S_{C}\rightarrow S_{C}+\frac{1}{2\,\pi}\int_{\Sigma} \sA^{\sa}_{\mu}\,j^{\sa}\,\dbar X^{\mu}\,.
\ee
The purely quadratic nature of the worldsheet model~\eqref{gfhAct} can be preserved by absorbing the explicit dependence on the background into the conjugate of $X^{\mu}$ (cf. equation~\eqref{Pidef} for the analogous procedure in the gravitational context):
\be\label{pidef}
P_{\mu}\rightarrow P_{\mu} + \sA^{\sa}_{\mu}\,j^{\sa}:=\Pi_{\mu}\,.
\ee
This leads to a worldsheet action
\be\label{wsa2}
S=\frac{1}{2\,\pi}\int_{\Sigma}\Pi_{\mu}\,\dbar X^{\mu}+\frac{1}{2}\,\psi_{\mu}\,\dbar\psi^{\mu} +S_{C}\,,
\ee
with free OPEs of the worldsheet fields (in the $k\rightarrow 0$ limit):
\be\label{OPEs}
X^{\mu}(z)\,\Pi_{\nu}(w)\sim \frac{\delta^{\mu}_{\nu}}{z-w}\,, \qquad \psi^{\mu}(z)\,\psi^{\nu}(w)\sim \frac{\eta^{\mu\nu}}{z-w}\,, \qquad j^{\sa}(z)\,j^{\mathsf{b}}(w)\sim \frac{f^{\mathsf{abc}}\,j^{\mathsf{c}}(w)}{z-w}\,,
\ee
for $\eta_{\mu\nu}$ is the $d$-dimensional Minkowski metric.

The price for this simplicity (which is in stark contrast to the complicated interacting 2d CFT obtained by coupling the ordinary heterotic string to a background) is that $\Pi_{\mu}$ is not invariant under local gauge transformations. Under an infinitesimal gauge transformation with parameter $\varepsilon^{\sa}$,
\be\label{gt1}
\delta \sA^{\sa}_{\mu}= f^{\mathsf{abc}}\,\varepsilon^{\mathsf{b}}\,\sA^{\mathsf{c}}_{\mu}-\partial_{\mu}\varepsilon^{\sa}\,, \qquad \delta j^{\sa}= f^{\mathsf{abc}}\,\varepsilon^{b}\,j^{\mathsf{c}}\,,
\ee
which indicates that 
\be\label{gt2}
\delta\Pi_{\mu} =-j^{\sa}\,\partial_{\mu}\varepsilon^{\sa}\,.
\ee
This is a common feature of all curved $\beta\gamma$-systems, of which \eqref{wsa2} is an example~\cite{Nekrasov:2005wg}, and is problematic only if $\Pi_{\mu}$ has a singular OPE with itself after a gauge transformation. Fortunately, it is easy to check that
\be\label{gt3}
(\Pi_{\mu}+\delta\Pi_{\mu})(z)\,(\Pi_{\nu}+\delta\Pi_{\nu})(w)\sim 0\,,
\ee
so the structure of the OPEs \eqref{OPEs} is preserved under gauge transformations. Note that infinitesimal gauge transformations can be implemented by a local operator $\cO_{\varepsilon}=-\varepsilon^{\sa} j^{\sa}$, which obeys
\be\label{gtop}
\cO_{\varepsilon}(z)\,\cO_{\lambda}(w)\sim\frac{f^{\mathsf{abc}}\varepsilon^{\sa}\,\lambda^{\mathsf{b}}\,j^{\mathsf{c}}}{z-w}=\frac{\cO_{[\lambda,\varepsilon]}}{z-w}\,,
\ee
and acts correctly on all worldsheet fields.


\subsection{Yang-Mills equations as an anomaly}

The worldsheet action \eqref{wsa2} has additional symmetries beyond holomorphic re\-pa\-ra\-metri\-sa\-tion invariance. Indeed, the action is invariant under the transformations
\be\label{ftran1}
\delta X^{\mu}=-\epsilon\,\psi^{\mu}\,, \qquad \delta\psi^{\mu}=\epsilon\left(\Pi^{\mu}-\sA^{\mu\,\sa}j^{\sa}\right)\,, \qquad \delta\Pi_{\mu}=\epsilon \psi^{\nu}\partial_{\mu} \sA^{\sa}_{\nu} j^{\sa}\,,
\ee
where $\epsilon$ is a constant fermionic parameter of conformal weight $(-\frac{1}{2},0)$. These transformations are generated by a fermionic current
\be\label{Gcurr}
\sG=\psi^{\mu}\left(\Pi_{\mu}-\sA^{\sa}_{\mu}\,j^{\sa}\right)\,,
\ee
on the worldsheet, which is the extension of $\psi\cdot P$ to the non-trivial gauge background. This current is a holomorphic conformal primary of dimension $3/2$ and is invariant under gauge transformations of the background field.

The OPE of $\sG$ with itself has only simple poles, generating a bosonic current:
\be\label{Hcurr0}
\sG(z)\,\sG(w)\sim \frac{\sH}{z-w}\,,
\ee
where $\sH$ is the extension of $P^2$ to the non-trivial gauge background and $F_{\mu\nu}^\sa$ denotes the Yang-Mills field strength as usual:
\be\label{Hcurr}
\sH = \Pi^2 - 2\, \Pi^\mu \sA_{\mu}^{\sa} j^\sa + \sA_\mu^\sa \sA^{\mu \sb} j^\sa j^\sb + \psi^\mu \psi^\nu F_{\mu\nu}^\sa j^\sa - \partial\left( \partial_\mu \sA^{\mu \sa} j^\sa \right) + f^{\sa\sb\sc} j^\sc \sA^{\mu \sb} \partial \sA_\mu^\sa\,.
\ee
As both $\sG$ and $\sH$ are composite operators on the worldsheet, their definition as currents in the fully quantum mechanical regime requires normal ordering to remove singular self-contractions. We use a point-splitting prescription to do this; for example, the explicit normal ordering of the second term in $\sH$ is given by:
\be\label{no1}
- 2\, \Pi^\mu \sA_{\mu}^{\sa} j^\sa(z):=\frac{\im}{\pi}\,\oint \d w\, \frac{\Pi^{\mu}(w)\,\sA_{\mu}^{\sa}(z) j^{\sa}(z)}{z-w}\,,
\ee
where the integral is taken on a small contour in $w$ around $z$. From now on, we assume this normal ordering implicitly. It is straightforward to check that the normal-ordered current $\sH$ is a conformal primary of dimension $2$ and gauge invariant.

Gauging the symmetries associated with these currents leads to a worldsheet action
\be\label{wsa3}
S=\frac{1}{2\,\pi}\int_{\Sigma}\Pi_{\mu}\,\dbar X^{\mu}+\frac{1}{2}\,\psi_{\mu}\,\dbar\psi^{\mu} +\chi\,\sG +e\,\sH +S_{C}\,,
\ee
where the gauge fields $\chi$, $e$ are fermionic of conformal weight $(-\frac{1}{2},1)$ and bosonic of conformal weight $(-1,1)$, respectively. The worldsheet symmetries associated with $\sG,\sH$ can now be gauge-fixed, along with holomorphic reparametrisation invariance; choosing conformal gauge with $\chi=e=0$ leads to a free gauge-fixed action
\be\label{wsa4}
S=\frac{1}{2\,\pi}\int_{\Sigma}\Pi_{\mu}\,\dbar X^{\mu}+\frac{1}{2}\,\psi_{\mu}\,\dbar\psi^{\mu} +b\,\dbar c + \bar{b}\,\dbar \bar{c} +\beta\,\dbar\gamma +S_{C}\,,
\ee
and associated BRST charge
\be\label{BRST}
Q=\oint c\,T+ bc \partial c +\gamma\,\sG + \bar{c}\,\sH +\frac{\bar{b}}{2}\,\gamma^2\,.
\ee
Here, $(c,b)$ and $(\bar{c},\bar{b})$ are fermionic ghost systems for which $c,\bar{c}$ have conformal weight $(-1,0)$, while $(\gamma,\beta)$ are a bosonic ghost system for which $\gamma$ has conformal weight $(-\frac{1}{2},0)$. In the BRST charge, $T$ denotes the (appropriately normal-ordered) holomorphic stress tensor of the worldsheet CFT, including all contributions from the worldsheet current algebra and all ghosts, except the $(c,b)$ system.

This gauge fixing is anomaly-free if and only if the BRST charge is nilpotent: $Q^2=0$. Given the free OPEs \eqref{OPEs}, this calculation can be performed \emph{exactly} -- there is no need for a background field expansion as in the analogous calculation for the heterotic string~\cite{Callan:1985ia}. It is straightforward to show that $Q^2=0$ only if the central charge of the worldsheet current algebra obeys $\mathfrak{c}=41-\frac{5d}{2}$, and
\be\label{anom1}
\sG(z)\,\sH(w)\sim 0\,.
\ee
The first of these conditions is familiar from flat space and kills the holomorphic conformal anomaly; similar to the type II case described in chapter~\ref{CurvedTypeII}, it is completely independent of the background fields. The second condition -- that the OPE between $\sG$ and $\sH$ be non-singular -- is trivially satisfied in a flat background, but becomes non-trivial in the presence of $\sA^{\sa}_{\mu}$. Making use of the identity
\be\label{nocurr}
j^{\sa}j^{\sb}(z) - j^{\sb}j^{\sa}(z)=f^{\mathsf{abc}}\,\partial j^{\mathsf{c}}(z)\,,
\ee
for normal-ordered products of the worldsheet current, which is derived in appendix~\ref{CA} along with another useful equation, one finds:
\be\label{anom2}
\sG(z)\,\sH(w)\sim -3 \frac{\psi^{\nu} j^{\sa}\,D^{\mu}F^{\sa}_{\mu\nu}}{(z-w)^2}-\frac{\partial\left(\psi^{\nu}j^{\sa}\,D^{\mu}F^{\sa}_{\mu\nu}\right)}{z-w}-\frac{\psi^{\mu}\psi^{\nu}\psi^{\sigma}j^{\sa}\,D_{\mu}F^{\sa}_{\nu\sigma}}{z-w}\,,
\ee
where $D_{\mu}$ is the gauge-covariant derivative with respect to $\sA_{\mu}^{\sa}$.

Requiring \eqref{anom1} to hold then imposes the constraints
\be\label{eoms}
D^{\mu}F^{\sa}_{\mu\nu}=0\,, \qquad D_{[\mu}F^{\sa}_{\nu\sigma]}=0\,,
\ee
on the background gauge field, which are precisely the Yang-Mills equations and Bianchi identity. Furthermore, these are the \emph{only} constraints placed on the background gauge field by anomaly cancellation in the 2d worldsheet theory. Thus, non-linear Yang-Mills theory (in any dimension) is described by an \emph{exact} anomaly in a \emph{free} 2d chiral CFT, unlike the analogous calculations in the heterotic~\cite{Callan:1985ia} or type I~\cite{Abouelsaood:1986gd} superstring.


\subsection{Recoupling gravity and the gauge anomaly}

Gravitational degrees of freedom (in the guise of multi-trace terms) can be recoupled by reinstating the level $k$ (now assumed to be a positive integer). In the ordinary heterotic string, the interplay between gravitational and gauge theoretic degrees of freedom, mediated by the B-field, leads to the Green-Schwarz anomaly cancellation mechanism~\cite{Green:1984sg}. Naively, one might expect a similar phenomenon to arise in the heterotic ambitwistor string, especially since coupling to the background gauge field (through the left-moving worldsheet current algebra) is the same as in string theory (cf., \cite{Moore:1984ws,Hull:1985jv}). This means that we should encounter the usual gauge anomaly associated with chiral fermions on the worldsheet.


In heterotic string theory, resolving this gauge anomaly relies crucially on the form of the B-field coupling to the worldsheet CFT (cf., \cite{Witten:1999eg}). This leads to a modified gauge transformation of the B-field and the shift of its field strength by a Chern-Simons term for the background gauge field. In a chiral model such as the ambitwistor string, the coupling of other background fields (including the B-field) is different and the resolution of the anomaly is no longer clear.

\medskip

To see this, it suffices to consider an abelian background gauge field $\sA_{\mu}$, now coupled to the worldsheet through an (abelian) current algebra of level $k\in\Z_{+}$. Under a gauge transformation $\sA_{\mu}\rightarrow \sA_{\mu}-\partial_{\mu}\varepsilon$, the worldsheet field $\Pi_{\mu}$ transforms as $\Pi_{\mu}\rightarrow\Pi_{\mu}-j\partial_{\mu}\varepsilon$. The gauge-transformed $\Pi_{\mu}$ now has a singular OPE with itself:
\be\label{ganom1}
(\Pi_{\mu}-j\,\partial_{\mu}\varepsilon)(z)\,(\Pi_{\nu}-j\,\partial_{\nu}\varepsilon)(w)\sim k\,\frac{\partial_{\mu}\varepsilon\,\partial_{\nu}\varepsilon}{(z-w)^2}-k\,\frac{\partial_{\mu}\varepsilon\,\partial(\partial_{\nu}\varepsilon)}{z-w}\,,
\ee
proportional to the level $k$. Such anomalous OPEs can be removed in chiral CFTs of $\beta\gamma$-type by compensating for the gauge transformation with a shift of the $\Pi_{\mu}$ field (e.g., \cite{Nekrasov:2005wg}). To remove the anomalous OPE \eqref{ganom1}, define the gauge transformation of $\Pi_{\mu}$ by:
\be\label{ganom2}
\Pi_{\mu}\rightarrow \Pi_{\mu}-j\partial_{\mu}\varepsilon +\frac{k}{2}\partial_\mu\varepsilon\,\partial\varepsilon\,.
\ee
It is straightforward to check that this level-dependent shift removes the singularities from \eqref{ganom1}, ensuring that $\Pi_{\mu}$ has a non-singular OPE with itself in any choice of gauge.

Unfortunately, there is still a gauge anomaly at the level of the worldsheet currents $\sG$ and $\sH$. It is straightforward to see that these currents are no longer gauge invariant; for instance
\be\label{ganom3}
\sG \rightarrow \sG +\frac{k}{2} \psi^{\mu} \partial_{\mu}\varepsilon\,\partial\varepsilon\,,
\ee
under a gauge transformation with the proviso \eqref{ganom2}. Furthermore, the OPE of $\sG$ with itself now has a triple pole contribution 
\be\label{ganom4}
\sG(z)\,\sG(w)\sim \frac{k\,\sA^{\mu} \sA_{\mu}}{(z-w)^3}+\cdots \,,
\ee
which must vanish. In other words, the $\sG$, $\sH$ current algebra imposes the gauge-dependent, algebraic equation of motion $\sA^2=0$ on the background gauge field.

A potential remedy for this situation would be to modify the current $\sG$. Such modifications are constrained by conformal weight and fermionic statistics to take the form
\be\label{ganom5}
\sG\rightarrow \sG + \psi^{\mu}\psi^{\nu}\psi^{\rho}\,H_{\mu\nu\rho} + \psi^{\mu}\,C_{\mu\nu}\,\partial X^{\nu}\,,
\ee
for some $H_{\mu\nu\rho}$ and $C_{\mu\nu}$ that depend only on $X$. Corrections proportional to $H_{\mu\nu\rho}$ are reminiscent of the standard Green-Schwarz mechanism, but it is easy to see that they cannot remove the gauge-dependent triple pole \eqref{ganom4}. While we have been unable to use corrections proportional to $C_{\mu\nu}$ to remove the gauge anomaly completely, there are some suggestive hints which emerge.

Consider the modification of $\sG$ given by
\be\label{gmod1}
\sG\rightarrow \sG +\frac{k}{2}\,\psi^{\mu}\, \sA_{\mu}\sA_{\nu}\,\partial X^{\nu}\,.
\ee
It is worth noting that analogous terms appear in descriptions of anomaly cancellation for the Green-Schwarz formalism of the heterotic string~\cite{Hull:1986xn}, although the precise connection (if any) is unclear. This modification removes the triple pole \eqref{ganom4} entirely, and leads to $O(k)$ modifications of $\sH$:
\begin{multline}\label{hmod1}
 \sH\rightarrow \sH + k \left(\Pi^{\mu}\,\sA_{\mu}\sA_{\nu}\,\partial X^{\nu}-\psi^{\mu}\psi^{\nu}\,\sA_{\mu}\partial \sA_{\nu}-\sA^{2}\,\sA_{\mu}\partial X^{\mu}\,j  -\frac{1}{2} \partial \sA_{\mu}\,\partial \sA^{\mu} \right. \\
 \left. +\frac{1}{2} \partial\left(\partial_{\mu}(\sA^{\mu} \sA_{\nu})\partial X^{\nu}\right) - \psi^{\mu}\psi^{\nu}\,\partial_{\mu}(\sA_{\nu}\sA_{\sigma})\partial X^{\sigma}\right) +\frac{k^2}{4}\,\sA^2\,(\sA_{\mu}\partial X^{\mu})^2\,.
\end{multline}
The modified $\sG$ and $\sH$ are worldsheet conformal primaries of the appropriate dimension, although both are still gauge-dependent. 

While these modifications do not remove all gauge-dependence, the OPE of $\sG$ with $\sH$ takes a remarkably simple form. Indeed, on the support of the Maxwell equations for the background gauge field, one finds the structure:
\begin{multline}\label{ganom6}
 \sG(z)\,\sH(w)\sim k\left[\frac{\psi^\mu \psi^{\nu}\psi^{\rho}}{(z-w)^2} (\cdots) + \frac{\psi^{\mu}\psi^{\nu}\psi^{\rho} \partial X^{\sigma}}{z-w} (\cdots) + \frac{\psi^{\mu}\psi^{\nu}\partial\psi^{\rho}}{z-w} (\cdots) \right. \\ 
 \left. +\frac{\psi^{\mu}\,\partial X^{\nu}}{z-w} (\cdots) + \frac{\psi^{\mu}\partial X^{\nu} \partial X^{\rho}}{z-w} (\cdots) + \frac{\partial \psi^{\mu} \partial X^{\nu}}{z-w} (\cdots)\right]\,,
\end{multline}
where the $(\cdots)$ stand for gauge-dependent tensors constructed from the background gauge field. Although far from satisfactory, these modifications do kill all $O(k^2)$ contributions to the OPE, as well as terms proportional to $\Pi_{\mu}$ and $j$ -- all of which occur for generic modifications \eqref{ganom5}. Furthermore, many of the terms appearing in \eqref{ganom6} actually have a surprisingly simple form; for instance, the double pole is
\be\label{ganom7}
\sG(z)\,\sH(w)\sim -\frac{3k}{2}\,\frac{\psi^\mu \psi^{\nu}\psi^{\rho}}{(z-w)^2}\,\sA_{\mu} F_{\nu\rho} + \frac{1}{z-w} (\cdots)\,,
\ee
namely, a Chern-Simons term. 

We expect that a full resolution of the gauge anomaly for the heterotic ambitwistor string requires a full knowledge of its coupling to other background fields. These fields obey higher-derivative equations of motion, and there are additional fields (such as a massless 3-form) which do not appear in the standard heterotic string~\cite{Mason:2013sva,Azevedo:2017lkz,Azevedo:2017yjy}. A recent description of the effective free theory on spacetime for these fields should be a useful tool in this regard~\cite{Berkovits:2018jvm}. We hope that future work will lead to a full resolution of these issues.

\newpage

\section{Vertex operators for heterotic and type II am\-bi\-twis\-tor strings in curved backgrounds}\label{C3.VertexOps}

Ambitwistor strings~\cite{Mason:2013sva,Berkovits:2013xba} have many surprising properties; while much attention has rightly been paid to their utility for computing scattering amplitudes they can also be defined on non-linear background fields as has been shown in the previous chapter and~\cite{Adamo:2014wea,Adamo:2018hzd}. On such curved backgrounds the ambitwistor string is described by a chiral worldsheet CFT with free OPEs (for details see chapters~\ref{CurvedTypeII} and~\ref{C3.Heterotic}). This allows for many \emph{exact} computations in these backgrounds, in stark contrast to conventional string theories where an expansion in the inverse string tension is needed (cf., \cite{Fradkin:1985ys,Callan:1985ia,Abouelsaood:1986gd}).

Thus far, only a RNS formalism for the ambitwistor string has been shown to be quantum mechanically consistent at the level of the worldsheet. While pure spinor and Green-Schwarz versions of the ambitwistor string (or deformations thereof) have been defined on curved backgrounds~\cite{Chandia:2015sfa,Chandia:2015xfa,Azevedo:2016zod,Chandia:2016dwr}, it is not clear that they are anomaly-free since only classical worldsheet calculations have been done in these frameworks. In this chapter we study the heterotic and type II ambitwistor strings in the RNS formalism, at the expense of only working with NS-NS backgrounds. These backgrounds will be non-linear, and generic apart from constraints imposed by nilpotency of the BRST operator (i.e., anomaly cancellation): the Yang-Mills equations in the heterotic case and the NS-NS supergravity equations in the type II case.

For each of these models, we construct vertex operators in the $(-1,-1)$ picture for all NS-NS perturbations of the backgrounds and investigate the constraints imposed on the operators by BRST closure. In the heterotic model we consider only one such vertex operator whose BRST closure imposes the linearised gluon equations of motion (as well as gauge-fixing conditions) on the perturbation around a Yang-Mills background. In the type II model we consider three vertex operator structures, corresponding to symmetric rank-two tensor, skew-symmetric rank-two tensor, and scalar perturbations. With a background metric (obeying the vacuum Einstein equations), BRST closure fixes the two tensorial perturbations to be a linearised graviton and B-field respectively. On a general NS-NS background (composed of a non-linear metric, B-field and dilaton), the three structures are combined into a single vertex operator, whose BRST closure imposes the linearised supergravity equations of motion on the perturbations. 

We comment on the descent procedure for obtaining vertex operators in picture number zero, as well as the prospects for obtaining integrated vertex operators. We also mention some unresolved issues regarding the GSO projection in curved background fields. The work presented in this chapter has first been published in~\cite{Adamo:2018ege}.

\subsection{Heterotic ambitwistor string}

As a warm up we first describe the vertex operator for a gluon in the heterotic ambitwistor string on a generic Yang-Mills background field since the calculations here are mostly straightforward. This model was defined in a gauge background in chapter~\ref{C3.Heterotic} and~\cite{Adamo:2018hzd}. As before we take the formal limit $k\rightarrow 0$ to decouple gravitational degrees of freedom from the model~\cite{Berkovits:2004jj,Adamo:2018hzd}.


\subsubsection{Gluon vertex operator}
\label{glVO}

Our goal is now to describe perturbations of the Yang-Mills background $A_{\mu}^{\sa}$ at the level of vertex operators in the worldsheet CFT. Let $a_\mu^\sa(X)$ be a perturbation of the background. A natural ansatz for an associated vertex operator in the `fixed' picture (i.e., picture number $-1$) is
\begin{align}
\label{CurvedGluonVO}
V = c\bar{c}\, \delta(\gamma)\, \psi^\mu\, a_\mu^\mathsf{a}\, j^\mathsf{a}\,.
\end{align}
This is an admissible vertex operator if it is annihilated by the BRST operator $Q$. Since $V$ is a conformal primary of spin zero, the only interesting contributions to $QV$ come from higher poles in OPEs with the currents \eqref{Gcurr} and \eqref{Hcurr}. Using the free OPEs \eqref{OPEs}, it is straightforward to show that
\begin{align}
\mathsf{G}(z)V(w)\sim - \frac{c\bar{c}\, \delta(\gamma)\,D^\mu a_\mu^\mathsf{a}\,j^{\sa}(w)}{(z-w)^2}+\cdots\,,
\end{align}
and
\begin{align}
\mathsf{H}(z)V(w)\sim \frac{c\bar{c}\, \delta(\gamma)\,\psi^\nu j^{\sa}}{(z-w)^2}\left(D^\mu D_\mu a_\nu^\mathsf{a} + 2  f^{\mathsf{a}\mathsf{b}\mathsf{c}} a^{\mathsf{b} \mu} F^\mathsf{c}_{\mu \nu}\right)(w) + \cdots\,,
\end{align}
where the $+\cdots$ represent single pole terms in the OPE which will not contribute to the action of the BRST charge. 

In particular, these OPEs indicate that
\be\label{gQV}
QV=c\bar{c}\,\delta(\gamma)\,j^{\sa}\left[\partial\bar{c}\,\psi^{\nu}\left(D^{\mu}D_{\mu} a_{\nu}^{\sa}+2 f^{\mathsf{abc}}\,a^{\mathsf{b}\mu}\,F^{\mathsf{c}}_{\mu\nu}\right)-\partial\gamma\,D^{\mu}a_{\mu}^{\sa}\right]\,.
\ee
So requiring $QV=0$ imposes the Lorenz gauge condition ($D^{\mu}a_{\mu}^{\sa}=0$) as well as the linearised Yang-Mills equations
\be\label{linYM}
D^{\mu}D_{\mu} a_{\nu}^{\sa}+2 f^{\mathsf{abc}}\,a^{\mathsf{b}\mu}\,F^{\mathsf{c}}_{\mu\nu}=0\,
\ee
on the perturbation. In other words, the vertex operator lies in the BRST cohomology if and only if $a_{\mu}^{\sa}$ describes an on-shell gluon fluctuation on the non-linear Yang-Mills background.

The standard descent procedure (cf., \cite{Friedan:1985ge,Verlinde:1987sd,Witten:2012bh}) can be used to obtain the gluon vertex operator in zero picture number. To do this, we simply use the standard picture changing operator $\delta(\beta)\mathsf{G}$ to get
\begin{align}
 c\tilde{c}U(w) & =\lim_{z\rightarrow w}\delta(\beta)\mathsf{G}(z)\,V(w) \\
 & =c\tilde{c}\left(\Psi^\mu\Psi^\nu D_\nu a^\mathsf{a}_\mu j^\mathsf{a}+(\Pi^\mu-\mathsf{A}^{\mu\mathsf{a}}j^{\mathsf{a}}) a_\mu^{\mathsf{b}}j^{\mathsf{b}}-f^{\mathsf{abc}}a_{\mu}^{\mathsf{b}}\,j^{\mathsf{c}}\,\partial A^{\mu\mathsf{a}}\right)(w)\,. \label{Dgluon}
\end{align}
An equivalent way to derive $U(w)$ is by linearising the current $\mathsf{H}$ around a Yang-Mills background, keeping in mind that the perturbation $a_{\mu}^{\sa}$ obeys the Lorenz gauge condition.

Further descent into an integrated vertex operator using the $b$-ghost and the stress-energy tensor can be carried out as in the usual string. How to perform the descent using the $\bar{b}$-ghost and $\mathsf{H}$ current remains an open question, although it is well-known how to do so in a flat background~\cite{Mason:2013sva,Adamo:2013tsa,Ohmori:2015sha}.


\subsection{Type II ambitwistor string}\label{TypeIIVO}

We now move on to define vertex operators for the type II ambitwistor string on a curved NS-NS background composed of a metric $g_{\mu\nu}$, B-field $B_{\mu\nu}$ and dilaton $\Phi$, which was introduced in~\cite{Adamo:2014wea} and described in chapter~\ref{CurvedTypeII}.


\subsubsection{Graviton vertex operator}

To begin, consider the type II model with \emph{only} a background metric $g_{\mu\nu}$ turned on, and let $h_{\mu\nu}(X)$ be a symmetric, traceless perturbation of this metric. A fixed picture vertex operator associated to this perturbation is given by
\begin{align}\label{gravityop}
 V_{h}=c\bar{c}\,\delta(\gamma)\delta(\bar\gamma)\,\cO_{h}=c\bar{c}\,\delta(\gamma)\delta(\bar\gamma)\left(\bar{\psi}_\mu\psi^\nu h^\mu{}_\nu-\frac{1}{2}(\partial g_{\mu\nu})h^{\mu\nu}\right).
\end{align}
Note that this contains a quantum correction term proportional to a worldsheet derivative. While this quantum correction vanishes for flat or certain highly symmetric backgrounds (e.g., a plane wave metric written in Brinkmann coordinates~\cite{Adamo:2017sze}), it plays a crucial role on a general background.

For $V_{h}$ to be an admissible vertex operator, it must be annihilated by the BRST operator \eqref{IIcQ}. Since $V_h$ is a conformal primary of spin 0 on the worldsheet, any potential obstructions to its $Q$-closure arise from OPEs between the operator $\cO_h$ and the currents \eqref{GeneralG}, \eqref{GeneralGbar} and \eqref{GeneralH} with $H_{\mu\nu\rho}=0=\Phi$. One finds:

\begin{align}
 &\mathcal{G}(z)\,\cO_{h}(w)\sim-\frac{\psi^\nu\,\nabla_{\mu} h^{\mu}{}_{\nu}}{(z-w)^2}(w)+\cdots\,, \\
 &\bar{\mathcal{G}}(z)\,\cO_{h}(w)\sim\frac{g^{\rho\sigma}\bar{\psi}_\mu\,\nabla_\rho h^\mu{}_\sigma}{(z-w)^2}(w)+\cdots\,,
\end{align}
and 
\begin{multline}\label{gback1}
 \frac{\mathcal{H}(z)}{2}\,\cO_h(w)\sim \frac{h^{\mu\nu}R_{\mu\nu}}{(z-w)^3}(w)+\frac{\bar\psi_\alpha\psi^\beta}{2\,(z-w)^2}\left(\nabla_{\kappa}\nabla^{\kappa} h^{\alpha}_{\beta} -  2R^{\alpha}{}_{\sigma \gamma \beta} h^{\sigma \gamma} \right. \\
 \left. -R^{\sigma\alpha} h_{\sigma \beta}  -R^{\sigma}{}_{\beta} h^{\alpha}_{\sigma}+2h^\lambda{}_\beta R_\alpha{}_\lambda\right)(w) +\frac{\partial X^\gamma}{(z-w)^2}\left(\frac{1}{2}h_{\mu\nu}\partial_\gamma R^{\mu\nu}\right. \\
 \left.+\frac{1}{4}\partial_\gamma g^{\mu\nu}(\nabla_{\alpha}\nabla^{\alpha} h_{\mu \nu} - 2 R_{\mu \alpha \beta \nu} h^{\alpha \beta} -R^\lambda_{\; \mu} h_{\lambda \nu}  -R^\lambda_{\; \nu} h_{ \mu \lambda})\right)(w)+\cdots\,,
\end{multline}
where the $+\cdots$ stand for terms which do not contribute to the action of the BRST operator. 

Since the background metric obeys the vacuum Einstein equations ($R_{\mu\nu}=0$), these OPEs imply that
\begin{multline}\label{ggraviton}
QV_{h}=c\bar{c}\,\delta(\gamma)\delta(\bar\gamma)\bigg[\partial\gamma\,\bar{\psi}_\mu\,\nabla^{\nu} h^\mu{}_\nu-\partial\bar\gamma\, \psi^\nu\,\nabla_{\mu} h^{\mu}{}_{\nu} \\
 \left.+\frac{\partial\bar{c}\,\bar\psi_\mu\psi^\nu}{2}\left(\nabla_{\alpha}\nabla^{\alpha} h^{\mu}_{\nu} -  2R^{\mu}{}_{\alpha \beta \nu} h^{\alpha \beta}\right) +\frac{\partial\bar{c}\,\partial g^{\mu\nu}}{4}\,\left(\nabla_{\alpha}\nabla^{\alpha} h_{\mu \nu} - 2 R_{\mu \alpha \beta \nu} h^{\alpha \beta}\right)\right]\,.
\end{multline}
Thus, the OPEs between the vertex operator and the currents $\cG$, $\bar{\cG}$ impose the de Donder gauge condition
\be\label{deDonder}
\nabla^{\mu}h_{\mu\nu}=0\,,
\ee
which is consistent with expectations from the flat background case~\cite{Mason:2013sva}. The OPE between the vertex operator and the current $\cH$ leads to the linearised Einstein equation for a metric perturbation on a vacuum Einstein background:
\be\label{linEin}
\nabla_{\alpha}\nabla^{\alpha} h_{\mu \nu} - 2 R_{\mu \alpha \beta \nu} h^{\alpha \beta}=0\,.
\ee
In other words, requiring $QV_{h}=0$ imposes precisely the physical gauge-fixing and linearised equation of motion for a graviton on the perturbation $h_{\mu\nu}$.

\medskip

What happens when the background B-field and dilaton are switched on? Keeping the form \eqref{gravityop} for the vertex operator, it remains to check the action of the \emph{full} (i.e., with $g_{\mu\nu}$, $H_{\mu\nu\rho}$ and $\Phi$) BRST operator \eqref{IIcQ} on $V_h$. The additional background fields do not change the fact that $QV_{h}$ is governed entirely by the OPEs between $\cO_h$ and the currents \eqref{GeneralG}, \eqref{GeneralGbar} and \eqref{GeneralH}, although these OPEs are now substantially more complicated. One finds that
\begin{align}
 &\mathcal{G}(z)\,\cO_{h}(w)\sim-\frac{\psi^\nu}{(z-w)^2}\left(\nabla_\mu h^\mu{}_\nu-2h^\mu{}_\nu\partial_\mu\Phi\right)+\cdots \,,
\\
 &\bar{\mathcal{G}}(z)\,\cO_{h}(w)\sim\frac{g^{\rho\sigma}\bar{\psi}_\mu}{(z-w)^2}\left(\nabla_\rho h^\mu{}_\sigma-2h^\mu{}_\rho\partial_\sigma\Phi\right)+\cdots\,,
\end{align}
while the OPE between $\cH$ and $\cO_h$ is 
\begin{multline}
 \frac{\cH(z)}{2}\,\cO_{h}(w)\sim \frac{h^{\mu\nu}}{(z-w)^3}\left(R_{\mu\nu}+2\nabla_\mu\nabla_\nu\Phi-\frac{1}{4}H_{\mu\rho\sigma}H_\nu{}^{\rho\sigma}\right) 
\\
 +\frac{\bar\psi_{\alpha}\psi^{\beta}}{(z-w)^2}\left[h^{\lambda}_{\beta}\left(R^{\alpha}{}_{\lambda}+2\nabla^\alpha\nabla_\lambda\Phi-\frac{1}{4}H^{\alpha}{}_{\rho\sigma}H_\lambda{}^{\rho\sigma}\right) + \frac{1}{2}\left(\nabla_{\lambda}\nabla^{\lambda}h^{\alpha}_{\beta}-2R^{\alpha}{}_{\sigma\rho\beta}h^{\sigma\rho}\right. \right. 
\\
-R^{\sigma\alpha}h_{\sigma\beta} -R^{\sigma}{}_{\beta}h^{\alpha}_{\sigma}-h^{\rho}_{\sigma} H_{\beta\rho\kappa}H^{\alpha\sigma\kappa}-2(h^{\alpha}_{\sigma}\nabla_{\beta}\partial^{\sigma}\Phi+h_{\beta\sigma}\nabla^{\alpha}\partial^{\sigma}\Phi+\nabla_{\sigma}h^{\alpha}_{\beta} \partial^{\sigma}\Phi)\Big)\bigg] 
\\
 +\frac{1}{(z-w)^2}\left[\frac{h_{\mu\nu}}{2}\,\partial\!\left(R^{\mu\nu}+2\nabla^\mu\nabla^\nu\Phi-\frac{1}{4}H^{\mu}{}_{\rho\sigma}H^{\nu\rho\sigma}\right)+\frac{\partial g^{\mu\nu}}{4}\left(\nabla_{\lambda}\nabla^{\lambda}h_{\mu\nu}-2R_{\mu\alpha\beta\nu}h^{\alpha\beta} \right.\right. 
\\
 -R^{\lambda}{}_{\mu}h_{\lambda\nu}-R^{\lambda}{}_{\nu}h_{\lambda\mu}-h^{\lambda}_{\sigma}H_{\mu\lambda\alpha}H_{\nu}{}^{\sigma\alpha}-2\left(h_{\mu\sigma}\nabla_{\nu}\partial^{\sigma}\Phi-h_{\nu\sigma}\nabla_{\mu}\partial^{\sigma}\Phi+\nabla_{\sigma}h_{\mu\nu}\partial^{\sigma}\Phi\right)\Big)\bigg] 
\\
 +\frac{\psi^{\rho}\psi^{\sigma}}{2\,(z-w)^2}\left(\nabla_{\nu}h_{\lambda\sigma}\,H_{\rho}{}^{\nu\lambda}+\frac{h^{\alpha\beta}}{2}\,\nabla_{\alpha}H_{\beta\sigma\rho}\right) \qquad\qquad\qquad\qquad\qquad
\\
-\frac{\bar\psi_{\rho}\bar{\psi}_{\sigma}}{2\,(z-w)^2}\left(\nabla_{\nu}h_{\lambda}^{\sigma}\,H^{\rho\nu\lambda}+\frac{h^{\alpha\beta}}{2}\,\nabla_{\alpha}H_{\beta}{}^{\sigma\rho}\right) \qquad\qquad
\\ 
+\cdots\,, \qquad\qquad\qquad\qquad\qquad\qquad\qquad\qquad\qquad
\end{multline}
where all numerators are evaluated at $w$ on the worldsheet, and $+\cdots$ again denotes terms which will not contribute to the action of the BRST operator.

Using the fact that the background fields obey the non-linear equations of motion \eqref{SugraEOM}, this means that
\begin{multline}\label{NSgraviton}
 QV_{h}=c\bar{c}\,\delta(\gamma)\delta(\bar\gamma)\bigg[\partial\gamma\,\bar{\psi}_\mu\,(\nabla^{\nu} h^\mu{}_\nu-2h^{\mu}{}_{\nu}\partial^{\nu}\Phi)-\partial\bar\gamma\, \psi^\nu\,(\nabla_{\mu} h^{\mu}{}_{\nu}-2h^{\mu}{}_{\nu}\partial_{\mu}\Phi) 
\\
+ \frac{\partial\bar{c}}{4}\,\left(2\bar\psi^{\mu}\psi^{\nu}+\partial g^{\mu\nu}\right)\left(\nabla_{\lambda}\nabla^{\lambda}h_{\mu\nu}-2R_{\mu\rho\sigma\nu}h^{\rho\sigma}-R^{\lambda}{}_{\mu}h_{\lambda\nu}-R^{\lambda}{}_{\nu}h_{\lambda\mu}\right. 
\\
 \left.-h^{\lambda}_{\sigma}H_{\mu\lambda\alpha}H_{\nu}{}^{\sigma\alpha}-2\left(h_{\mu\sigma}\nabla_{\nu}\partial^{\sigma}\Phi+h_{\nu\sigma}\nabla_{\mu}\partial^{\sigma}\Phi+\nabla_{\sigma}h_{\mu\nu}\partial^{\sigma}\Phi\right)\right) 
\\
 +\frac{\partial\bar{c}}{2}\left(\psi^{\mu}\psi^{\nu}-\bar{\psi}^{\mu}\bar{\psi}^{\nu}\right)\left(\nabla_{\rho}h_{\lambda\nu}\,H_{\mu}{}^{\rho\lambda}-\frac{h^{\rho\sigma}}{2}\,\nabla_{\rho}H_{\sigma\mu\nu}\right)\bigg]\,,
\end{multline}
where indices are raised and lowered with the background metric. The requirement $QV_{h}=0$ therefore imposes the generalised de Donder gauge condition
\be\label{gdeDonder}
\nabla^{\mu}h_{\mu\nu}=2\,h_{\mu\nu}\partial^{\mu}\Phi\,,
\ee
as well as the linearised equation of motion
\begin{multline}\label{linNSEin}
\nabla_{\lambda}\nabla^{\lambda}h_{\mu\nu}-2R_{\mu\rho\sigma\nu}h^{\rho\sigma}-R^{\lambda}{}_{\mu}h_{\lambda\nu}-R^{\lambda}{}_{\nu}h_{\lambda\mu}-h^{\lambda}_{\sigma}H_{\mu\lambda\alpha}H_{\nu}{}^{\sigma\alpha} \\
-2\left(h_{\mu\sigma}\nabla_{\nu}\partial^{\sigma}\Phi+h_{\nu\sigma}\nabla_{\mu}\partial^{\sigma}\Phi+\nabla_{\sigma}h_{\mu\nu}\partial^{\sigma}\Phi\right)=0\,.
\end{multline}
As desired, this is precisely the linearisation of the symmetric tensor equation from \eqref{SugraEOM} for a metric perturbation, for details of the derivation of this equation see appendix~\ref{LinSugraG}.

However, we also obtain an \emph{antisymmetric} constraint from the last line of \eqref{NSgraviton}:
\be\label{skewgrav}
\nabla_{\rho}h_{\lambda[\nu}\,H_{\mu]}{}^{\rho\lambda}-\frac{h^{\rho\sigma}}{2}\,\nabla_{\rho}H_{\sigma\mu\nu}=0\,.
\ee
From a spacetime perspective, this is unexpected: given a symmetric, traceless perturbation $h_{\mu\nu}$, one only expects to obtain the symmetric equation of motion \eqref{linNSEin}. The antisymmetric equation \eqref{skewgrav} arises because the background fields $\{g,H,\Phi\}$ are still treated as fluctuating quantum fields by the worldsheet theory. Indeed, these background fields are functionals of the worldsheet field $X^{\mu}(z)$, which is a full quantum field contributing to all OPEs. 

This means that the perturbation $h_{\mu\nu}$ can backreact on the background geometry, leading to additional constraints. In particular, a metric perturbation sources terms in the antisymmetric equation of motion for the background fields \eqref{SugraEOM}\footnote{The metric perturbation can also source a scalar constraint, but it is easy to see that this vanishes on the support of the background equations of motion, see appendix~\ref{LinSugraG}.}. At the level of a spacetime variational problem, this corresponds to evaluating the spacetime action on $\{g+h,H,\Phi\}$ and varying it with respect to all these fields. Projecting the resulting equations of motion onto the parts linear in $h$ will yield the symmetric equation \eqref{linNSEin} and the antisymmetric equation \eqref{skewgrav} as well as the trivial scalar constraint.

Consequently, the graviton vertex operator only makes sense in the BRST cohomology in the presence of a background metric. When a full NS-NS background is turned on, $QV_h=0$ leads to the physical gauge-fixing condition \eqref{gdeDonder} and correct equation of motion \eqref{linNSEin}, but also an additional backreaction constraint \eqref{skewgrav}. We will see the resolution of this issue in chapter~\ref{NSNSvertex}.


\subsubsection{B-field vertex operator}

Consider a B-field perturbation $b_{\mu\nu}(X)$, which is anti-symmetric ($b_{\mu\nu}=b_{[\mu\nu]}$). As in the graviton case, initially we seek a vertex operator to describe this perturbation on a background metric $g_{\mu\nu}$ alone. Using consistency with the flat space GSO projection as a guide, the candidate vertex operator in the fixed picture is:
\be\label{0bvertex}
V_{b}^{(0)}=\frac{c\bar{c}}{2}\,\delta(\gamma)\delta(\bar\gamma)\,\left(\psi^{\mu}\psi^{\nu}\,b_{\mu\nu}-\bar{\psi}_{\mu}\bar{\psi}_{\nu}\,b^{\mu\nu}\right)\,.
\ee
It is straightforward to compute the action of the BRST operator $Q$ on $V^{(0)}_{b}$; since the operator is a conformal primary of spin zero with a canonical ghost structure, $QV^{(0)}_b$ is controlled entirely by the OPEs between the terms in brackets in \eqref{0bvertex} and the currents $\cG$, $\bar{\cG}$, $\cH$ (with $H_{\mu\nu\rho}=0=\Phi$).

This leads to
\begin{multline}\label{0bfield}
QV^{(0)}_{b}=c\bar{c}\,\delta(\gamma)\delta(\bar\gamma)\,\bigg[\partial\gamma\,\bar{\psi}_\nu\,\nabla_{\mu} b^{\mu\nu}+\partial\bar\gamma\, \psi^\nu\,\nabla^{\mu} b_{\mu\nu} \\
 +\frac{\partial\bar{c}}{4}\left(\psi^{\mu}\psi^{\nu}-\bar{\psi}^{\mu}\bar{\psi}^{\nu}\right)\left(\nabla_{\lambda}\nabla^{\lambda} b_{\mu\nu} -  2R_{\sigma\mu\nu\rho} b^{\sigma\rho}+2R^{\sigma}{}_{\mu} b_{\nu\sigma} \right)\bigg]\,.
\end{multline}
Using the vacuum Einstein equations for the background, $QV^{(0)}_b=0$ imposes the gauge-fixing constraint
\be\label{0bgauge}
\nabla^{\mu}b_{\mu\nu}=0\,,
\ee
as well as the equation of motion
\be\label{0bEOM}
\nabla_{\lambda}\nabla^{\lambda} b_{\mu\nu} -  2R_{\sigma\mu\nu\rho}\,b^{\sigma\rho}=0\,
\ee
on the perturbation. Sure enough, \eqref{0bEOM} is precisely the linearised equation of motion for a B-field propagating on a vacuum Einstein background.

\medskip

From our experience with the graviton vertex operator, we know that a B-field perturbation in a general NS-NS background will source the linearised scalar and symmetric tensor equations of motion, leading to unwanted constraints on the perturbation. Nevertheless, it is instructive to see how this arises by constructing a vertex operator for the perturbation $b_{\mu\nu}$ with a background metric, B-field and dilaton.

It is easy to see that $V^{(0)}_{b}$ is no longer correct in this case; we claim that it must be supplemented by additional terms with non-standard worldsheet ghost structure. To write these terms down, we must bosonise the worldsheet ghost systems $(\beta, \gamma)$ and $(\bar\beta,\bar\gamma)$~\cite{Friedan:1985ge}. Let $\phi$ be a chiral scalar on the worldsheet, and $(\eta,\xi)$ be a pair of fermions of spin $+1$ and $0$, respectively. These fields have OPEs 

\be\label{bgs}
 \phi(z)\,\phi(w)\sim -\ln(z-w)\,, \qquad \eta(z)\,\xi(w)\sim \frac{1}{z-w},
\ee
and are related to the ghosts $(\beta,\gamma)$ by
\be\label{bgs1}
 \gamma=\eta\, e^{\phi}\,, \qquad \beta=e^{-\phi}\,\partial\xi\,,
\ee
using the fact that an exponential of the chiral scalar $\e^{k\phi}$ has spin $-(k+\frac{k^2}{2})$. An additional copy of each system, $\bar\phi$, $(\bar{\eta},\bar{\xi})$ is introduced (with identical statistics) for the $(\bar\beta,\bar\gamma)$ ghost system.

With these bosonised ghost systems, the B-field vertex operator on a general NS-NS background is given by
\be\label{bvertex}
V_{b}=V^{(0)}_{b}+\cO_{b}^{(1)}+\bar{\cO}_{b}^{(1)}\,,
\ee
where the additional operators are
\begin{align}
\begin{split}
\mathcal{O}_b^{(1)} & = \frac{c \bar{c}}{4} \partial \bar{c} \, \partial \xi \, \e^{-2 \phi} \e^{-\bar{\phi}} \,\psi^\mu H_{\mu\rho\sigma} b^{\rho \sigma}\,,
\\
\bar{\mathcal{O}}_b^{(1)} &= \frac{c \bar{c}}{4} \partial \bar{c} \, \partial \bar{\xi} \, \e^{-2 \bar{\phi}} \e^{-{\phi}} \,\bar{\psi}_\mu H^{\mu\rho\sigma} b_{\rho \sigma}\,.
\end{split} 
\end{align}
The fact that these additional operators are required is perhaps not surprising, since the background B-field couples to the BRST operator in a manner that is distinctly different to the background metric.

We must now check the action of the BRST operator on $V_b$. While $QV_{b}^{(0)}$ was governed entirely by the OPEs between the currents $\cG$, $\bar{\cG}$ and $\cH$, the same is not true of $QV_b$. This is due to the non-standard ghost structure of $\cO_{b}^{(1)}$, $\bar{\cO}^{(1)}_{b}$. For instance, there are now non-trivial OPEs with the structure constant terms in \eqref{IIcQ} that must be accounted for:
\begin{align}
-2 \bar{b} \gamma \bar{\gamma}(z)\, \mathcal{O}^{(1)}_b(w) &\sim \frac{c \bar{c} e^{-\bar{\phi}} \eta }{z-w} \frac{\bar{\psi}_\mu H^{\mu\rho\sigma} b_{\rho \sigma}}{2}+\cdots\,, \label{gauge_cancel1}
\\
-2 \bar{b} \gamma \bar{\gamma} (z) \bar{\cO}^{(1)}_b(w) &\sim \frac{c \bar{c} e^{-{\phi}} \bar{\eta} }{z-w} \frac{{\psi}^\mu H_{\mu\rho\sigma} b^{\rho \sigma}}{2}+\cdots\,, \label{gauge_cancel2}
\end{align}
making use of the general rule
\begin{align}
\e^{\pm\phi}(z)\,\e^{k\phi}(w)= (z-w)^{\mp k}:\e^{\pm\phi}(z)\,\e^{k\phi}(w):
\end{align}
for OPEs between exponentials of the chiral scalar. Note that keeping track of contributions from the expansion of $\e^{\pm\phi}(z)$ is going to be of crucial importance. Equations~\eqref{gauge_cancel1} and~\eqref{gauge_cancel2} cancel algebraic contributions to the OPEs
\begin{align}
\bar{\gamma} \mathcal{G}(z)\, V^{(0)}_b(w) &\sim -\frac{c \bar{c} \e^{-\phi} \bar{\eta}}{z-w} \left( \bar{\psi}_\beta \left(\nabla_\alpha b^{\alpha \beta} - 2b^{\alpha \beta} \partial_\alpha \Phi \right) + \frac{\psi^\mu H_{\mu\rho\sigma} b^{\rho \sigma}}{2}  \right)\,, \label{gauge_cond1}
\\
\gamma \bar{\mathcal{G}}(z)\, V^{(0)}_b(w) &\sim -\frac{c \bar{c} \e^{-\bar{\phi}} {\eta}}{z-w} \left( {\psi}^\beta \left(\nabla^\alpha b_{\alpha \beta} - 2b_{\alpha \beta} \partial^\alpha \Phi \right) + \frac{\bar{\psi}_\mu H^{\mu\rho\sigma} b_{\rho \sigma}}{2}  \right)\,.\label{gauge_cond2}
\end{align}
Similarly, at every stage of this calculation it is crucial to consider all possible contributions from ghosts to the OPEs. Note that contributions from the stress-energy tensor terms in $Q$ remain trivial, since both $\cO_{b}^{(1)}$ and $\bar{\cO}^{(1)}_{b}$ are conformal primaries of spin zero -- despite their non-trivial ghost structure.

The final result of these calculations is
\begin{multline}\label{bfield}
 QV_b= \frac{c \bar{c}}{4} \partial \bar{c}\, \e^{-\phi} \e^{-\bar{\phi}}\bigg[\bar{\psi}_\rho {\psi}^\sigma \left( H^{\mu \alpha \rho} (\d b)_{\mu\alpha\sigma} +  H_{\mu \alpha\sigma} (\d b)^{\mu\alpha\rho}  \right)+ \partial g^{\rho\sigma} \left( H^{\mu \beta}{}_\rho (\d b)_{\mu\beta\sigma} \right) \\
+(\psi^\mu \psi^\nu - \bar{\psi}^\mu \bar{\psi}^\nu)\left(\nabla_{\lambda}\nabla^{\lambda} b_{\mu\nu} -   2R_{\alpha\mu\nu\beta} b^{\alpha\beta} + 2R^\alpha{}_{\mu} b_{\nu \alpha }- 2\partial^\alpha \Phi  \nabla_\alpha b_{\mu\nu} +4 b_{\alpha \mu} \nabla_\nu \partial^\alpha \Phi\right)\bigg] \\
-c \bar{c}\, \e^{-\phi} \bar{\eta}\bar{\psi}_\beta \left(\nabla_\alpha b^{\alpha \beta} - 2b^{\alpha \beta} \partial_\alpha \Phi \right)-c \bar{c}\, \e^{-\bar{\phi}} \eta \psi^\beta \left(\nabla^\alpha b_{\alpha \beta} - 2b_{\alpha \beta} \partial^\alpha \Phi \right) \\
+\frac{c \bar{c}}{12} \partial \bar{c}\, \e^{-\phi}\, \partial \e^{-\bar{\phi}}\,H^{\mu\nu\rho} (\d b)_{\mu\nu\rho}- \frac{c \bar{c}}{12} \partial \bar{c}\, \partial \e^{-\phi}\, \e^{-\bar{\phi}}\,H^{\mu\nu\rho} (\d b)_{\mu\nu\rho}\,,
\end{multline}
where $(\d b)_{\mu\alpha\sigma}=\partial_\mu b_{\alpha\sigma}+\partial_\alpha b_{\sigma\mu}+\partial_\sigma b_{\mu\alpha}$ and all terms proportional to the background equations of motion \eqref{SugraEOM} have been set to zero. As desired, setting $QV_b=0$ enforces the gauge condition
\be\label{bgauge}
\nabla^\mu b_{\mu\nu} = 2 b_{\mu\nu}\, \partial^\mu \Phi\,,
\ee
along with the linearised equation of motion for a B-field perturbation on a NS-NS background as derived in appendix~\ref{LinSugraB}:
\be\label{bEOM}
 \nabla_{\lambda}\nabla^{\lambda} b_{\mu\nu} -2 R_{\rho\mu \nu\sigma}\, b^{\rho\sigma}+2R^\sigma{}_{[\mu} b_{\nu] \sigma} -2 \partial^\sigma \Phi\, \nabla_\sigma b_{\mu\nu} + 4 b_{\sigma [\mu}\, \nabla_{\nu]} \partial^\sigma \Phi=0\,.
\ee
We also obtain additional scalar and symmetric backreaction constraints on the perturbation:
\be\label{sbfield}
H_{\mu}{}^{\rho\sigma}\, (\d b)_{\nu\rho\sigma}=0=H\cdot(\d b)\,.
\ee
So as expected, $V_b$ only makes sense in the BRST cohomology on a purely metric background.

\subsubsection{Dilaton vertex operator}

In usual superstring theory, the form of the dilaton vertex operator~\cite{Kataoka:1990ga} is complicated by the fact that the dilaton couples to the worldsheet action through the Fradkin-Tseytlin term~\cite{Fradkin:1985ys}. A similar mechanism is in play in the ambitwistor string, visible at the level of the BRST charge through the last term in the matter stress-energy tensor \eqref{stress_tensor}. For a scalar perturbation on spacetime $\varphi(X)$, the associated ambitwistor string vertex operator is composed of four terms: 
\begin{align}
\label{DilatonVO}
 V_\varphi =  \cO^{(1)}_\varphi + \bar{\cO}^{(1)}_\varphi + \cO^{(2)}_\varphi + \bar{\cO}^{(2)}_\varphi\,,
\end{align}
where
\begin{align}
\cO^{(1)}_\varphi &= - c \bar{c}\, \partial \bar{c}\, \partial \xi \,   \e^{-2 \phi}\, \e^{- \bar{\phi}}\,{\psi}^\mu \partial_\mu \varphi\,,
\\
\bar{\cO}^{(1)}_\varphi &= - c \bar{c}\, \partial \bar{c}\, \partial \bar{\xi} \, \e^{-2 \bar{\phi}}\, \e^{- {\phi}}\, \bar{\psi}_\mu \partial^\mu \varphi\,,
\\
\cO^{(2)}_\varphi &= 2\, c \bar{c}\, \partial \e^{-\phi}\, \e^{-\bar{\phi}}\, \varphi\,,
\\
\bar{\cO}^{(2)}_\varphi &=  -2\, c \bar{c}\, \e^{-\phi}\,  \partial \e^{-\bar{\phi}}\, \varphi\,.
\end{align}
Note that unlike the graviton and B-field vertex operators, \eqref{DilatonVO} differs in the flat space limit from other formulae appearing in the literature~\cite{Berkovits:2018jvm}. This is due to our use of a complex fermion system for the spin $\frac{1}{2}$ matter fields on the worldsheet, as opposed to the real fermion system used elsewhere.

Unlike the previous cases, not all constituents of $V_{\varphi}$ are conformal primaries. In particular, the operators $\cO^{(2)}_\varphi$ and $\bar{\cO}^{(2)}_\varphi$ are not primary, so when calculating $QV_{\varphi}$ care must be taken to account for contributions from their OPEs with stress tensor terms in the BRST operator. The relevant OPEs are
\begin{align}
\begin{split}
(c T  + bc\partial c)(z)\, \cO^{(2)}_\varphi(w) &\sim - 2\,\frac{c \partial c\, \bar{c}\, \e^{-\phi } \e^{-\bar{\phi} } }{(z-w)^2}\,\varphi+\cdots\,,
\\
(c T  + bc\partial c)(z)\, \bar{\cO}^{(2)}_\varphi(w) &\sim 2\,\frac{c \partial c\, \bar{c}\, \e^{-\phi } \e^{-\bar{\phi} } }{(z-w)^2}\,\varphi+\cdots\,,
\end{split} 
\end{align} 
so the anomalous conformal weight contributions cancel between the two operators. 

The non-trivial ghost structure of all four contributions in \eqref{DilatonVO} necessitates a careful treatment of the ghost contributions to the action of the BRST operator. On a general NS-NS background, the result is
\begin{multline}\label{dilaton}
 QV_{\varphi}=2c\bar{c}\,\partial\bar{c}\left(\partial\e^{-\phi}\,\e^{-\bar{\phi}}-\e^{-\phi}\,\partial\e^{-\bar{\phi}}\right)\left(\nabla_{\mu}\partial^{\mu}\varphi-2\,\partial_{\mu}\Phi\,\partial^{\mu}\varphi\right) \\
 -c\bar{c}\,\partial\bar{c}\,\e^{-\phi}\e^{-\bar\phi}\left[\left(\partial g^{\mu\nu}+2\bar{\psi}^{\mu}\psi^{\nu}\right)\,\nabla_{\mu}\partial_{\nu}\varphi+\frac{1}{2}\left(\psi^{\mu}\psi^{\nu}-\bar{\psi}^{\mu}\bar{\psi}^{\nu}\right)\,H_{\mu\nu\sigma}\,\partial^{\sigma}\varphi\right]\,.
\end{multline}
Requiring $QV_{\varphi}=0$ therefore imposes scalar, symmetric and anti-symmetric equations of motion on the perturbation:
\be\label{scalardilaton}
\nabla_{\mu}\partial^{\mu}\varphi-2\,\partial_{\mu}\Phi\,\partial^{\mu}\varphi=0
\ee
\be\label{tensordilaton}
\nabla_{\mu}\partial_{\nu}\varphi=0\,, \qquad  H_{\mu\nu\sigma}\,\partial^{\sigma}\varphi=0\,.
\ee
As expected, only the scalar equation~\eqref{scalardilaton} is the desired one; the two tensor equations~\eqref{tensordilaton} arise from the backreaction of the scalar perturbation on the metric and B-field sectors of the background as seen in appendix~\ref{LinSugraPhi}.

However, the situation for the dilaton vertex operator is worse than for the graviton or B-field: even with a pure metric background, we still obtain a tensor equation $\nabla_{\mu}\partial_{\nu}\varphi=0$, which over-constrains the perturbation. Although the vertex operator \eqref{DilatonVO} gives the correct scalar equation of motion, its inclusion in the BRST cohomology enforces unphysical constraints on the spectrum.

\subsubsection{NS-NS vertex operator}\label{NSNSvertex}

For each of the graviton, B-field and dilaton vertex operators, we have seen that the associated vertex operator is not in the BRST cohomology of the type II ambitwistor string on a general NS-NS background. While the graviton \eqref{gravityop} and B-field \eqref{0bvertex} vertex operators are BRST-closed on the support of the appropriate linearised field equations on a pure gravity background, the dilaton operator is only BRST-closed on the support of additional, unphysical equations for \emph{any} sector of background fields.

These issues are overcome by combining the graviton, B-field and dilaton vertex operators into a single NS-NS vertex operator, which simultaneously perturbs each sector of the background. Indeed, from the spacetime perspective this is much more natural than exciting a perturbation of one of the fields on its own, since the non-linear equations of motion \eqref{SugraEOM} intertwine all three. This `fat graviton,' sometime expressed heuristically as $h_{\mu\nu}\oplus b_{\mu\nu}\oplus\varphi$, is the natural perturbation of the NS-NS sector of type II supergravity.

The candidate vertex operator is given by summing together each of three vertex operators constructed above:
\be\label{NSvertex}
V_{\mathrm{NS}}=V_{h}+V_{b}+V_{\varphi}\,,
\ee
where $V_{h}$ is given by \eqref{gravityop}, $V_b$ by \eqref{bvertex}, and $V_{\varphi}$ by \eqref{DilatonVO}. Computing $QV_{\mathrm{NS}}$ is straightforward: We simply add together the results for the BRST operator acting on each of the three components, \eqref{NSgraviton}, \eqref{bfield} and \eqref{dilaton}. The distinct ghost structures in the result impose different constraints on the background fields.

From the terms proportional to $c\bar{c} \e^{-\phi}\bar{\eta}$ and $c \bar{c}\e^{-\bar{\phi}}\eta$, we obtain the gauge conditions
\be\label{NSgf}
\nabla^{\mu}h_{\mu\nu}=2\,h_{\mu\nu}\,\partial^{\mu}\Phi\,, \qquad \nabla^{\mu}b_{\mu\nu}=2\,b_{\mu\nu}\,\partial^{\mu}\Phi\,.
\ee
Terms proportional to $c\bar{c}\partial\bar{c} \e^{-\phi}\e^{-\bar\phi}$ encode tensorial equations of motion. The symmetric equation, which appears contracted into $(2\bar\psi^{(\mu}\psi^{\nu)}+\partial g^{\mu\nu})$, is
\begin{multline}\label{NSsym}
 \nabla_{\lambda}\nabla^{\lambda}h_{\mu\nu}-2R_{\mu\rho\sigma\nu}\,h^{\rho\sigma}-2R^{\lambda}{}_{(\mu}\,h_{\nu)\lambda}-h^{\rho}_{\sigma}\,H_{\mu\rho\lambda}H_{\nu}{}^{\sigma\lambda} \\
-4\left(h_{\sigma(\mu}\,\nabla_{\nu)}\partial^{\sigma}\Phi+\frac{1}{2}\nabla_{\sigma}h_{\mu\nu}\,\partial^{\sigma}\Phi\right)+H_{\rho\sigma(\mu}\,(\d b)_{\nu)}{}^{\rho\sigma}-4\nabla_{(\mu}\partial_{\nu)}\varphi=0\,,
\end{multline}
while the anti-symmetric equation, which appears contracted into $(\psi^{\mu}\psi^{\nu}-\bar{\psi}^{\mu}\bar\psi^{\nu})$, is
\begin{multline}\label{NSasym}
 \nabla_{\lambda}\nabla^{\lambda}b_{\mu\nu}-2R_{\rho\mu\nu\sigma}\,b^{\rho\sigma}+2R^{\sigma}{}_{[\mu}\, b_{\nu]\sigma} +4\left(b_{\sigma[\mu}\,\nabla_{\nu]}\partial^{\sigma}\Phi-\frac{1}{2}\nabla_{\sigma}b_{\mu\nu}\,\partial^{\sigma}\Phi\right) \\
 +2\nabla_{\rho}h_{\sigma[\nu}\,H_{\mu]}{}^{\rho\sigma}-h^{\rho\sigma}\,\nabla_{\rho}H_{\sigma\mu\nu}-2 H_{\mu\nu\sigma}\,\partial^{\sigma}\varphi=0\,.
\end{multline}
Finally, a scalar equation of motion 
\be\label{NSscalar}
\nabla_{\mu}\partial^{\mu}\varphi-2\partial_{\mu}\Phi\,\partial^{\mu}\varphi - \frac{H\cdot \d b}{24}=0\,
\ee
is imposed by terms proportional to the ghost structure $c\bar{c}\partial\bar{c}(\e^{-\phi}\partial\e^{-\bar\phi}-\partial\e^{-\phi}\e^{-\bar\phi})$.

Sure enough, equations \eqref{NSgf} are the generalised de Donder gauge conditions for graviton and B-field perturbations, while equations \eqref{NSsym} -- \eqref{NSscalar} are precisely the linearised equations of motion for the NS-NS sector of type II supergravity as obtained in appendix~\ref{LinSugra}. Thus, $V_{\mathrm{NS}}$ is in the BRST cohomology of the type II ambitwistor string if and only if it encodes a physical, on-shell perturbation for the NS-NS sector of supergravity on spacetime.


\subsection{Discussion}
\label{DiscussionVO}

In this chapter, we found vertex operators for the heterotic and type II ambitwistor strings with curved background fields. In the heterotic case, we gave the gluon vertex operator on any Yang-Mills background: BRST closure imposes the physical equations of motion and gauge-fixing constraint on the gluon perturbation. For the type II model things are more subtle. In a pure gravity background, we found graviton and B-field vertex operators which are BRST closed when the appropriate physical constraints are imposed on the perturbations. On a general NS-NS background (composed of a metric, B-field and dilaton), a fully consistent vertex operator is given by simultaneously encoding perturbations to all three sectors. BRST closure then imposes the appropriate physical constraints on these perturbations, given by the linearised equations of motion and a generalised de Donder gauge.

The fact that these vertex operators can be determined \emph{exactly} -- without recourse to any background field expansion -- points to a significant difference between ambitwistor string and ordinary string theory, where such calculations on a general background would be impossible. It should be noted that a generalisation of the vertex operators given here allows for \emph{any} gauge-fixing condition on the perturbations -- the procedure is a straightforward extension of what is done on a flat background~\cite{Berkovits:2018jvm}. The Lorenz or (generalised) de Donder conditions obtained here are, in a sense, the `minimal' such gauge-fixing constraints. 

Of course, one hopes to use these vertex operators to compute physical observables in non-trivial backgrounds. At three-points, this requires knowing the operators in both the fixed (i.e., negative picture number) picture emphasised here, as well as the descended vertex operators (i.e., picture number zero). In the heterotic theory, the descended vertex operator \eqref{Dgluon} is easy to obtain through the standard procedure or linearising the constraint $\mathsf{H}$. 

In the type II case, one can again follow the standard procedure by colliding $V_{\mathrm{NS}}$ with the picture changing operators $\delta(\bar\beta)\cG$ and $\delta(\beta)\bar{\cG}$, respectively. Some terms in the resulting operator will be $Q$-exact and not contribute to correlation functions; these pure gauge contributions can be isolated by applying the picture changing operators in different order, and then comparing the results. Equivalently, the descended vertex operator can be computed by linearising the $\cH$ current \eqref{GeneralH} around the chosen background. 

On a general NS-NS background, the resulting vertex operator is complicated, but in highly symmetric backgrounds (usually those of interest for perturbative calculations) the descended vertex operator can be quite tractable. For instance, the three-point graviton amplitude on a vacuum plane wave spacetime has been computed directly from ambitwistor strings~\cite{Adamo:2017sze}, as we will see in chapter~\ref{C4.Worldsheet}. We expect the descent procedure to be manageable enough for explicit calculation of 3-point functions around other highly symmetric backgrounds.

To obtain genus zero, $n$-point worldsheet correlations functions (for $n>3$), the analogue of descent with respect to the $\cH$ current must be understood. In flat backgrounds, where $\cH^{\mathrm{flat}}=\Pi^2$, this procedure is understood and leads to the appearance of the scattering equations~\cite{Mason:2013sva,Adamo:2013tsa,Ohmori:2015sha}. However, on general backgrounds $\cH$ has complicated $X$-dependence which obstructs a straightforward evaluation of the path integral. In deformations of the ambitwistor string, where $\cH$ has $X$-dependence even in flat backgrounds, it is still not understood how to perform descent with respect to $\cH$~\cite{Azevedo:2017yjy,Jusinskas:2016qjd,Casali:2016atr}. Clearly, a resolution of this issue is required if ambitwistor strings are to be a useful tool in the study of perturbative QFT on curved backgrounds.

Finally, we note that the fate of the GSO projection (which ensures that the spectrum of the type II ambitwistor string is equivalent to that of type II supergravity) in curved space remains unclear. Indeed, in the graviton vertex operator \eqref{gravityop} the term proportional to a worldsheet derivative does not obey the na\"ive GSO projection, but is clearly required to ensure that $QV_h=0$ yields covariant equations. Other terms in the B-field and dilaton vertex operators also na\"ively seem to be in the GSO-odd sector, but dropping them yields non-covariant or unphysical (algebraic and first derivative) equations of motion.

One potential way to address the issue of the GSO projection is to formulate the curved space worldsheet theory with two real fermion systems, rather than the complex fermion system used here. The price to pay is that the action is no longer free and a true background field expansion must be used. OPEs would be calculated order-by-order in perturbation theory, but we expect that calculations of the nilpotency of $Q$ and $Q$-closure of vertex operators will become trivial after a certain low loop order. This follows from the fact that the non-perturbative calculations using the complex fermion model give only a finite number of low order poles in the OPEs.

%% file: chapter4.tex
\chapter{Scattering on plane waves}\label{chapter4}


\section{Results from quantum field theory in curved spacetimes}\label{C4.Spacetime}

 
The double copy is a precise conjecture about how, in a specific class of representations, momentum space formulae for gravity scattering amplitudes are related to those of gauge theory. This is known to be true at tree level. While there is currently no general proof at higher loop orders in perturbation theory, a growing body of evidence suggests that the double copy also holds at loop level, at the time of writing up to 5 loops. The remarkable success of the double copy prescription has led to an oft-repeated slogan in the amplitudes community: Gravity = $(\mbox{Gauge Theory})^2$. Despite the vast array of evidence, the geometric and fully non-linear origins of the double copy remain mysterious. Most clear proofs thus far are expressed in momentum space for perturbations around a flat background.  While first examples of the double copy at the level of classical non-linear solutions in gauge theory and gravity have been explored, this work is restricted to algebraically special solutions and does not probe dynamics in the same way as scattering amplitudes.  


In this chapter, first published in~\cite{Adamo:2017nia}, we address the question as to whether the double copy relationship between gauge theory and gravity holds for perturbation theory on curved backgrounds. To do this, we consider the simplest curved backgrounds for which there is a well-defined notion of S-matrix: \emph{sandwich plane waves}~\cite{Bondi:1958aj}. These are metric or gauge field backgrounds which are flat in the asymptotic past and future in generic directions but contain a compactly supported region of curvature. This curvature can be thought of as a burst of unidirectional radiation (gravitational or electromagnetic) which is turned on and then switched off at some finite retarded times. The possibility of scattering on a plane wave background may seem controversial in light of the fact that such spacetimes are not in general globally hyperbolic~\cite{Penrose:1965rx}. Nevertheless, we will see that the evolution of massless fields is unitary without leakage, so the S-matrix does indeed make sense.


The relationship Gravity =(Yang Mills$)^2$ is already nicely manifest in the underlying gravitational and electromagnetic plane waves, written in Brinkmann coordinates. With coordinates $X^{\mu}=(u,v,x^a)$, $a=1,\ldots , d-2$, the Brinkmann form of the metric is Kerr-Schild, given by 
$$
\d s^2=\d s^2_{\rm flat}  -H_{ab}(u)\,x^a\,x^b\, \d u^2\, , \qquad\mbox{ where } \qquad \d s^2_{\rm flat}=2 \d u\, \d v  - \delta_{ab}\,\d x^a \d x^b\, ,
$$
whereas the corresponding electromagnetic potential is
$$
\sA=F(u)_a\, x^a \d u\, ,
$$
so that the metric perturbation from flat space is naturally a sum of terms of the form $A\odot A$.  Here $H_{ab}(u)$ and $F_a(u)$ are curvatures and are freely prescribable functions of $u$ subject to $H_{ab}$ being trace-free for the Einstein equations to be satisfied (this restriction disappears if a dilaton is allowed).\footnote{Note that this classical double copy differs from that for the more general Kerr-Schild pp-waves considered in~\cite{Monteiro:2014cda}. There, if the Maxwell field is $\phi k_\mu$, the metric is $\d s^{2}_{\mathrm{flat}}+\phi k_\mu k_\nu$ where $k_\mu$ is a null vector and $\phi$ a solution to the transverse wave equation. Such solutions can often be considered to be longitudinal with $\phi$ playing the role of a Coulomb-like source term that is analogous to a propagator and therefore not squared. We consider plane waves with a radiative Maxwell term, so the whole Maxwell field must be squared to obtain a gravitational field.} For a sandwich wave, $H_{ab}$ and $F_{a}$ are supported in some interval $u\in[u_1,u_2]$ so that spacetime and connection are flat for $u\rightarrow\pm\infty$.  
For both types of plane wave we will see that it is possible to find complete sets of polarisation states for in and out momentum eigenstates for linear massless fields of integral spins.

The flat `in' and `out' regions of sandwich plane waves allow us to define the S-matrix. We focus on the special case of 3-point amplitudes; remember that in flat space, this is where the slogan Gravity = $(\mbox{Gauge Theory})^2$ of the double copy is literally~\cite{Kawai:1985xq}:
\begin{equation*}
\cM_{3}^{\mathrm{flat}} = \left(\cA^{\mathrm{flat}}_{3}\right)^2,
\end{equation*}
where $\cM_{3}^{\mathrm{flat}}$ and $\cA^{\mathrm{flat}}_{3}$ are the 3-point gravity and gauge-theory amplitudes in Min\-kow\-ski space, stripped of overall momentum conserving delta functions and coupling constants. Hence, we expect that \emph{if} there is a notion of double copy which holds in curved backgrounds, it should be most easily found at the level of 3-point amplitudes for which propagators are not yet required.

We consider such 3-point amplitudes for scalars, gauge theory and gravity on a gravitational plane wave background, and for charged scalars and gauge theory on a Yang-Mills plane wave background in any number of spacetime dimensions. In each case, the computation reduces to an integral which depends on the background field; it turns out that the \emph{integrand}\footnote{This `tree level integrand' is the equivalent of `stripping off momentum conserving delta functions' in the flat space amplitudes.} of the resulting expression carries sufficient information to determine if there is a double copy. 

We find that the 3-point amplitudes for gluons on a plane wave gauge background and for gravitons on plane wave spacetimes have two parts written symbolically as
$$
\mathcal{\cA}^{\mathrm{pw}}_3=F+C\, , \qquad \mathcal{M}_3^{\mathrm{pw}}=\mathcal{F}^2 - \mathcal{C}\, .
$$
Here, $F$ is precisely the flat spacetime integrand for three gluon scattering, whereas $\mathcal{F}$ is the 3-gluon integrand on the gravitational plane wave background.  Thus, there is a correction term between the square of the gluon 3-point amplitude and the graviton 3-point amplitude on a plane wave metric. The flat space $F$ can be mapped to $\mathcal{F}$ after some replacements of momenta and polarisation vectors by their curved (and non-constant) counterparts. These replacements are non-local on spacetime and are fixed by finding solutions to the Hamilton-Jacobi equations that allow one to bring momentum eigenstates into the interior of spacetime from future or past infinity in the curved case.  That it is non-local on a curved spacetime is not a surprise as the double copy is only  expressed locally  on momentum space.


The correction terms $C$ and $\mathcal{C}$ arise from the `tails' formed by the linearised free fields backscattering off the background. Scalar waves propagate cleanly on a plane wave background subject to Huygens' principle~\cite{Friedlander:1975eqa}, but spin one and spin two do not~\cite{Mason:1989}. The tails of momentum eigenstates in the past pick up terms encoding the `memory' of the field through which they have passed (i.e., the integral of the field strength in the electromagnetic case). Remarkably, we find that $C^{2}\rightarrow\mathcal{C}$ with an extension of the same replacements used to relate $F$ and $\mathcal{F}$. 

Define $\widetilde \cA_{3}=F-C$ to be the gluon 3-point integrand on a gauge background with flipped sign (or colour charge) for the background gauge field, and let $\rho$ to be the replacement maps from flat to curved kinematics and gauge to gravitational background fields.  Then our double copy can be written as
$$
\cM_3 =\rho(\cA_3\widetilde\cA_3).
$$
This is strong evidence that a notion of double copy persists more generally in the presence of background curvature.

Our formulae therefore also allow a study of the\emph{ memory effect} for plane waves on the amplitude. The key ingredient in the integrand is a vielbein whose non-trivial change from past to future exemplifies the memory effect~\cite{Braginsky:1986ia,Braginsky:1987,Ludvigsen:1989kg}, which has been studied in detail for sandwich plane waves (e.g., \cite{Zhang:2017rno,Zhang:2017geq}). For a charged field on a gauge background, it gives a momentum shift from past to future infinity proportional to the integral of the field.  On a gravitational background, the linear planes that are wave fronts of a standard momentum eigenstate in the past become  diverging quartic surfaces, \emph{Dupin cyclides}, in the future \cite{Friedlander:1975eqa}.  This memory effect will also give rise to new infrared divergences that have been studied in the case of a charged field on an electromagnetic plane wave background \cite{Dinu:2012tj,Ilderton:2012qe}.

\medskip

Non-linear plane wave backgrounds for both gravity and gauge theory have been reviewed in chapter~\ref{Background}. Free fields on these backgrounds are constructed in chapter~\ref{FF}, where we also confirm that (for scalars, gauge theory and gravity) the S-matrix for these states is well-defined in the sense that scattering is unitary and there is no particle creation. We close this chapter with a brief discussion of Huygens' principle and tails. Chapter~\ref{GravB} contains the calculation of 3-point amplitudes and integrands for scalars, gauge theory and gravity on the gravitational plane wave background; chapter~\ref{GaugeB} contains the analogous calculations for charged scalars and Yang-Mills theory on a background plane wave gauge field. In chapter~\ref{TDC}, these two calculations are mapped onto each other; this map defines the double copy for 3-point amplitudes on plane wave backgrounds. We also show how the gauge theory 3-point functions on the two backgrounds are related by a double copy map which acts only on the background. Chapter~\ref{Discuss} concludes. In appendix~\ref{Impulse}, we provide explicit amplitude formulae for the special case of the impulsive plane wave background. Appendix~\ref{S-matrix-integrands} contains the operational definitions of tree level amplitude and integrand used throughout the chapter.


\subsection{Free fields on plane wave backgrounds and inner products}
\label{FF}

Amplitudes in flat space are functionals of free fields and are usually expressed as functions of momenta after being evaluated on momentum eigenstates. In curved space, such solutions are not so obviously available and it is here that we must use the special structure of plane waves. Friedlander showed that Huygens' principle remains valid for the scalar wave equation in plane wave spacetimes: there exist solutions with delta-function support on null hypersurfaces through every null direction~\cite{Friedlander:1975eqa}. These null hypersurfaces are level surfaces of solutions to the Hamilton-Jacobi equation, which provide curved space analogues of the function $k\cdot X$ for null vectors $k$ in Minkowski space. 

Such functions provide analogues of momentum eigenstates, and also lead to integral formulae for general solutions to the wave equation~\cite{Ward:1987ws}. Generalising~\cite{Mason:1989}, we can raise the spin to obtain free fields of spin one and two with arbitrary polarisations, but Huygens' principle no longer holds and tails appear. Furthermore, a consequence of the memory effect will be that, unlike in flat spacetime, a momentum eigenstate in the past will not evolve into one in the future. Nevertheless,  we can show that, despite the lack of global hyperbolicity of plane waves \cite{Penrose:1965rx}, the scattering problem is well-defined on a plane wave background, featuring unitary evolution without leakage or particle creation. 


\subsubsection{Scalar wave equation}

The plane progressing waves of Friedlander are obtained from solutions to the Ha\-mil\-ton-Jacobi equation for null geodesics 
$$
g^{\mu\nu}(\p_\mu\phi)(\p_\nu \phi)=0\, ,
$$
such that arbitrary functions of $\phi$ satisfy the wave equation (when multiplied by a fixed pre-factor). Solutions are most easily obtained in Einstein-Rosen coordinates where they can be separated using the explicit symmetries leading to 
$$
\phi_k=k_0\, v+ k_i\,y^i +\frac{k_ik_jF^{ij}(U)}{2\,k_0}\, , 
$$
where $(k_0,k_i)$ are constants and $F^{ij}=\int \gamma^{ij}(s)\d s$ as in \eqref{Fij}. The wave equation in Einstein-Rosen coordinates is
\be\label{sweqER}
\frac{1}{\sqrt{-|g|}}\partial_{\mu}\left(\sqrt{-|g|}\,g^{\mu\nu}\,\partial_{\nu}\,\Phi\right)=\left(2\partial_{U}\,\partial_{V} +(\p_U \sqrt{\gamma})\p_V - \gamma^{ij}\partial_{i}\,\partial_{j}\right)\Phi=0\,,
\ee
and it can be seen directly that this is solved by~\cite{Friedlander:1975eqa, Ward:1987ws}
\be\label{scalsol}
\Phi(X)=\Omega(U)\,\e^{\im\,\phi_{k}}\,, \qquad \Omega(U):=|\gamma^{-1}(U)|^{1/4} = |E(u)|^{-\frac{1}{2}}\,,
\ee


In Brinkmann coordinates, the wave equation is 
\be\label{sweq}
\left(2\partial_{u}\,\partial_{v}+H(u,\mathbf{x})\,\partial^{2}_{v} - \partial_{a}\,\partial^{a}\right)\Phi=0\,,
\ee
and of course this is solved by the same $\Phi$. Using \eqref{diffeo1}, it can be expressed in Brinkmann coordinates as: 
\be\label{phi}
\phi_{k}:= \frac{k_{0}}{2}\sigma_{ab}\,x^{a}x^{b}+k_{i}E^{i}_{a}\,x^{a} + k_{0}\,v + \frac{k_{i}\,k_{j}}{2\,k_{0}} F^{ij}\,,
\ee
with $F^{ij}(u)$ and $(k_0,k_i)$ as before, and $\sigma_{ab}=\dot{E}^{i}_{a}\,E_{b\,i}$ the deformation tensor defined by \eqref{shear}. The natural momentum associated with $\phi_{k}$ is:
\begin{multline}\label{momentum}
K_{\mu}\,\d X^\mu := \d\phi_{k}=\\
 k_0\,\d v
 +\left( \frac{k_0}{2}\,\dot{\sigma}_{bc}\,x^{b}x^{c}+k_{i}\dot{E}^{i}_{b}x^{b}+\frac{k_{i}k_{j}}{2k_0}\gamma^{ij}\right)\d u+(k_{i}E^{i}_{a}+k_{0}\,\sigma_{ab}x^{b})\d x^a\,.
\end{multline}
Although $K_{\mu}$ is a $(u,x^a)$-dependent generalisation of the constant momentum familiar from flat space, it is nevertheless null by construction from the Hamilton-Jacobi equation. To see this explicitly, note that $\dot{\sigma}_{bc}=\dot{E}^{i}_{b}\dot{E}_{c\,i}-H_{bc}$.

The solutions $\Phi=\Omega\e^{\im\phi_k}$ clearly reduce to on-shell momentum eigenstates when the background is Minkowski space, and hence can be chosen to do so in one or the other asymptotic region. We can use this to characterise in and out scattering states in terms of $\phi_{k}$: An in state $\Phi^{-}$ is one which looks like a plane wave $\e^{\im k\cdot X}$ in the in-region ($u<u_1$), while an out state $\Phi^{+}$ looks like a plane wave in the out-region ($u>u_2$). This comes down to requiring the vielbein to become trivial in the past or the future:
\be\label{vbbc}
\lim_{u\rightarrow\pm\infty}E^{a\,\pm}_{i}(u) = \delta^{a}_{i}\,.
\ee
In terms of the solution to the Hamilton-Jacobi equations, $\phi_k$, the distinction becomes:
\be\label{phibc}
\phi^{-}_{k}|_{\mathrm{in}}= k_{0}\,v+k_{i}\delta^{i}_{a}\,x^{a}+u\,\delta^{ij}\, \frac{k_{i}k_{j}}{2k_0}=\phi^{+}_{k}|_{\mathrm{out}}\,.
\ee
The positive frequency condition on these states is simply that $k_{0}\geq0$.

\medskip

Even at the level of the free theory, some interesting facts about the S-matrix on a plane wave spacetime can be derived by making use of the natural inner product between two solutions to the free equation of motion. This uses complex conjugation to turn the standard symplectic form on the space of solutions of the wave equation into an inner product:
\be\label{sip1}
\left\la \Phi_1 | \Phi_{2}\right\ra = \im \int_{\Sigma} \left(\Phi_{1}\wedge * \d \bar{\Phi}_{2} - \bar{\Phi}_{2}\wedge * \d\Phi_{1}\right)\,,
\ee
where $\Sigma$ is an arbitrary Cauchy surface. Plane wave spacetimes do not admit a Cauchy hypersurface~\cite{Penrose:1965rx}, but one can instead choose the foliation by hypersurfaces $\Sigma_u$ of constant $u$. In this case, the inner product gives:
\be\label{sip2}
\left\la \Phi_1 | \Phi_{2}\right\ra = \im \int_{\Sigma_{u}} \d v\,\d^{d-2}x\,\left(\Phi_{1}\,\partial_{v}\bar{\Phi}_{2} - \bar{\Phi}_{2}\,\partial_{v}\Phi_{1}\right)\,,   
\ee
evaluated at some fixed $u$. 

Consider the inner product between two positive frequency in states, say $\Phi^{-}_{1}$ and $\Phi^{-}_{2}$ with constant momentum components $\{k_{0},k_i\}$ and $\{l_{0},l_i\}$ respectively. Using \eqref{sip2}, this gives
\be\label{sbog1}
\la\Phi^{-}_{1}|\Phi^{-}_{2}\ra=2\,k_{0}\,\delta(k_{0}-l_{0})\,\delta^{d-2}(k_{i}-l_{i})\,,
\ee
with all $u$-dependence dropping out. As desired, the evolution problem underlying the scattering theory is unitary, since there is no `leakage' of momentum -- at any value of $u$ -- between the two in states.

Similarly, the inner product between a positive frequency in state and a negative frequency out state (namely $\la \Phi^{+}_{1}|\bar{\Phi}^{-}_{2}\ra$) encodes the presence of `particle creation' in the plane wave background. Without loss of generality, the inner product can be evaluated at $u=0>u_{2}$, leading to:
\begin{multline}\label{sbog2}
\left\la \Phi^{+}_{1}|\bar{\Phi}^{-}_{2}\right\ra= \delta(k_{0}+l_{0})\,(k_{0}-l_{0})\,\Omega^{-}(0)\int \d^{d-2}x\,\exp\left[\im\left(\frac{l_{0}}{2}\sigma_{ab}^{-}(0)\,x^{a}x^{b}\right.\right. \\
\left.\left.+(k_{a}+l_{i}E^{-\,i}_{a}(0))x^{a}+\frac{l_i l_j}{2l_0}F^{ij}_{-}(0)\right)\right]\,.
\end{multline}
However, the assumption of positive frequency means that $k_{0}+l_{0}\geq0$, so on the support of the overall delta function this inner product vanishes:
\be\label{sbog2*}
\left\la \Phi^{+}_{1}|\bar{\Phi}^{-}_{2}\right\ra=0\,,
\ee
confirming the well-known result that there is no particle creation for scalar QFT in plane wave spacetimes~\cite{Gibbons:1975jb,Garriga:1990dp}. Equivalently: positive frequency in states do not develop a negative frequency part in the out-region.

The final independent inner product is between positive frequency in and out states, $\la\Phi^{+}_{1}|\Phi^{-}_{2}\ra$. This quantity encodes the amplitude for in-to-out scattering in the plane wave spacetime~\cite{Garriga:1990dp}. The inner product can again be evaluated at $u=0$:
\begin{multline}\label{sbog3}
\la\Phi^{+}_{1}|\Phi^{-}_{2}\ra=2\,k_{0}\,\delta(k_{0}-l_{0})\,\e^{-\im{s}_{l}}\,\Omega^{-}(0) \\
\times \int \d^{d-2}x\,\exp\left[\im\left((k_{a}-l_{i} E^{-\,i}_{a}(0))\,x^{a} -\frac{l_0}{2}\sigma_{ab}^{-}(0)\,x^{a}x^{b}\right)\right]\,,
\end{multline}
where the (constant) phase ${s}_l$ is defined as
\begin{equation*}
{s}_{l}:=\frac{l_i\, l_j}{2l_0}F^{ij}_{-}(0)\,.
\end{equation*}
Now, by \eqref{newvb} it follows that
\be\label{vout}
E^{-}_{ia}(u)=u\,b_{ia}+c_{ia}\,, \quad \forall u>u_2\,,
\ee
where $b$, $c$ are constant, invertible $(d-2)\times(d-2)$ matrices. This leaves a Gaussian integral to do in \eqref{sbog3}, with the result:
\be\label{bog3*}
\la\Phi^{+}_{1}|\Phi^{-}_{2}\ra=2\,k_{0}\left(\frac{2\pi}{\im\, l_{0}}\right)^{\frac{d-2}{2}}\,\delta(k_{0}-l_{0})\,\frac{\e^{-\im({s}_{l}+{r}_{k,l})}}{\sqrt{|b|}}\,,
\ee
after using the fact that $\Omega^{-}(0)=\sqrt{|c^{-1}|}$ and defining another phase
\begin{equation*}
{r}_{k,l}:=-\frac{1}{2 l_{0}}(k_{a}-l_{i}c^{i}_{a})\, c^{ak} (b^{-1})^{b}_{k}\,(k_{b}-l_{j}c^{j}_{b})\,.
\end{equation*}
As expected, this matches the result in the literature~\cite{Garriga:1990dp}.


\subsubsection{Spin one}

The  action for free gauge fields propagating on a plane wave spacetime is
\be\label{fYM1}
S^{\mathrm{free}}[A]=\frac{1}{g^{2}}\int_{M}\d u\,\d v\,\d^{d-2}x\, \tr\!\left(\nabla_{[\mu}\,A_{\nu]}\,\nabla^{\mu}\, A^{\nu}\right)\,,
\ee
where $A_{\mu}$ is the gauge field and $\nabla$ the Levi-Civita connection. We will see that on a plane wave it is consistent to simultaneously impose both a Lorenz gauge $\nabla_{\mu}A^{\mu}=0$ and a light-cone gauge $A_{v}=0$, since $\p_v$ is Killing.  With this, the linearised equations of motion for the gauge connection are
\be\label{ymeom}
g^{\rho\sigma}\nabla_{\rho}\nabla_{\sigma} A_{\mu}=0\,, \qquad \partial_{\mu} A^{\mu}=0=A_{v}\,.
\ee
These can be solved using the $d-2$ spin-raising operators
\be\label{hraise}
\mathcal{R}^{a}:=\d u\,\delta^{ab}\frac{\partial}{\partial x^{b}} +\d x^{a}\,\frac{\partial}{\partial v}\,,
\ee
where the free index labels different possible polarisation states. As tensors, the $\mathcal{R}^a$ are covariantly constant. Acting on a solution to the wave equation, $\Phi$, it is easily checked that $\mathcal{R}^a\Phi$ satisfies \eqref{ymeom}, so $\mathcal{R}^a$ is naturally a spin-raising operator (this generalises the four-dimensional, twistorial approach in~\cite{Mason:1989}).  Thus with $\Phi$ the scalar wave \eqref{scalsol} we construct the free gauge field
\be\label{ymsol}
A_{\mu}\,\d X^\mu=\frac{1}{k_0}\epsilon_a\,\mathcal{R}^a\Phi = \frac{1}{k_0}\epsilon_a \mathcal{R}^a \left(\Omega\,\e^{\im\,\phi_k}\right)\,,
\ee
where $\phi_{k}$ and $\Omega$ are as before and the polarisation vector $\epsilon^a$ is constant. We can also define a `curved' $\varepsilon_\mu$ so that
\be\label{polar}
A_\mu= \varepsilon_\mu\, \Phi\,, \qquad \mbox{ where }\qquad \varepsilon_\mu\, \d X^\mu =\epsilon^a  \left(\frac{k_{j}}{k_0}E^j_a+\sigma_{ab}\,x^{b}\right) \d u +\epsilon_a \,\d x^a\,.
\ee
This satisfies the free equation of motion and gauge-fixing conditions. Similarly to its flat space counterpart, the curved polarisation vector obeys
\be\label{ponshell}
\vepsilon\cdot K=g^{\mu\nu} \vepsilon_{\mu} K_{\nu}=0\,,
\ee
where $K$ is as defined in \eqref{momentum}. In the flat space limit,  $A_{\mu}$ reduces to a standard linearised plane wave $\varepsilon^{\mathrm{flat}}_{\mu}\e^{\im k\cdot X}$, with the non-trivial constant components of $\varepsilon^{\mathrm{flat}}_{\mu}$ being $\epsilon_a$.

In and out states are defined in the same way as for the scalar: an in state looks like a Minkowski plane wave in the in-region, while an out state looks like a Minkowski plane wave in the out-region.

As in the scalar case, an inner product on free gauge fields is induced by the boundary term of the action~\cite{Crnkovic:1986ex}. Restricted to a constant $u$ hypersurface, this inner product is:
\be\label{gip2}
\left\la A_{1}|A_{2}\right\ra:= \im \int_{\Sigma_u}\d v\,\d^{d-2}x\,\left(A_{1}^{\mu}\,\bar{F}_{2\,v\mu}-\bar{A}^{\mu}_{2}\,F_{1\,v\mu}\right)\,,
\ee
which is easily used to compute the three cases of interest. Assuming positive frequency for all (un-conjugated) fields, one finds:
\begin{subequations}\label{gbog}
\begin{eqnarray}
\left\la A^{-}_{1} | A^{-}_{2}\right\ra &= &2\,k_{0}\,\epsilon_1 \cdot \epsilon_2\,\delta(k_{0}-l_{0})\,\delta^{d-2}(k_{i}-l_{i})\,,\\
\left\la A^{+}_{1} | \bar{A}^{-}_{2}\right\ra &= &0\,,\\
\left\la A^{+}_{1} | A^{-}_{2}\right\ra&= &2\, k_{0}\,\left(\frac{2\pi}{\im\, l_{0}}\right)^{\frac{d-2}{2}}\,\epsilon_1 \cdot \epsilon_2\,\delta(k_{0}-l_{0})\,\frac{\e^{-\im({s}_{l}+{r}_{k,l})}}{\sqrt{|b|}}\,,
\end{eqnarray}
\end{subequations}
where $\epsilon_1 \cdot\epsilon_2 = \epsilon_1^a \epsilon_2^b \delta_{ab}$ and the phases ${s}_{l}$, ${r}_{k,l}$ are the same as the scalar case. Unsurprisingly, \eqref{gbog} indicate that the evolution problem is unitary and that there is no particle creation for gauge fields propagating on the plane wave spacetime.


\subsubsection{Spin two}
\label{RaisingGraviton}

Finally, consider linearised metric fluctuations $h_{\mu\nu}$ on the plane wave background. Assuming that the background is a solution to the vacuum Einstein equations and choosing a transverse-traceless gauge for the perturbations
\be\label{gravgfgeneric}
\nabla_{\mu}h^{\mu}_{\sigma}=0=h_{\mu}^{\mu} \, ,
\ee
the linearised Einstein equation is:
\be\label{graveom}
\nabla_{\sigma}\nabla^{\sigma}h_{\mu\nu}-2 R^{\rho}_{\:\:\:\mu\nu\sigma}\,h_{\rho}^{\sigma}=0\,,
\ee
with $R^{\rho}_{\:\:\:\mu\nu\sigma}$ the background curvature tensor. For a vacuum plane wave in Brinkmann coordinates (i.e., $H_{a}^{a}=0$), the gauge for $h_{\mu\nu}$ can be further fixed by requiring the vanishing of the $v$-components $h_{v \mu}=0$. With these conditions, the linearised equation is:
\be\label{graveom2}
g^{\mu\nu}\partial_{\mu}\partial_{\nu}h_{\rho\sigma}+4\,\delta^{u}_{(\rho} \partial_{|v|} h_{\sigma)a}\,H^{a}_{b}\,x^{b}-2\,\delta^{u}_{\rho}\delta^{u}_{\sigma}\, H^{ab}\, h_{ab}=0\,,
\ee   
where all Christoffel symbols have been written out explicitly in Brinkmann coordinates as given in equation~\eqref{Christoffel}.

Solutions to \eqref{graveom2} can be constructed by acting on the massless scalar twice with the spin-raising operator \eqref{hraise}. This leads to:
\be\label{grsol2}
h_{\mu\nu}\,\d X^{\mu}\,\d X^{\nu} = \frac{1}{k_0^2}\epsilon_{a}\,\mathcal{R}^{a}\left(\epsilon_{b}\,\mathcal{R}^{b}\,\Phi\right)=
\left((\varepsilon\cdot \d X)^2 - \frac{\im}{k_{0}} \epsilon_a \epsilon_b\,\sigma^{ab} \d u^2\right)\,\Phi\,,
\ee
where $\epsilon_a$ is chosen to be null with respect to $\delta^{ab}$ to ensure that the gauge condition $h_{\mu}^{\mu}=0$ is obeyed. Note in particular the `tail' term proportional to $\epsilon_a \epsilon_b \,\sigma^{ab}$: unlike in Minkowski spacetime, metric perturbations on a plane wave background do not carry a polarisation which is simply the `square' of a gauge field's polarisation. The reason for this is that the second spin raising operator in \eqref{grsol2} acts not only on the scalar solution (which contributes a second copy of $\varepsilon_{\mu}$) but also on the first spin raising operator (or equivalently, on the first copy of $\vepsilon_{\mu}$, which -- unlike in Minkowski space -- is not a constant vector).

Thus the perturbative double copy for plane wave backgrounds involves subtleties not present in Minkowski space. For linear perturbations around flat space,  $h_{\mu\nu}\sim A_{\mu}\odot A_{\nu}$ for  momentum eigenstates, whereas in plane wave spacetimes we have $h_{\mu\nu}\sim A_{\mu}\odot A_{\nu} +C_{\mu\nu}$, with  correction  $C_{\mu\nu}$ given by the last term proportional to $\sigma^{ab}$ in \eqref{grsol2}. 

The boundary term in the linearised Einstein-Hilbert action induces an inner product on metric fluctuations~\cite{Crnkovic:1986ex}:
\be\label{grip}
\left\la h_{1}|h_{2}\right\ra=\im \int_{\Sigma_u} \d v\,\d^{d-2}x\,\left(h_{1}^{\mu\sigma}\,\partial_{v}\bar{h}_{2\,\mu\sigma}-\bar{h}_{2}^{\mu\sigma}\,\partial_{v}h_{1\,\mu\sigma}\right)\,.
\ee
Once again calculating the inner products between incoming and outgoing states gives:
\begin{eqnarray}\left\la h^{-}_{1} | h^{-}_{2}\right\ra& =& 2\,k_{0}\, (\epsilon_1 \cdot \epsilon_2)^2\,\delta(k_{0}-l_{0})\,\delta^{d-2}(k_{i}-l_{i})\,,\nonumber\\
\left\la h^{+}_{1} | \bar{h}^{-}_{2}\right\ra &=& 0\,,\nonumber \\
\left\la h^{+}_{1} | h^{-}_{2}\right\ra&=& 2\, k_{0}\,\left(\frac{2\pi}{\im l_{0}}\right)^{\frac{d-2}{2}}\,(\epsilon_1 \cdot \epsilon_2)^{2}\,\delta(k_{0}-l_{0})\,\frac{\e^{-\im({s}_{l}+{r}_{k,l})}}{\sqrt{|b|}}\,.\label{grbog1}
\end{eqnarray}
So despite the `correction' term in $h_{\mu\nu}$, the physical properties of unitary evolution and no particle creation are preserved.


\subsubsection{Charged free fields in plane wave gauge fields}

Although we assume that the background gauge potential in \eqref{gBr1} is valued in the Cartan algebra\footnote{The Cartan case was considered in~\cite{Adamo:2017nia} because the amplitudes there were originally calculated as a check for the ambitwistor string results of~\cite{Adamo:2017sze}. When the latter paper was written, it was not yet known how to couple heterotic ambitwistor strings to non-abelian background gauge fields, this issue was only resolved later in~\cite{Adamo:2018hzd}. There is no deeper reason for working with this restriction here.}, it couples non-trivially to free fields which are charged under the gauge group. Consider a free, charged scalar: 
\be\label{fcscal1}
S^{\mathrm{free}}[\Phi]= \frac{1}{2}\int \d u\,\d v\,\d^{d-2}x\,D_{\mu}\Phi\,\overline{ D^{\mu}\Phi}\,,
\ee
where $D_{\mu}=\partial_{\mu}-\im e\sA_{\mu}$, with $\sA_{\mu}$ the background gauge field \eqref{gBr1} and $e$ the charge of $\Phi$.  In the first instance, we will take $e$ to be a standard $\U(1)$ charge, but more generally, $\sA$ takes values in the Cartan subalgebra of some gauge group, $\Phi$ in some root space, and $e$ will then be the corresponding root and $e\sA$ the corresponding contraction with $\sA$ encoding the commutator.  The free equation of motion for the charged scalar is thus
\be\label{csceom}
D_{\mu}D^{\mu}\Phi(X)=\left(2\partial_{u}\,\partial_{v} -\partial_{a}\,\partial^{a}-2\im\,x^{a}e\,\dot{\sA}_{a}\,\partial_{v}\right)\Phi(X)=0\,.
\ee
Solutions to this `charged' wave equation are given by:
\be\label{cssol}
\Phi(X)=\e^{\im\,\tilde{\phi}_{k}}\,,
\ee
where
\be\label{cssol2}
\tilde{\phi}_{k}=k_{0}\,v +(k_{a}+e\sA_{a})\,x^{a}+\frac{1}{2\,k_{0}}\,f(u)\,.
\ee
The function $f(u)$ is the analogue of the $F^{ij}(u)$ which appeared in the gravitational case:
\be\label{ffunc}
f(u):=\int^{u}\d s\,\left(k_{a}+e\sA_{a}(s)\right)\,\left(k^{a}+e\sA^{a}(s)\right)\,.
\ee
When the background gauge field is turned off, it is easy to see that these solutions become the usual momentum eigenstates of Minkowski space.

The natural momentum associated with these scalars is defined by 
\begin{align}\label{csmom1}
{\sK}_{\mu}\,\d X^\mu & :=-\im\e^{-\im\tilde \phi_k}\,D_{\mu}\,\e^{\im\tilde{\phi}_{k}}\,\d X^\mu \nonumber \\
& = k_0 \d v+ \frac{1}{2\,k_{0}}(k_{a}+e\sA_{a})(k^{a}+e\sA^{a})\d u +(k_{a}+e\sA_{a})\d x^a\,.
\end{align}
The components of $\sK_{\mu}$ are functions of $u$, but it is easy to see that this momentum is null.

The distinction between in and out states for the charged scalar is in direct analogy with the definitions on the gravitational background. An incoming state is one which looks like a Minkowski plane wave in the in-region, while an outgoing state looks like a Minkowski plane wave in the out-region. This distinction manifests itself in the boundary conditions on $\sA$:
\be\label{gfbc}
\lim_{u\rightarrow\pm\infty}\sA^{\pm}_{a}(u)=0\,.
\ee
Note that unlike the massless scalar in the gravitational background, the exponential dependence on $x^a$ for the charged scalar is at most linear in any region.

The inner product on the charged scalars is given by
\be\label{csip1}
\left\la \Phi_{1}|\Phi_2\right\ra=\im\int_{\Sigma_u}\d v\,\d^{d-2}x\,\left(\Phi_{1}\,\partial_{v}\bar{\Phi}_{2}- \bar{\Phi}_{2}\,\partial_{v}\Phi_{1}\right)\,,
\ee
and once again there are three inner products of physical interest. These are:
 \begin{eqnarray}
 \left\la \Phi^{-}_{1}|\Phi^{-}_{2}\right\ra&=&2k_{0}\,\delta(k_{0}-l_{0})\,\delta^{d-2}\!\left(k_{a}-l_{a}\right)\,,
\nonumber \\
 \left\la \Phi^{+}_{1}|\bar{\Phi}^{-}_{2}\right\ra&=&0\,,\nonumber \\
 \left\la \Phi^{+}_{1}|\Phi^{-}_{2}\right\ra&=&2k_{0}\,\delta(k_{0}-l_{0})\,\delta^{d-2}\!\left(k_{a}-l_{a}+c_{a}\right)\,\e^{\im\,\tilde{{s}}_{l}}\,,\label{csbog1}
 \end{eqnarray}
where $c_{a}$ is the inner product of $\sA^{-}_{a}(0)$ in the Cartan subalgebra with the charge of the field.  The momentum conservation then indicates the `kick' received by the field from the memory effect.  
The phase $\tilde{{s}}_{l}$ is defined by
\begin{equation*}
 \tilde{{s}}_{l}:=\frac{f_{-}(0)}{2\,l_{0}}\,. 
\end{equation*}
The equations \eqref{csbog1} indicate that the classical S-matrix associated with this charged scalar is unitary with no particle production.


\subsubsection{Spin one on a gauge background}
\label{RaisingGluon}

The linearised equation of motion for a gauge field $a_{\mu}$ charged under the same gauge group as the background $\sA_{\mu}$ is:
\be\label{cgfeom}
D_{\mu}\left(D^{\mu}a^{\nu}-D^{\nu}a^{\mu}\right)+a_{\mu}\left(\partial^{\mu}\sA^{\nu}-\partial^{\nu}\sA^{\mu}\right)=0\,.
\ee
Solutions to this equation are simplified by choosing a Lorenz gauge $D_{\mu}a^{\mu}=0$ along with\footnote{This is of course not possible on a general background, but is possible here because $\p_v$ is a symmetry.} $a_{v}=0$; the latter condition actually reduces the Lorenz condition to $\partial_{\mu}a^{\mu}=0$. Solutions are then found by acting on the charged scalar solution with $\mathcal{R}^a$ as before in the gravitational case. This leads to
\be\label{cgfsol}
a_{\mu}\,\d X^{\mu}=\tilde{\epsilon}_{a}\left( \d x^a + \frac{1}{k_0}(k^a+e\sA^a)\,\d u\right)\,\e^{\im\tilde{\phi}_{k}}\,,
\ee
where $\tilde{\epsilon}_{a}$ is a (constant) $(d-2)$-dimensional vector which we will take to be null. As in the gravitational case, we define a polarisation $d$-vector $\tvepsilon_\mu$ as 
\be\label{gfpol}
\tvepsilon_\mu\, \d X^\mu = \tilde{\epsilon}_{a}\,\left(\d x^a + \frac{1}{k_0}(k^a+e\sA^a)\,\d u\right). 
\ee
This polarisation is on-shell in the sense that $\sK\cdot \tvepsilon=0$.

With these gauge choices, the inner product is essentially equivalent to \eqref{gip2} giving:
\begin{eqnarray}
 \left\la a^{-}_{1}|a^{-}_{2}\right\ra&=&2k_{0}\,\tilde{\epsilon}_1\cdot \tilde{\epsilon}_2\,\delta(k_{0}-l_{0})\,\delta^{d-2}\!\left(k_{a}-l_{a}\right)\,,
 \nonumber \\
 \left\la a^{+}_{1}|\bar{a}^{-}_{2}\right\ra&=&0\,,
 \nonumber \\
 \left\la a^{+}_{1}|a^{-}_{2}\right\ra&=&2\,k_{0}\,\tilde{\epsilon}_1\cdot \tilde{\epsilon}_2\,\delta(k_{0}-l_{0})\,\delta^{d-2}\!\left(k_{a}-l_{a}+c_{a}\right)\,\e^{\im\,\tilde{{s}}_{l}}\,.
\label{gfbog}
\end{eqnarray}
So we again have a unitary classical S-matrix with no particle creation, as before.


\subsubsection{Huygens' principle and tails}

The wave equation in flat and plane wave spacetimes satisfies Huygens' principle~\cite{Friedlander:1975eqa}.  In intuitive terms, the principle states that waves can propagate in all directions without  scattering off the background metric and generating a \emph{tail}. The sharp definition is that there should exist solutions to the wave equation with delta-function support along null hypersurfaces tangent to every null direction through every point. These are simply given in the above by $\Omega\,\delta(\phi_k -c)$ where $c$ is a constant. 

This principle fails for linear fields of spin one and spin two~\cite{Mason:1989}, however.  We can construct these fields by spin raising as above. At spin one, to get a field with delta function support along $\phi_k=0$, we must start by raising the spin of a solution to the scalar wave equation of the form $\Omega\,\phi_k\,\Theta(\phi_k)$ where $\Theta$ is the Heaviside step function. With this, the corresponding spin-one potential is 
$$
A =\Theta(\phi_k)\,\frac{\epsilon_a}{k_0}\,\mathcal{R}^a \left(\Omega\,\phi_k\,\Theta(\phi_k)\right)= \Omega\,\epsilon^a \left(\d x_a +\left(\frac{k_j}{k_0}\,E^j_a + \sigma_{ab}x^b\right)\d u\right)\,\Theta(\phi_k)\, ,
$$
and the field strength is
\begin{multline}
F =\d A= \delta(\phi_k)\,\Omega\,\epsilon^a \left(\d x_a +\left(\frac{k_j}{k_0}E^j_a +\sigma_{ab}x^b\right)\,\d u\right)\wedge\d\phi_k \\
+\Theta(\phi_k)\,\Omega\,\epsilon^{a}\,\left(\sigma_{ab}\,\d x^b\wedge \d u -\sigma^{b}_{b}\,\d x_{a}\wedge\d u\right)\, .
\end{multline}
We see that the field strength has developed a tail in the second line, which is not supported at $\phi_k=0$. This tail can be thought of as the consequence of the interaction between the impulsive electromagnetic field and the gravitational background. There is a similar story for the spin-two field where one starts with $\Phi= \phi_k^3\, \Theta(\phi_k)$.  

In these examples, the tail is proportional to the shear of the $\partial_{U}$ null geodesic congruence (i.e., trace-free part of $\sigma_{ab}$). So tails are generally identified by the part of the field in which the shear appears explicitly. In the free solutions constructed above, terms contributing to the tails are readily identified: $\sigma_{ab}\,x^{b} \d u$ from $\varepsilon\cdot \d X$ at spin one and two, and the spin two correction term $C=-\frac{\im}{ k_{0}}\epsilon_a \epsilon_b\sigma^{ab}\d u^2$. However, we will see that the contributions to the tail from $\varepsilon_{\mu}$ alone actually drop out of amplitude calculations. So for spin one fields on a plane wave spacetime, the tail terms do not effect the amplitude -- even though they appear explicitly in the scattering states.

This observation is perhaps related to a different definition of tails for the propagation of gauge fields on a plane wave spacetimes, in terms of a Green's function in~\cite{Gunther:1974,Gunther:1988}. That discussion does not give tails for gauge fields but \emph{does} for graviton propagation~\cite{Harte:2013dba}, and indeed we will see that it is the extra correction term $C$ that is important for  graviton amplitudes. 

Note that this treatment of tails does not simply extend to fields propagating on the gauge theory plane wave background because we cannot simply obtain solutions from  arbitrary functions of $\tilde \Phi$ as it now has charge. So, in the gauge background case, we will simply take the tail to be those terms in a curved polarisation vector that depend explicitly on the potential $\sA$. This is consistent with the fact that such potential terms encode the memory in the asymptotic regions via \eqref{gmem}, just as the deformation tensor $\sigma_{ab}$ does on a gravitational background.


\subsection{3-point amplitudes on the gravitational background}
\label{GravB}

We now consider the 3-point amplitudes of scalars, gauge fields and gravitons on the gravitational sandwich plane wave background. In each case, this calculation is performed by evaluating the cubic part of the action on solutions to the linearised equations of motion on the background. For each theory, the amplitude formulae are presented in terms of an integral over the $u$ variable (in Brinkmann coordinates), which cannot be done explicitly for general spacetimes. Stripping off the integration underlying the action integral, together with the three $\Phi$s associated with the three on-shell fields, we are left with a tree level integrand expression which is sufficient for exploring the double copy structure of the amplitudes. See appendix~\ref{S-matrix-integrands} for further discussion of the scattering amplitudes and tree level integrand.


\subsubsection{Scalars}

Consider the cubic scalar theory
\be\label{scal1}
S[\Phi]=\frac{1}{2}\int_{M} \d u\,\d v\,\d^{d-2}x\,\left(g^{\mu\nu}\partial_{\mu}\Phi\,\partial_{\nu}\Phi - \frac{\lambda}{3}\Phi^{3}\right)\,,
\ee
where $g^{\mu\nu}$ is the inverse of the plane wave metric \eqref{Br1} in Brinkmann coordinates. The 3-point amplitudes of interest are given by evaluating the cubic portion of the action\footnote{A similar calculation has been done for scalar contact interactions of arbitrary valence in certain homogeneous plane wave backgrounds~\cite{Papadopoulos:2002bg}.}
\be\label{scubic}
-\frac{\lambda}{6}\int_{M}\d u\,\d v\,\d^{d-2}x\,\Phi_{1}(X)\,\Phi_{2}(X)\,\Phi_{3}(X)\,,
\ee
where $\Phi_{{r}}(X)$ are solutions to the linearised equations of motion of \eqref{scal1} for ${r}=1,2,3$. When evaluating \eqref{scubic}, there are basically two distinct configurations which need to be considered: three in states, or one out and two in states (the other configurations are easily related to these). 

The case when all three states are incoming is the easiest. This gives
\begin{multline}\label{s3p1}
-\frac{\lambda}{6}\int_{M}\d u\,\d v\,\d^{d-2}x\,\Phi^{-}_{1}(X)\,\Phi^{-}_{2}(X)\,\Phi^{-}_{3}(X)  \\
=-\frac{\lambda}{6}\, \delta^{d-1}\!\left(\sum_{{r}=1}^{3}k_{{r}}\right)\int \d u\,|E^{-}|\,(\Omega^{-})^{3}\,\exp\left(\im F^{ij}\sum_{{s}=1}^{3}\frac{k_{{s}\,i}k_{{s}\,j}}{2k_{{s}\,0}}\right) \\
=-\frac{\lambda}{6}\,\delta^{d-1}\!\left(\sum_{{r}=1}^{3}k_{r}\right)\,\int \frac{\d u}{\sqrt{|E^{-}|}}\,\exp\left(\im F^{ij}\sum_{{s}=1}^{3}\frac{k_{{s}\,i}k_{{s}\,j}}{2k_{{s}\,0}}\right)\,. 
\end{multline}
where
\begin{equation*}
 \delta^{d-1}\!\left(\sum_{{r}=1}^{3}k_{{r}}\right):= \delta\!\left(\sum_{r=1}^{3} k_{r\,0}\right)\, \delta^{d-2}\!\left(\sum_{r=1}^{3} k_{r\,i}\right)\,.
\end{equation*}
The delta functions arise from performing the integrations in $\d v$ and $\d^{d-2}x$, with $|E^{-}|$ an overall Jacobian factor appearing in the second line. Using the relationship \eqref{scalsol} between $\Omega(u)$ and $|E|$, the various $u$-dependent factors left inside the integral can be slightly simplified in passing to the third line.

The other configuration is a bit more complicated. In this case one has
\begin{multline}\label{s3p2}
-\frac{\lambda}{6}\int_{M}\d u\,\d v\,\d^{d-2}x\,\Phi^{-}_{1}(X)\,\Phi^{-}_{2}(X)\,\Phi^{+}_{3}(X)  
=-\frac{\lambda}{6}\,\delta\!\left(\sum_{{r}=1}^{3}k_{{r}\,0}\right) \int \d u\,\d^{d-2}x\, (\Omega^{-})^{2}\Omega^{+} \times\\ 
\exp\left(\im \frac{k_{3\,0}}{2}(\sigma_{ab}^{-}-\sigma_{ab}^{+}) x^{a} x^{b} \right. 
\left. +\im\,(k_{1\,i}+k_{2\,i})E^{i\,-}_{a} x^{a} + \im\, k_{3\,i}E^{i\,+}_{a} x^{a} + \sum_{{s}=1}^{3}\frac{k_{{s}\,i}k_{{s}\,j}}{2k_{{s}\,0}} F^{ij}_{{s}}\right)\,.
\end{multline}
Due to the mixed asymptotic conditions, momentum conservation in the $v$-direction no longer eliminates the quadratic $x$-dependence from the exponential, leaving a $(d-2)$-dimensional Gaussian integral. Performing this integral leaves:
\begin{multline}\label{s3p3}
-\frac{\lambda}{6\,(k_{3\,0})^{\frac{d-2}{2}}}\,\delta\!\left(\sum_{{r}=1}^{3}k_{{r}\,0}\right) \int \d u\, (\Omega^{-})^{2}\Omega^{+}\, \sqrt{\frac{(2\pi\im)^{d-2}}{|A|}} \\
\times \exp\left(-\frac{\im}{2\,k_{3\,0}}J_{a} J_{b} (A^{-1})^{ab} +\im \sum_{{s}=1}^{3}\frac{k_{{s}\,i}k_{{s}\,j}}{2k_{{s}\,0}} F^{ij}_{{s}}\right)\,,
\end{multline}
where
\begin{equation*}
A_{ab}:= \sigma_{ab}^{-}-\sigma_{ab}^{+}\,, \qquad J_{a}:=(k_{1\,i}+k_{2\,i})\,E^{i\,-}_{a} + k_{3\,i}\,E^{i\,+}_{a}\,.
\end{equation*}
Nevertheless, applying the definition of the tree level integrand to these results (see earlier or appendix~\ref{S-matrix-integrands}), somewhat tautologically gives the extremely simple answer
\be\label{s3pi1}
\cM_{3}(\Phi^{-}_{1}, \Phi^{-}_{2}, \Phi^{\pm}_{3})=1\,,
\ee
after stripping off a power of the coupling, overall delta-functions, and `universal' $u$-dependent functions that depend on the choice of $\Phi$'s.

This is a general feature. Although the precise form of the \emph{amplitude} will vary significantly between different configurations of incoming and outgoing states -- as in \eqref{s3p1} versus \eqref{s3p3}, the \emph{integrands} will be the same. This is the closest thing to CPT symmetry in flat spacetime -- interpreted here as the ability to exchange incoming and outgoing states while simultaneously conjugating polarisations and charges -- which survives on a sandwich plane wave background.


\subsubsection{Gauge theory}
     
The Yang-Mills action on a curved background is:
\be\label{YM1}
S[A]=\frac{1}{g^{2}}\int_{M} \tr\left(F\wedge * F\right)\,,
\ee
where $*$ is the Hodge star and $F=[D,D]$ is the curvature of the connection $D=\nabla + A$, for $\nabla$ the Levi-Civita connection. The 3-point amplitude is given by the cubic portion of the action \eqref{YM1} evaluated on linearised states of the form \eqref{ymsol}. In the Lorenz gauge of chapter~\ref{FF}, the 3-point amplitude reads:
\be\label{ym3p1}
g\,f^{\sa_{1}\sa_{2}\sa_3} \int \d u\,\d v\,\d^{d-2}x\left(A_{3}^{b}\,A^{\mu}_{2}\,\partial_{\mu}A_{1\,b} - A_{2}^{b}\,A_{3}^{\mu}\,\partial_{\mu}A_{1\,b} + \mathrm{cyclic}\right)\,,
\ee
where $f^{\sa_{1}\sa_{2}\sa_3}$ are the structure constants of the gauge group. As before, there are essentially two independent configurations in which this amplitude can be evaluated: three in states or two in states and one out state. 

However, some simplifications occur in the amplitude even before the asymptotic behaviour of the states has been specified. Evaluated on general linearised free fields, \eqref{ym3p1} becomes
\be\label{ym3p2}
 g\,f^{\sa_{1}\sa_{2}\sa_3} \int \d u\,\d v\,\d^{d-2}x\,\left(\vepsilon_{1}\cdot\vepsilon_{3}\,(K_{1}\cdot\vepsilon_{2}-K_{3}\cdot\vepsilon_{2}) +  \mathrm{cyclic} 
 \right)\,\prod_{{r}=1}^{3}\Omega_{{r}}\,\e^{\im\phi_{{r}}}\,,
\ee
where the $\Omega_{{r}}$ and $\phi_{{r}}$ (${r}=1,2,3$) depend on whether the state is incoming or outgoing. Since the functional form of the integrand (i.e., the portion of this expression in the parentheses) is independent of the state configuration, it suffices to identify the integrand in the simplest configuration. As in the scalar example, this will be the all incoming configuration, since there are more delta functions in this case.

Even for the three-incoming configuration, the integrand of \eqref{ym3p2} is \emph{a priori} a function of the $x^a$ through the polarisations \eqref{polar} and momenta \eqref{momentum}. However, thanks to the identities: 
\begin{eqnarray}
\label{polrels}
K_{{r}}\cdot \varepsilon_{{s}}&=&\left\{\begin{array}{c c}
                                          0 & \mathrm{if}\;\; {r}={s} \\
                                          E^{i\,a}(k_{{r}\,0}\frac{k_{{s}\,i}}{k_{s\, 0}}\epsilon_{{s}\,a}-k_{{r}\,i}\epsilon_{{s}\,a})\quad & \mathrm{otherwise}
                                         \end{array}\right. \,,
\\
\label{polrels2}
\vepsilon_{{r}}\cdot\varepsilon_{{s}}&=&\left\{\begin{array}{c c}
                \quad    \qquad  \qquad  0 \qquad\quad\qquad& \quad \mathrm{if}\;\; {r}={s} \\
-\epsilon_{r} \cdot \epsilon_{s} &
\qquad  \mathrm{otherwise}
                                         \end{array}\right.\,,
\end{eqnarray}
it follows that the integrand is actually \emph{independent} of the $x^a$. 
This allows the $\d v$ and $\d^{d-2}x$ integrals to be done as the only dependence on these variables is in the exponential:
\begin{multline}\label{ym3p3}
 g\,f^{\sa_{1}\sa_{2}\sa_3}\,\delta^{d-1}\!\left(\sum_{{r}=1}^{3}k_{r}\right)\,\int \frac{\d u}{\sqrt{|E^-|}}\,\left(\varepsilon_{1}\cdot\vepsilon_{3}\,(K_{1}\cdot\vepsilon_{2}-K_{3}\cdot\vepsilon_{2}) + \mathrm{cyclic}
\right)\,\\ \times\, \exp\left(\im F^{ij}\sum_{{s}=1}^{3}\frac{k_{{s}\,i}k_{{s}\,j}}{2k_{{s}\,0}}\right)\,.
\end{multline}
On the support of the momentum conserving delta functions, this simplifies to 
\be\label{ym3p4}
 2g\,f^{\sa_{1}\sa_{2}\sa_3}\,\delta^{d-1}\!\left(\sum_{{r}=1}^{3}k_{r}\right)\,\int \frac{\d u}{\sqrt{|E^-|}}\,\left(\vepsilon_{1}\cdot\vepsilon_{3}\,K_{1}\cdot\vepsilon_{2} +  \mathrm{cyclic}
 \right)\, \exp\left(\im F^{ij}\sum_{{s}=1}^{3}\frac{k_{{s}\,i}k_{{s}\,j}}{2k_{{s}\,0}}\right)\,.
\ee
 As we saw for the scalar, the amplitude boils down to a $u$-integration which depends on the particulars of the background plane wave geometry. The integrand, though, is easily identified as:
\be\label{ym3p5}
\boxed{\cM_{3}(A_1,A_2,A_3)=\vepsilon_{1}\cdot\vepsilon_{3}\,K_{1}\cdot\vepsilon_{2}+   \mathrm{cyclic} 
\,.}
\ee
Note that although this has the same functional form as the flat space 3-point integrand for Yang-Mills theory, it is \emph{not} equal to the flat space result. Indeed, the integrand in this case is a function of $u$, given explicitly by
\begin{equation}
\label{ym3p6}
 \cM_{3}(A_1,A_2,A_3)= -\,\epsilon_1 \cdot \epsilon_3\,E^{i}_{a}\left(\frac{k_{1\,0}}{k_{2\, 0}}\,k_{2\,i}\,\epsilon^a_{2}-k_{1\,i}\,\epsilon^a_{2}\right) +  \mathrm{cyclic}
\end{equation}
after using \eqref{polrels}--\eqref{polrels2}. Note that the tails associated with the asymptotic states do not contribute to the amplitude, as a result of the identities \eqref{polrels}--\eqref{polrels2}.

\smallskip

The other configuration -- two incoming states and one outgoing state -- is more complicated. The primary reason for this is that the $x$-dependence of the integrand does not drop out. Assuming that the scattering states are $A^{-}_{1}$, $A^{-}_{2}$ and $A_{3}^{+}$ we now have
\begin{eqnarray}
\vepsilon_{{r}}\cdot\vepsilon_{3}&=&-\epsilon_{r} \cdot \epsilon_3\,,
\nonumber\\
K_{{r}}\cdot\vepsilon_{3}&=&\epsilon_3^a\, \left(k_{{r}\,0}\frac{k_{3\,i}}{k_{3\, 0}}E^{+\,i}_{a}-k_{{r}\,i}E^{-\,i}_{a}\right)+k_{{r}\,0}\epsilon_3^a x^{b}\,(\sigma^{+}_{ab}-\sigma^{-}_{ab})\,,\nonumber \\
K_{3}\cdot\vepsilon_{{r}}&=&\epsilon_{r}^a\, \left(k_{3\,0}\frac{k_{{r}\,i}}{k_{r\,0}}E^{-\,i}_{a}-k_{3\,i}E^{+\,i}_{a}\right)+k_{3\,0}\epsilon^a_{{r}}x^{b}\,(\sigma^{-}_{ab}-\sigma^{+}_{ab})\,,
\label{mixpol}
\end{eqnarray}
for ${r}=1,2$. The integration over $\d^{d-2}x$ is now a rather involved Gaussian integral, which has the rough structure of \eqref{s3p3} plus a derivative of this result. Since the integrand is the primary object of interest here, we will only  consider \eqref{ym3p5}.


\subsubsection{Gravity}

The 3-point amplitude for gravitons on the plane wave background is encoded by extracting the cubic portion of the Einstein-Hilbert action,
\be\label{EH1}
S[g]=\frac{1}{\kappa^2}\int_{M} \d^{d}X\,\sqrt{-|g|}\,R\,,
\ee
perturbed around the plane wave background metric. To do this, a recent perturbative re-writing of the Einstein-Hilbert action is useful~\cite{Cheung:2016say}. For perturbations $h_{\mu\nu}$ around a fixed background geometry $g_{\mu\nu}$, this action takes the form:
\be\label{CR1}
S[h]=\frac{1}{4\,\kappa^2}\int \d^{d}X\,\sqrt{-|g|}\left[\nabla_{\mu}\sigma_{\nu\rho}\,\nabla_{\lambda}\sigma^{\kappa\rho}\left(\sigma^{\mu\lambda}\delta^{\nu}_{\kappa} - 2\,\sigma^{\nu\lambda}\delta^{\mu}_{\kappa}\right) +\sigma^{\mu\nu}\,R_{\mu\nu}\right]\,,
\ee
where the perturbations are encoded in
\begin{equation*}
\sigma_{\mu\nu}=g_{\mu\nu}+\kappa\,h_{\mu\nu}+\frac{\kappa^2}{2}h^{2}_{\mu\nu} +\cdots\,, \qquad \sigma^{\mu\nu}=g^{\mu\nu}-\kappa\,h^{\mu\nu}+\frac{\kappa^2}{2}h^{\mu\nu}-\cdots\,,
\end{equation*}
and indices are raised and lowered with the background metric (e.g., $h^{2}_{\mu\nu}=h_{\mu\rho}g^{\rho\sigma}h_{\sigma\nu}$). On the vacuum plane wave background in Brinkmann coordinates, $|g|=-1$ and $R_{\mu\nu}=0$ so expanding \eqref{CR1} to cubic order is straightforward. This leads to the cubic term:
\be\label{CR2}
\frac{\kappa}{4}\int \d u\,\d v\,\d^{d-2}x\left(h^{\mu\nu}\nabla_{\mu}h_{\rho\sigma}\nabla_{\nu}h^{\rho\sigma}-2\,h^{\rho\nu}\nabla_{\mu}h_{\rho\sigma}\nabla_{\nu}h^{\mu\sigma}\right)\,.
\ee
We have checked that this matches the cubic contribution from expanding the standard Einstein-Hilbert action around a plane wave background.

The 3-point amplitude is given by evaluating \eqref{CR2} on three of the linearised perturbations \eqref{grsol2}. With the transverse-traceless gauge conditions on $h_{\mu\nu}$, the covariant derivatives in \eqref{CR2} reduce to partial derivatives, leaving:
\be\label{gr3p1}
\frac{\kappa}{4}\int \d u\,\d v\,\d^{d-2}x\left(h_{1}^{\mu\nu}\partial_{\mu}h_{2\,\rho\sigma}\partial_{\nu}h_{3}^{\rho\sigma}-2\,h_{1}^{\rho\nu}\partial_{\mu}h_{2\,\rho\sigma}\partial_{\nu}h_{3}^{\mu\sigma}\right) + \mathrm{all}\:\:\mathrm{permutations}\,.
\ee 
A computation gives a typical term in the sum over permutations of external states to be:
\begin{multline}\label{gr3p2}
h_{1}^{\mu\nu}\partial_{\mu}h_{2\,\rho\sigma}\partial_{\nu}h_{3}^{\rho\sigma}-2\,h_{1}^{\rho\nu}\partial_{\mu}h_{2\,\rho\sigma}\partial_{\nu}h_{3}^{\mu\sigma}= \\
\Bigg(\left(2\vepsilon_{3}\cdot K_{2}\,\vepsilon_{1}\cdot K_{3}\,\vepsilon_{1}\cdot\vepsilon_{2}-\vepsilon_{1}\cdot K_{2}\, \vepsilon_{1}\cdot K_{3}\,\vepsilon_{2}\cdot\vepsilon_{3}\right)\,(\vepsilon_{2}\cdot\vepsilon_{3})
\\
-\im \,\vepsilon_{2}\cdot\vepsilon_{3}\,\sigma^{ab}\left(\frac{k_{2\,0}k_{3\,0}}{k_{1\,0}}\vepsilon_{2}\cdot\vepsilon_{3}\, \epsilon_{1\,a}\epsilon_{1\,b} -2k_{2\,0}\,\vepsilon_{1}\cdot\vepsilon_{2}\,\epsilon_{1\,b}\epsilon_{3\,a}\right)\Bigg)\,\e^{\im(\phi_{1}+\phi_{2}+\phi_{3})}\,.
\end{multline}
To proceed further, the configuration of the external states must be specified. Building on the scalar and gauge theory calculations, it is clear that the easiest configuration to treat is the one with all three states incoming.

In this configuration, identities of the form \eqref{polrels}--\eqref{polrels2} ensure that the only $x$-dependence in terms  like \eqref{gr3p2} is in the overall exponential. This allows the $\d v$ and $\d^{d-2}x$ integrations to be done explicitly, resulting in momentum conserving delta functions. On the support of these delta functions, the 3-point amplitude for incoming states reads:
\begin{multline}\label{gr3p3}
\frac{\kappa}{2}\,\delta^{d-1}\!\left(\sum_{{r}=1}^{3}k_{r}\right)\,\int \frac{\d u}{\sqrt{|E^-|}}\,\left[\left(\vepsilon_{1}\cdot\vepsilon_{3}\,K_{1}\cdot\vepsilon_{2}+ \mathrm{cyclic}
\right)^{2}\right. 
-\im\, k_{1\,0}k_{2\,0}k_{3\,0}\,\sigma^{ab}\mathcal{C}_{a}\mathcal{C}_{b}
\Big]\\
\times\, \exp\left(\im F^{ij}\sum_{{s}=1}^{3}\frac{k_{{s}\,i}k_{{s}\,j}}{2k_{{s}\,0}}\right)\,,
\end{multline}
where the quantity $\mathcal{C}_{a}$ is defined as
\be\label{curvcor}
\mathcal{C}_{a}:= \vepsilon_{2}\cdot\vepsilon_{3}\,\frac{\epsilon_{1\,a}}{k_{1\, 0}} + \vepsilon_{1}\cdot\vepsilon_{3}\, \frac{\epsilon_{2\,a}}{k_{2\, 0}}+\vepsilon_{1}\cdot\vepsilon_{2}\, \frac{\epsilon_{3\,a}}{k_{3\, 0}} \;.
\ee
The upshot is that the 3-point integrand for gravity on a plane wave spacetime is given by
\begin{multline}\label{gr3p4}
\cM_{3}(h_{1},h_{2},h_{3})=\left(\vepsilon_{1}\cdot\vepsilon_{3}\,K_{1}\cdot\vepsilon_{2}+\vepsilon_{1}\cdot\vepsilon_{2}\,K_{2}\cdot\vepsilon_{3} +\vepsilon_{2}\cdot\vepsilon_{3}\,K_{3}\cdot\vepsilon_{1}\right)^{2} 
\\
-\im\, k_{1\,0}k_{2\,0}k_{3\,0}\,\sigma^{ab}\,\mathcal{C}_{a}\,\mathcal{C}_{b}\,.
\end{multline}
This structure mirrors what one might have guessed based solely on the structure of the linearised perturbations \eqref{grsol2}. 

So it seems that 3-point amplitudes on a plane wave spacetime do not simply obey double copy as they do in flat space. Indeed, we find that
\be\label{dc1}
\boxed{\cM_{3}(h_{1},h_{2},h_{3})=\left(\cM_{3}(A_{1},A_{2},A_{3})\right)^{2}-\im\, k_{1\,0}k_{2\,0}k_{3\,0}\,\sigma^{ab}\,\mathcal{C}_{a}\,\mathcal{C}_{b}\,.}
\ee
Unlike the gluon amplitudes, the tails associated to graviton perturbations \emph{do} contribute to the amplitude. Note that they do so in an intrinsically geometric way: the tail contribution couples via the deformation tensor associated with the background geometry. To find the `square root' of perturbative gravity on a plane wave background, one must instead turn to Yang-Mills theory in the presence of a background plane wave gauge field.


\subsection{3-point amplitudes on the gauge field background}
\label{GaugeB}

The 3-point amplitudes for charged scalars and Yang-Mills theory in a plane wave background gauge field are now computed. As in the gravitational setting, these amplitudes reduce to an integral over the $u$-coordinate which depends on the particulars of the background, but the tree level integrands are easily identified.


\subsubsection{Charged scalars} 

To obtain a gauge invariant cubic scalar interaction that carries charge with respect to the background gauge field, the charges of the three fields must add up to zero.  
\be\label{cscal1}
S_{\mathrm{int}}[\Phi]= \int \d u\,\d v\,\d^{d-2}x\,\left( \Phi_1\Phi_2\Phi_3\right)\,,
\ee
where $D_{\mu}\Phi_{r}=(\partial_{\mu}-\im e_{r}\sA_{\mu})\Phi_{r}$, with $\sA_{\mu}$ the background gauge field \eqref{gBr1}. The charges $e_{r}$ as roots encode the commutators.

Armed with the linearised solutions \eqref{cssol}, we can compute the 3-point amplitudes by evaluating the cubic portion of the action \eqref{cscal1}. This means that the amplitude can be reduced to a $u$-integration fairly straightforwardly in an arbitrary configuration:
\be\label{cs3p1}
\delta^{d-1}\!\left(\sum_{{r}=1}^{3}k_{{r}}\right)\,\int \d u\,\exp\left(\im\sum_{{s}=1}^{3}\frac{f_{{s}}}{2k_{{s}\,0}}\right)\,.
\ee 
Note that in the all incoming case, the translation action of the gauge field on the total momentum has cancelled because the charges must add up to zero by gauge invariance.  In the other case, this translation will manifest itself in an additional memory term analogous to $c_a$ in the last line of equations~\eqref{csbog1}, which we ignored here as it is irrelevant for our purposes.

From equation~\eqref{cs3p1} it is easy to read off the tree level integrand for the 3-point scattering of charged scalars on the plane wave gauge field background:
\be\label{cs3p2}
\cA_{3}(\Phi_{1},\Phi_{2},\Phi_{3})=1\,.
\ee
This is independent of the specifics of the configuration as for  the gravitational background.


\subsubsection{Gauge theory}

Now consider a dynamical gauge field $a$ on the fixed plane wave background $\sA$. Although the background gauge field $\sA$ is valued in the Cartan of the gauge group, the dynamical gauge field carries arbitrary colour structure. The dynamical gauge field is governed by the action
\be\label{cgf1}
S[a]=\frac{1}{g^2}\int\tr\left(\cF\wedge*\cF-\d\sA\wedge*\d\sA\right)\,,
\ee
where $\cF$ is the curvature of $\sA+a$ and the kinetic term for the non-dynamical background field is subtracted. 

The cubic term in the action \eqref{cgf1} is
\be\label{cgf3p1}
\int \d u\,\d v\,\d^{d-2}x\,\tr\left(a_{\mu}\,a_{\nu}\left(\partial^{\mu}a^{\nu}-\partial^{\nu}a^{\mu}+[\sA^{\mu},a^{\nu}]\right)\right)\ \,.
\ee
We must choose the colour structure so as to obtain a non-trivial trace. All non-trivial examples are essentially the same and are equivalent to taking the $\SU(2)$ case with $a_3$ in the Cartan, and $a_1$, $a_2$ respectively of charge $\pm 1$ with respect to the Cartan generator. In particular the three charges add up to zero. Together with the gauge choices made in \eqref{cgfsol}, the 3-point amplitude reduces to
\be\label{cgf3p2}
g\,f^{\sa_{1}\sa_{2}\sa_3} \int \d u\,\d v\,\d^{d-2}x\,\left(a_{2}^{\mu}\,a^{\nu}_{3}\partial_{\mu}a_{1\,\nu}-a_{2}^{\mu}\,a_{3}^{\nu}\partial_{\nu}a_{1\,\mu} + \mathrm{cyclic}\right)\,.
\ee
Evaluating on the states \eqref{cgfsol} with arbitrary asymptotics, subject to the same caveat about the memory dependent momentum shifts in the two incoming and one outgoing state case as in equation~\eqref{cs3p1}, leads to
\be\label{cgf3p3}
\im g\,f^{\sa_{1}\sa_{2}\sa_3}\,\delta^{d-1}\!\left(\sum_{{r}=1}^{3}k_{{r}}\right) \int \d u\, 
\left[\tvepsilon_{1}\cdot\tvepsilon_{3}\,({\sK}_{1}\cdot\tvepsilon_{2}-{\sK}_{3}\cdot\tvepsilon_{2})+ \mathrm{cyclic}
\right]\,\exp\left(\im\sum_{{s}=1}^{3}\frac{f_{{s}}}{2\,k_{{s}\,0}}\right)\,.
\ee
Note that in both cases on the support of these delta functions (memory shifted for two incoming and one outgoing state), the identity ${\sK}_{1}\cdot\tvepsilon_{2} = -{\sK}_{3}\cdot\tvepsilon_{2}$ holds. Therefore the result further reduces to:
\begin{equation}
\label{cgf3p4}
2\im g\,f^{\sa_{1}\sa_{2}\sa_3}\,\delta^{d-1}\!\left(\sum_{{r}=1}^{3}k_{{r}}\right) \int \d u\,  
\left[\tvepsilon_{1}\cdot\tvepsilon_{3}\,{\sK}_{1}\cdot\tvepsilon_{2}+\mathrm{cyclic}
\right]\,\exp\left(\im\sum_{{s}=1}^{3}\frac{f_{{s}}}{2\,k_{{s}\,0}}\right)
\end{equation}
 Thus the integrand can be written in terms of on-shell data:
\be\label{cgf3p5}
\boxed{\cA_{3}(a_{1},a_{2},a_{3})=\tvepsilon_{1}\cdot\tvepsilon_{3}\,\sK_{1}\cdot\tvepsilon_{2}+\mathrm{cyclic}
\,,}
\ee
as expected.

This formula hides explicit dependence on the potential.
Using \eqref{csmom1} and \eqref{gfpol}, it follows that:
\begin{eqnarray}
\label{gpolrels}
\sK_{{r}}\cdot \tvepsilon_{{s}}&=&\left\{\begin{array}{c c}
                                          0 & \mathrm{if}\;\; {r}={s} \\
             \frac{ \tilde{\epsilon}_{{s}}^{a}}{k_{s 0}}(k_{{r}\,0}{k_{{s}\,a}}-{k_{s\, 0}}k_{{r}\,a}+ {k_{{r}\,0}} e_s\sA_{a}- {k_{s\, 0}}e_r\sA_{a}) \quad & \mathrm{otherwise}
                                         \end{array}\right. \,,\\
\label{gpolrels2}
\tvepsilon_{{r}}\cdot\tvepsilon_{{s}}&=&\left\{\begin{array}{c c}
                                          0 & \mathrm{if}\;\; {r}={s} \\
                                          -\tilde{\epsilon}_{{r}}\cdot\tilde{\epsilon}_{{s}} & \mathrm{otherwise}
                                         \end{array}\right.\,.
\end{eqnarray}
In particular, the background gauge field \emph{does} enter into the functional form of the integrand \eqref{cgf3p5}. The explicit form of the integrand is:
\begin{multline}\label{cgf3p6}
 \cA_{3}(a_{1},a_{2},a_{3})=-\frac{\tilde{\epsilon}_{1}\cdot \tilde{\epsilon}_{3}}{{k_{2\,0}}}\left[\left({k_{1\,0}}\,k_{2}\cdot \tilde{\epsilon}_{2}-{k_{2\,0}}\, k_{1}\cdot \tilde{\epsilon}_{2}\right) 
 +
\sA\cdot\tilde{\epsilon}_{2}\left({k_{1\,0}}e_2- {k_{2\, 0}}e_1\right)\right]
+\mathrm{cyclic}\,.
\end{multline}
Crucially, the terms linear in $\sA$ give a background-dependent correction to the flat space result analogous to the tail terms involving $\sigma_{ab}$ appearing in the gravity integrand \eqref{gr3p4}.  In both cases, they encode the memory. 



\subsection{The double copy}
\label{TDC}

Armed with explicit formulae for the 3-point integrands on both gravitational and gauge theory plane wave backgrounds, a precise statement of double copy can now be made. From \eqref{cgf3p6}, the 3-point integrand for gluons on the gauge theory plane wave background can be written compactly as:
\be\label{dc1*}
 \cA_{3}(a_{1},a_{2},a_{3})= F(\{k_{{r}\,0},k_{{r}\,a},\tilde{\epsilon}_{{r}}\}) + \,\mathsf{C}(\{k_{{r}\,0},k_{{r}\,a},\tilde{\epsilon}_{{r}}\}|\sA)\,,
\ee
where the function
\begin{equation}\label{flatamp}
F(\{k_{{r}\,0},k_{{r}\,a},\tilde{\epsilon}_{{r}}\}):=-\frac{\tilde{\epsilon}_{1}\cdot \tilde{\epsilon}_{3}}{k_{2\, 0}}\left(k_{1\,0}\,k_{2}\cdot \tilde{\epsilon}_{2}-k_{2\,0}\,k_{1}\cdot \tilde{\epsilon}_{2}\right) +\mathrm{cyclic}
\end{equation}
is the `flat' contribution to the integrand.\footnote{The spurious poles in $k_0$ are associated with our projection of the polarisation vectors $\epsilon_a$ to be orthogonal to both $\p_u$ and $\p_v$.} The tail-dependent correction term is
\begin{equation}\label{gcorr1}
\mathsf{C}(\{k_{{r}\,0},k_{{r}\,a},\tilde{\epsilon}_{{r}}\}| \sA):= \frac{\tvepsilon_{1}\cdot\tvepsilon_{3}}{k_{2\, 0}}\,\sA\cdot\tilde{\epsilon}_{2}(k_{1\,0}e_2-k_{2\,0}e_1)+ \mathrm{cyclic}
\end{equation}
Note that both $F$ and $\mathsf{C}$ are real functions, in the sense that they take real values provided the kinematic data is real-valued.

To double copy the integrand \eqref{dc1*}, one performs a sequence of simple steps:
\begin{enumerate}

 \item Flip the charge (i.e., the sign of the colour factor of $\sA$) to define 
 $\widetilde{\mathcal{A}}_3=F-\mathsf{C}$ and regard this as the conjugate of $\cA_{3}$:
 \be\label{dcs1}
|\cA_3|^2:= \cA_{3}\,\widetilde\cA_3=
  F^{2}(\{k_{{r}\,0},k_{{r}\,a},\tilde{\epsilon}_{{r}}\}) - \mathsf{C}^{2}(\{k_{{r}\,0},k_{{r}\,a},\tilde{\epsilon}_{{r}}\}|\sA)
 \ee
 
 \item Replace every spatial $(d-2)$-momentum 
by a curved version using the vielbein of the gravitational plane wave background (e.g., $k_{1\,a}\rightarrow k_{1\,i} E^{i}_{a}$). Replace the gauge background polarisations $\tilde{\epsilon}_a$ with gravitational background polarisations $\epsilon_a$. This yields\footnote{The latter operation is just a relabelling by removing all tildes. In particular, this replacement implies $\tvepsilon_{{r}}\cdot\tvepsilon_{{s}} \rightarrow \vepsilon_{{r}}\cdot\vepsilon_{{s}} $.}

 \be\label{dcs2}
 F^{2}(\{k_{{r}\,0},k_{{r}\,i} E^{i}_{a},\epsilon_{{r}}\}) - \mathsf{C}^{2}(\{k_{{r}\,0},k_{{r}\,i} E^{i}_{a},\epsilon_{{r}}\}|\sA)\,.
 \ee
 
 \item Replace the remaining (quadratic) dependence on the background gauge field with dependence on the background gravitational field using the rule:
 \be\label{dcs3}
 e_{r}e_{s}\,\sA^{a}\,\sA^{b} \rightarrow \left\{
 \begin{array}{c c}
   \im\, k_{{r}\,0}\,\sigma^{ab} & \mbox{if  } {r}={s} \\
   \im\, (k_{{r}\,0}+k_{{s}\,0})\,\sigma^{ab} & \mbox{otherwise}
 
 \end{array}\right.\,,
 \ee
 where $e_{{r}}$ is the charge under the background gauge field associated with external state ${r}=1,2,3$.
\end{enumerate}

The final step is motivated by dimensional considerations and suggested by the fact that $\sA_a$ encodes the gauge theory memory effect; if it is set to vanish in the in-region it will generically be a non-zero constant in the out-region remembering an integral of the field. Thus the quadratic combination $\sA_{a}\,\sA_{b}$ is where the memory effect can be seen in the amplitude.  In the gravitational case, the deformation tensor $\sigma_{ab}$ can be chosen to vanish in the past, but is then non-trivial in the future, although now  generically falling off asymptotically as  $u^{-1}$, by \eqref{newvb}. 
Therefore, the replacement \eqref{dcs3} identifies the fields responsible for memories, albeit with different functional dependence on $u$.  An additional power of momenta is needed on the gravitational side to ensure that the two combinations have the same mass dimension.    

Steps 1-3 result in an expression of the form
\be\label{dc2}
F^{2}(\{k_{{r}\,0},k_{{r}\,i} E^{i}_{a},\epsilon_{{r}}\}) - \mathsf{C}^{2}(\{k_{{r}\,0},k_{{r}\,i} E^{i}_{a},\epsilon_{{r}}\}|\sigma)\,.
\ee
Working on the support of momentum conservation in the $v$-direction -- which holds regardless of the asymptotic configuration of the three external states -- a bit of algebra reveals that
\be\label{dc3}
\mathsf{C}^{2}(\{k_{{r}\,0},k_{{r}\,i} E^{i}_{a},\epsilon_{{r}}\}|\sigma)=\im\, k_{1\,0}k_{2\,0}k_{3\,0}\,\sigma^{ab}\,\mathcal{C}_{a}\,\mathcal{C}_{b}\,,
\ee
and therefore that the expression \eqref{dc2} is in fact \emph{equal} to the 3-point integrand for gravitons on the gravitational plane wave background. 

\smallskip

There is also a canonical way to map the 3-point integrand for gluons on a gauge theory background to the 3-point integrand for gluons on a gravity background. This entails a `classical' double copy of the background (in the sense of~\cite{Monteiro:2014cda}) while leaving the functional form of the integrand unchanged. To see how this works,  use the integrand expression:
\be\label{sc1}
\cA_{3}(a_{1},a_{2},a_{3})=\tvepsilon_{1}\cdot\tvepsilon_{3}\,\sK_{1}\cdot\tvepsilon_{2}+\tvepsilon_{1}\cdot\tvepsilon_{2}\,\sK_{2}\cdot\tvepsilon_{3} +\tvepsilon_{2}\cdot\tvepsilon_{3}\,\sK_{3}\cdot\tvepsilon_{1}\,,
\ee
where $\sK_{{r}\,a}$ and $\tvepsilon_{{r}\,a}$ are given by \eqref{csmom1}, \eqref{gfpol} for ${r}=1,2,3$. Now perform the following replacements everywhere in \eqref{sc1}:
\be\label{sc2}
k_{{r}\,a}\rightarrow k_{{r}\,i}\,E^{i}_{a}\,, \qquad \tilde{\epsilon}_{{r}\,a}\rightarrow \epsilon_{{r}\,a} \,, \qquad e_{r}\,\sA_{a}\rightarrow k_{{r}\,0}\,\sigma_{ab}\,x^{b}\,.
\ee
The last of these replacements is motivated by the observation that the non-trivial component of the plane wave gauge field, namely $x^{a}\,\dot{\sA}_{a}$ is a linear function of $x$ while the non-trivial component of the plane wave metric, namely $-\ddot{E}^{i}_{a}E_{b\,i}\,x^{a} x^{b}$, is quadratic. 

After making the replacements \eqref{sc2}, the polarisation vectors in the gauge field background are mapped directly onto the polarisation vectors in the gravitational background: $\tvepsilon_{{r}\,\mu}\rightarrow\vepsilon_{{r}\,\mu}$. Although $\sK_{{r}\,\mu}$ is not quite mapped onto $K_{{r}\,\mu}$, it is easy to see that
\begin{equation*}
 \sK_{{r}}\cdot\tvepsilon_{{s}}\rightarrow K_{{r}}\cdot\vepsilon_{{s}}\,.
\end{equation*}
Calling this substitution map $\psi$, it follows immediately that
\be\label{sc3}
\psi\left(\cA_{3}(a_{1},a_{2},a_{3})\right)=\cM_{3}(A_1, A_2, A_3)\,,
\ee
where the two integrands have the same kinematic data but are defined on different backgrounds.


\subsection{Discussion}
\label{Discuss}

In this chapter we have made a preliminary investigation of how the notion of double copy generalises to curved scattering backgrounds starting with the three point amplitude on sandwich plane waves.  We find new features, but see that the double copy nevertheless does extend to this curved setting: 3-point graviton amplitudes on a plane wave spacetime can be obtained by taking the double copy of 3-point gluon amplitudes on a gauge theory plane wave background.

This statement can be expressed succinctly by encoding steps 2 and 3 of the double copy procedure in  a `replacement map' $\rho$, that acts on the spaces of $(d-2)$-kinematics and background gauge fields. The double copy for 3-point integrands on plane wave backgrounds is then simply:
\be\label{dc4}
\boxed{\cM_{3}(h_1, h_2, h_3)= \rho\left(|\cA_{3}(a_1, a_2, a_3)|^2\right)\,.}
\ee
This is consistent with the usual double copy on flat backgrounds expressed in  the KLT relations. In a flat background, $\rho$ acts trivially and this is the usual squaring relation.

We have only investigated the simplest scattering amplitudes (i.e., 3-point amplitudes), which are generated by contact interactions in the spacetime action. Higher-point amplitudes will involve propagator contributions; although explicit forms for propagators on plane wave backgrounds are known (e.g., \cite{Wolkow:1935zz,Ilderton:2012qe,Gibbons:1975jb,Harte:2012uw}), these are significantly more complicated that those arising from flat space. Nevertheless, the prescription given in chapter~\ref{TDC} seems universal: it dictates how to double copy the data for any $n$-point scattering amplitude.
Steps 1-3 do not depend on the number of external particles being three. So one can  optimistically conjecture a heuristic form of the double copy for $n$-point integrands on plane wave backgrounds:
\be\label{dc5}
\cM_{n}(h_{1},\ldots,h_{n}) = \rho\left(\sum_{\alpha,\beta\in S_{n}/\Z_{n}} \cA_{n}(\alpha)\,\mathcal{S}^{\sA}[\alpha|\beta]\,\widetilde{\cA}_{n}(\beta)\right)\,,
\ee
where the sum is over distinct colour-orderings for the $n$-point integrands on the gauge theory background, $\rho$ is the replacement map defined by steps 2 and 3 of the double copy, $\widetilde{\cA}_{n}$ is the integrand with opposite charges for the background and $\mathcal{S}^{\sA}[\alpha|\beta]$ is a plane wave analogue of the KLT matrix (perhaps obtained from the same replacement algorithm for the momenta).  However, now the $\cA$ and $\widetilde \cA$ must incorporate the non-trivial propagators on those backgrounds, and it is likely that these must also be subject to some replacement to work correctly on a gravitational background.

\smallskip

Our procedure is not a straightforward local identification of integrands.  It requires the replacement of certain structural functions appropriate for propagation on a gauge theory background by those for a gravitational background.  Indeed, colour/kinematics duality is usually expressed locally in momentum space, and so should not be expected to be local in spacetime.   Here we see evidence that a non-local procedure based on  Hamilton-Jacobi functions for propagation of momentum eigenstates from null infinity will do the trick.  Thus, the most optimistic message from this for the general curved colour-kinematic duality is that although a spacetime procedure cannot be local, it can work  by referring to null infinity, using Hamilton-Jacobi generating functions to create the identifications.

It would also be desirable to extend the double copy to other curved backgrounds. Although plane waves are a very special example of such backgrounds, there is some sense in which they are universal limits of \emph{all} spacetimes~\cite{Penrose:1976}. It would be interesting to see in what sense the results found here inform those for more general spacetimes.

Finally, we note that our original motivation for considering scattering on plane wave backgrounds was to provide a spacetime result to compare with an alternative calculation of these amplitudes using ambitwistor string theory~\cite{Mason:2013sva} adapted to a curved background~\cite{Adamo:2014wea}. As we will show in the next chapter, ambitwistor strings provide an alternative `stringy' approach to calculating amplitudes on curved backgrounds which gives pure field theory amplitudes without $\alpha'$ corrections.  The use of Hamilton-Jacobi functions to bring in momenta and polarisation vectors from null infinity should then tie in with the work in \cite{Adamo:2014yya,Geyer:2014lca,Adamo:2015fwa} where ambitwistor strings are formulated at null infinity.

\newpage

\section{Amplitudes from ambitwistor strings}\label{C4.Worldsheet}

Using perturbative string theory to study physics in the presence of curved background fields is a highly non-trivial task. When coupled to generic curved background fields, the string worldsheet action becomes a complicated interacting 2d CFT which can only be studied perturbatively (e.g., \cite{AlvarezGaume:1981hn,Braaten:1985is}). Quantum consistency of the worldsheet theory imposes an infinite tower of higher-derivative constraints on the background fields, which at lowest order are the two-derivative equations of motion of field theory~\cite{Fradkin:1985ys,Callan:1985ia,Banks:1986fu,Abouelsaood:1986gd}. It is therefore exceptionally difficult to tell if a given background field configuration satisfies the full string equations of motion, or to compute interesting target space quantities, such as scattering amplitudes, in the resulting worldsheet CFT.

Over the years, some notable exceptions to the first of these difficulties have been found. Vacuum plane wave metrics were argued to be admissible NS-NS backgrounds for string theory due to the vanishing of their higher curvature invariants~\cite{Amati:1988sa,Horowitz:1989bv}. Supergravity solutions based on AdS (times a compact space)~\cite{Freund:1980xh,Schwarz:1983qr} or pp-waves~\cite{Blau:2001ne,Blau:2002dy} supported by Ramond-Ramond flux were argued to be admissible backgrounds for type II string theory on the basis of uniqueness and symmetry constraints for the integrable sigma models with these target spaces~\cite{Metsaev:1998it,Metsaev:2001bj,Metsaev:2002re,Bena:2003wd,Berkovits:2004xu,Arutyunov:2008if,Stefanski:2008ik}. These examples play a central role in the concept of holography~\cite{Maldacena:1997re,Gubser:1998bc,Witten:1998qj} and its plane wave limit~\cite{Berenstein:2002jq}. The class of supersymmetric sigma models with curved target spaces can be expanded to include various integrable deformations (cf. \cite{Delduc:2013qra}), although it is not entirely clear if these deformations define consistent string theories~\cite{Arutyunov:2015mqj,Wulff:2016tju}.

Yet even with consistent string theories on curved backgrounds, writing explicit vertex operators or calculating worldsheet correlation functions has proved virtually impossible\footnote{A notable special case where progress \emph{has} been made is for AdS$_3$, where the worldsheet theory is a $\SL(2,\R)$ WZW model~\cite{Maldacena:2000hw,Maldacena:2000kv,Maldacena:2001km}.}. This is because the worldsheet model -- even if it is integrable -- remains an interacting CFT (as is the case for supersymmetric AdS backgrounds or vacuum plane waves), or because the worldsheet model is known only in Green-Schwarz form (as for the solvable pp-wave sigma models). Although some progress towards writing vertex operators on certain backgrounds has been made (cf. \cite{Jofre:1993hd,Dolan:1999dc,Berkovits:2000yr,Chandia:2013kja}), there is still no intrinsically stringy computation of the 3-point function in a curved background\footnote{\label{FN3Ptcorr}It should be noted that worldsheet methods \emph{have} been used to compute correlators in certain limits~\cite{Minahan:2012fh,Bargheer:2013faa,Minahan:2014usa} or with special configurations of external states~\cite{Berkovits:2012ps,Azevedo:2014rva} in AdS backgrounds. Cubic string field theory has been used to study interactions on pp-wave backgrounds~\cite{Constable:2002hw,Spradlin:2002ar}.}. This seems particularly remarkable given how much attention is paid to such backgrounds in the context of holography, where bulk observables are usually computed from field theory Witten diagrams rather than the string worldsheet.

\medskip

Our goal is to provide the first worldsheet calculation of 3-point functions on curved backgrounds using ambitwistor strings as alternative to standard string theory, making use of the curved background vertex operators found in chapter~\ref{C3.VertexOps}. Remember that in the context of curved background fields, heterotic and type II ambitwistor strings remain free worldsheet CFTs and yield a description of non-linear field theory in terms of a free 2d CFT as shown in chapters~\ref{C3.Heterotic} and~\ref{CurvedTypeII} respectively. 

This suggests that ambitwistor string theories can be used to study perturbative QFT on curved backgrounds. There are promising signs that this could be true: In the special case of four-dimensions, twistor string formulae for gravitational scattering amplitudes have a natural generalisation to gauged supergravity on AdS$_4$~\cite{Adamo:2015ina}. These formulae pass several consistency checks which indicate that they encode tree level physical observables in AdS$_4$ (at least up to boundary terms), but so far a direct link to standard expressions in general spacetime dimension is missing. 


\medskip

In this chapter, which has also been published in~\cite{Adamo:2017sze}, we quantise ambitwistor strings on plane wave backgrounds. Using that they encode the correct spectra of perturbations in terms of explicit (and computationally manageable) vertex operators, we show that they also encode the correct spacetime interactions by computing 3-point functions. Our focus is on the type II and heterotic ambitwistor strings, which describe gravitational and gauge theoretic degrees of freedom on spacetime, respectively. We study two simple examples of curved backgrounds in a RNS worldsheet formalism for the ambitwistor string: vacuum plane wave metrics (type II model) and plane wave abelian gauge fields (heterotic model). In both cases, explicit expressions for the vertex operators for gravitons or gluons from chapter~\ref{C3.VertexOps} are derived using the results from chapter~\ref{C4.Spacetime}. These vertex operators are then used to compute the 3-point worldsheet correlation functions, which match with the known field theory results from chapter~\ref{C4.Spacetime} in each case.  Indeed the computations of~\cite{Adamo:2017nia}, which we presented in chapter~\ref{C4.Spacetime}, were first executed in order to provide a standard spacetime computation against which to check the formulae of this chapter. This confirms the utility of ambitwistor strings in the study of perturbative field theory on curved backgrounds: Such calculations are impossible in ordinary string theory, even in the $\alpha'\rightarrow 0$ limit.

For details how type II and heterotic ambitwistor strings can be defined on curved background fields we refer to chapters~\ref{CurvedTypeII} and~\ref{C3.Heterotic}. Our focus is on background metric fields for the type II model and abelian background gauge fields for the heterotic model. In each case, quantum consistency of the model is equivalent to the usual field equations for the background. Chapter~\ref{grPW} considers the type II model on a vacuum plane wave metric background, and chapter~\ref{gtPW} considers the heterotic model on a plane wave gauge field background. In both cases, the worldsheet theory is anomaly-free because the backgrounds solve the (vacuum) equations of motion. We provide explicit expressions for vertex operators corresponding to gravitons and gluons, respectively, and compute their 3-point functions. These are seen to reproduce the known formulae for graviton and gluon scattering on gravitational and gauge field plane wave backgrounds.


\subsection{Type II model on a gravitational plane wave}
\label{grPW}

We now turn to a class of backgrounds which are among the simplest non-trivial solutions to the vacuum Einstein equations. When discussing gravitational waves, one often thinks about linear perturbations of a fixed background spacetime as first described by Einstein~\cite{einstein1916}. However, there also are exact, non-linear plane wave solutions to the Einstein equations. These have been known for over ninety years and studied in great detail, see e.g.~\cite{Baldwin:1926,Ehlers:1962zz,Griffiths:1991zp,Stephani:2003tm,Blau:2011}. Due to the vanishing of their higher-curvature invariants, it has long been known that certain plane wave metrics solve the vanishing beta-functional conditions for string theory to all orders in $\alpha'$~\cite{Amati:1988sa,Horowitz:1989bv}. Yet it has proven difficult to derive vertex operators for string theory on a plane wave background or indeed compute scattering amplitudes in such spacetimes\footnote{A notable exception is~\cite{Jofre:1993hd}, where candidate tachyon and graviton vertex operators are constructed in the bosonic string on a plane wave.}.

In this chapter, we study the type II ambitwistor string on vacuum plane wave spacetimes. As we know from chapter~\ref{CurvedTypeII} and~\ref{C3.VertexOps}, the worldsheet OPEs remain free and vertex operators can be constructed explicitly. The simplicity of the background spacetime renders the calculation of worldsheet correlation functions tractable. We start with a brief discussion of the worldsheet model and fixed vertex operators on plane waves in order to keep this chapter self-contained, after which we go on to calculate the descended vertex operator and show that the 3-point correlation functions on the Riemann sphere reproduce the known result for 3-point graviton amplitudes on a plane wave spacetime.


\subsubsection{Worldsheet model and fixed vertex operators}

As solutions of the vacuum Einstein equations ($H_{a}^{a}(u)=0$ in Brinkmann coordinates), vacuum plane wave metrics are admissible backgrounds for the type II ambitwistor string in the sense that $Q^2=0$, up to a conformal anomaly which can be killed by setting $d=10$, or ignored for our purposes at genus zero. However, a remarkable simplification occurs in the functional form of the BRST charge for any plane wave background in Brinkmann coordinates: all quantum corrections to the currents $\cG$, $\bar{\cG}$ and $\cH$ vanish! Indeed, a direct calculation from \eqref{GeneralG} -- \eqref{GeneralH} using the Christoffel symbols \eqref{Christoffel} leads to:
\begin{align}
\begin{split}
\label{BrinkmannG}
\mathcal{G} &= \psi^\mu\, \Pi_\mu\,,
\\
\bar{\mathcal{G}} &= g^{\mu \sigma}\, \bar{\psi}_\mu\, \Pi_\sigma + 2H^{a}_{b}\,x^{b}\, \bar{\psi}_v \,\bar{\psi}_a\,  \psi^u\,,
\end{split}
\end{align}
\begin{align}
\begin{split}
\label{BrinkmannH}
\mathcal{H} &= g^{\mu\sigma}\,\Pi_{\mu}\,\Pi_{\sigma} + \Pi_v \left(2 H^{a}_{b}\,x^{b}\, \bar{\psi}_a  \psi^u +2 H_{ab}\,x^{b}\,\bar{\psi}_{v}\psi^{a}+ \dot{H}\, \bar{\psi}_v \psi^u \right)
\\
& \qquad  -2\Pi_{a}\,H^{a}_{b}\,x^{b}\,\bar{\psi}_{v}\psi^{u} +2H^{a}_{b}\,\bar{\psi}_{a} \bar{\psi}_v\, \psi^b \psi^u\,.
\end{split}
\end{align}
In particular, all terms proportional to worldsheet derivatives in $\cG$, $\bar{\cG}$ and $\cH$ vanish in Brinkmann coordinates for the plane wave background.

Let us now consider a metric perturbation $h_{\mu\sigma}$ (i.e., a graviton) on the plane wave background. As in chapter~\ref{FF}, we find that it is consistent to further impose the gauge conditions that the $v$-components and trace of $h$ should vanish $h_{v \mu}=0=h^{\mu}_{\mu}$. Recall that we can make use of the graviton vertex operator~\eqref{gravityop} on a \emph{pure gravity} background and notice that the quantum correction term proportional to the worldsheet derivative will drop out in the gauge chosen here:
\begin{align}
\begin{split}
\label{FixedVOBrinkmann}
V= c\, \bar{c}\, \delta(\gamma)\, \delta(\bar{\gamma})\, \bar{\psi}_\mu\, \psi^\sigma h^\mu_\sigma\,
\end{split}
\end{align}

Remember that since $V$ has balanced conformal weight, the stress tensor part of the BRST charge acting on this vertex operator vanishes. The only non-trivial conditions for BRST-closure of $V$ arise from the currents $\cG$, $\bar{\cG}$ and $\cH$, which force the graviton to be on-shell. Due to the many cancellations in the currents and vertex operator, these calculations simplify considerably compared to the generic case in chapter~\ref{TypeIIVO}. The relevant OPEs are
\begin{align}
\begin{split}
\label{DoublePolesFermion}
\cG(z)\,\bar{\psi}_{\mu}\psi^{\sigma} h^{\mu}_{\sigma}(w) &\sim  -\frac{\partial_{\mu}h^{\mu}_{\sigma}\psi^{\sigma}}{(z-w)^2} + \frac{1}{z-w}\, (\cdots)\,,
\\
\bar{\cG}(z)\,\bar{\psi}_{\mu}\psi^{\sigma} h^{\mu}_{\sigma}(w) &\sim  \frac{g^{\rho\sigma}}{(z-w)^2} \partial_\rho h^\mu_\sigma \bar{\psi}_\mu + \frac{1}{z-w}\, (\cdots)\,
\end{split}
\end{align}
and
\begin{multline}
\cH(z)\,\bar{\psi}_{\mu}\psi^{\sigma} h^{\mu}_{\sigma}(w)\sim \frac{1}{(z-w)^2} \bigg[g^{\rho\lambda}\partial_\rho \partial_\lambda h^\mu_\sigma\, \bar{\psi}_\mu \psi^\sigma +2H^{a}_{b}x^{b}\,\partial_{v}h^{\mu}_{a}\,\bar{\psi}_{\mu}\psi^{u} \\
+2H^{a}_{b}x^{b}\,\partial_{v}h_{\sigma a}\,\bar{\psi}_{v}\psi^{\sigma} - 2 H^{ab}\,h_{ab}\,\bar{\psi}_{v}\psi^{u}\bigg] + \frac{1}{z-w}\,(\cdots)\,.
\end{multline}
As expected, vanishing of these equations enforces de Donder gauge~\eqref{deDonder} and the linearised equations of motion~\eqref{linEin} in their guise from chapter~\ref{FF}. Recall in particular that the linearised Einstein equations took the form~\eqref{graveom2} there.

A concrete realisation of $h_{\mu\sigma}$, analogous to the momentum eigenstate used in flat space \eqref{IIfixed} has been constructed in chapter~\ref{RaisingGraviton} by a spin-raising procedure~\cite{Mason:1989,Adamo:2017nia} applied to solutions of the scalar wave equation on a plane wave background, first constructed in~\cite{Ward:1987ws}. Remember that the key to this construction are solutions to the Hamilton-Jacobi equations of the form~\eqref{phi}:
\begin{equation*}
\phi_k=k_0\,v + \frac{k_0}{2}\,\sigma_{ab}\,x^{a}x^{b}+k_i\,E^{i}_{a}\,x^a-\frac{k_{i}\,k_j}{2\,k_0}\,F^{ij}\,,
\end{equation*}
where $(k_{0},k_i)$ were $d-1$ constants (which parametrise the non-trivial components of a null momentum), $E^i_a$ is the vielbein appearing in the relationship between Brinkmann and Einstein-Rosen coordinates \eqref{diffeo1}, $\sigma_{ab}=\dot{E}^{i}_{a} E_{b\,i}$ is the deformation tensor~\eqref{shear} and in equation~\eqref{Fij} we defined
\begin{equation*}
F^{ij}(u):=\int^{u}\d s\,\gamma^{ij}(s) = \int^{u} \d s\,E^{a\,(i}(s)\,E_{a}^{j)}(s)\,.
\end{equation*}
Recall that the choice of vielbein is not unique: given any $E^a_i$ for a particular plane wave metric, any other vielbein given in equation~\eqref{newvb} also represents the same metric. For a sandwich plane wave, the two particular choices of boundary condition given in equation~\eqref{vbbc0} are relevant:
\begin{equation*}
\lim_{u\rightarrow\pm\infty}E^{a\,\pm}_{i}=\delta^{a}_{i}\,.
\end{equation*}
These correspond to whether $\phi_k$ looks like $k\cdot X$ in the in- or out-regions of the sandwich plane wave; since we will always be considering amplitudes in which all external states have the same boundary conditions, we assume that $E^{a}_{i}=E^{a\,-}_{i}$ without loss of generality from now on.

Equipped with the function $\phi_k$, the graviton $h_{\mu\sigma}$ is given by equation~\eqref{grsol2}:
\begin{equation*}
h_{\mu\sigma}\,\d X^\mu\, \d X^\sigma= \left((\varepsilon\cdot \d X)^2-\frac{\im}{k_0}\,\epsilon^{a}\epsilon^{b}\,\sigma_{ab}\, \d u^2\right)\Omega\, \e^{\im \phi_k}
\end{equation*}
Here, $\varepsilon_{\mu}$ is the (non-constant) $d$-dimensional polarisation vector defined in equation~\eqref{polar}
\begin{equation*}
\varepsilon\cdot \d X=\epsilon_a\d x^a + \epsilon^a\left(\frac{k_j}{k_0}E^j_a +  \,\sigma_{ab}\, x^b\right)\d u
\end{equation*}
with $\epsilon_a$ a constant $(d-2)$-dimensional null vector. Remember that the function $\Omega(u)$ was defined in equation~\eqref{scalsol}
\begin{equation*}
\Omega(u):=|\gamma^{-1}(u)|^{1/4} = |E(u)|^{-\frac{1}{2}}\,.
\end{equation*}
It is straightforward to verify that this $h_{\mu\sigma}$ is traceless, satisfies $h_{v\mu}=0$, obeys the de Donder gauge conditions, and solves the linearised Einstein equations \eqref{graveom2}. In demonstrating this, it is useful to remember~\eqref{momentum}, which defines the momentum associated with the graviton:
\begin{multline*}
K_{\mu}\,\d X^\mu := \d\phi_{k}=\\
 k_0\,\d v
 +\left( \frac{k_0}{2}\,\dot{\sigma}_{bc}\,x^{b}x^{c}+k_{i}\dot{E}^{i}_{b}x^{b}+\frac{k_{i}k_{j}}{2k_0}\gamma^{ij}\right)\d u+(k_{i}E^{i}_{a}+k_{0}\,\sigma_{ab}x^{b})\d x^a\,.
\end{multline*}
This is null with respect to the plane wave metric ($K^2=g^{\mu\nu}K_{\mu}K_{\nu}=0$), and is also compatible with the polarisation vector~\eqref{polar} in the sense that $g^{\mu\nu}\vepsilon_{\mu} K_{\nu}=0$.


\subsubsection{Descent}

Having constructed fixed graviton vertex operators for the type II model on a plane wave background, one can now ask for vertex operators in other pictures. For the calculation of worldsheet correlation functions, it is particularly important to have the vertex operators of zero picture number (i.e., without any $\delta(\gamma)$ or $\delta(\bar{\gamma})$ insertions), usually obtained via the fermionic descent procedure from fixed vertex operators (cf. \cite{Friedan:1985ge}). This procedure entails successively extracting the simple poles between the fixed vertex operator and the currents $\cG$, $\bar{\cG}$, and results in the picture number zero vertex operator, defined up to gauge transformations. 

In flat space, the pure gauge contributions are proportional to $k\cdot P$ (i.e., the scattering equations), but on a curved background they can be much more subtle. To isolate the gauge-invariant portion of the descended vertex operator, we can exploit the fact that $\{\cG,\bar{\cG}\}=\cH$ quantum mechanically. First compute the descended operator by isolating the simple poles of $\bar{\cG}(\cG V)$, and then again by isolating the simple poles of $\cG(\bar{\cG}V)$. Since $\{\cG,\bar{\cG}\}=\cH$, it follows that the sum of the two resulting operators must be pure gauge, while the difference will be the gauge-invariant contribution to the descended operator.

To do this, first compute
\be\label{gsp1}
\cG(z)\,\bar{\psi}_{\mu}\psi^{\sigma}h^{\mu}_{\sigma}(w) \sim \frac{1}{z-w}\left(\Pi_{\mu}\,\psi^{\sigma}\,h^{\mu}_{\sigma} - \bar{\psi}_{\mu}\psi^{\sigma}\psi^{\rho}\,\partial_{\rho}h^{\mu}_{\sigma}\right)+\cdots\,,
\ee
\begin{multline}\label{gsp2}
\bar{\cG}(z)\,\bar{\psi}_{\mu}\psi^{\sigma}h^{\mu}_{\sigma}(w)\sim \frac{1}{z-w}\left(-g^{\sigma\rho}\,\Pi_{\rho}\,\bar{\psi}_{\mu}\,h^{\mu}_{\sigma} - g^{\lambda\rho}\,\bar{\psi}_{\lambda}\bar{\psi}_{\mu}\psi^{\sigma}\,\partial_{\rho}h^{\mu}_{\sigma}\right. \\
\left.-2 H^{a}_{b}\,x^{b}\,\bar{\psi}_{v}\bar{\psi}_{c}\psi^{u}\,h^{c}_{a}\right)+\cdots\,,
\end{multline}
where the $+\cdots$ indicate higher-order poles. Then take the simple pole of $\bar{\cG}$ with \eqref{gsp1} and of $\cG$ with \eqref{gsp2}, and compute the difference. Thanks to the current algebra $\{\cG,\bar{\cG}\}=\cH$ (which holds at the quantum level since the plane wave metric solves the vacuum equations of motion), the result is gauge invariant and BRST-closed.

A straightforward, if somewhat tedious, calculation then reveals the form of the picture number zero vertex operator:
\begin{multline}\label{dgvo}
c\,\bar{c}\,U=c\,\bar{c}\left[ \Pi_{\mu}\,\Pi_{\sigma}\,h^{\mu\sigma}-\Pi_{\mu}\,\bar{\psi}_{\rho}\psi^{\sigma}\,\partial^{\rho}h^{\mu}_{\sigma} +\Pi_{\sigma}\,\bar{\psi}_{\mu}\psi^{\rho}\,\partial_{\rho}h^{\mu\sigma} +\bar{\psi}_{\rho}\bar{\psi}_{\mu}\psi^{\sigma}\psi^{\lambda}\, \partial^{\rho}\partial_{\lambda} h^{\mu}_{\sigma}\right. \\
+H_{ab}\,x^{b}\,\Pi_{v}\,\bar{\psi}_{v}\psi^{\sigma}\,h^{a}_{\sigma} + H^{a}_{b}\,x^{b}\,\Pi_{v}\,\bar{\psi}_{c}\psi^{u}\,h^{c}_{a}-H^{a}_{b}\,x^{b}\,\Pi_{v}\,\bar{\psi}_{v}\psi^{u}\,h^{v}_{a}-2H^{a}_{b}\,x^{b}\,\Pi_{c}\,\bar{\psi}_{v}\psi^{c}\,h^{c}_{a} \\
+H^{a}_{b}\,\bar{\psi}_{v}\psi^{u}\left(x^{b}\,\bar{\psi}_{c}\psi^{d}\,\partial_{a}h^{c}_{d}-\bar{\psi}_{c}\psi^{b}\,h^{c}_{a}-\bar{\psi}_{a}\psi^{c}\,h^{b}_{c}-2x^{b}\,\bar{\psi}_{c}\psi^{\rho}\,\partial_{\rho}h^{c}_{a}\right) \\
+H^{a}_{b}\,x^{b}\,\bar{\psi}_{a}\bar{\psi}_{\mu}\psi^{u}\psi^{c}\,\partial_{v}h^{\mu}_{c}-\frac{\dot{H}}{2}\,\bar{\psi}_{v}\bar{\psi}_{a}\psi^{u}\psi^{\sigma}\,\partial_{v}h^{a}_{\sigma}-H_{bc}\,x^{c}\,\bar{\psi}_{v}\bar{\psi}_{a}\psi^{b}\psi^{\sigma}\,\partial_{v}h^{a}_{\sigma} \\
\left.\frac{\partial H}{2}\,\bar{\psi}_{\mu}\psi^{\sigma}\,\partial^{2}_{v}h^{\mu}_{\sigma}-\partial(H_{ab}\,x^{b}\,\bar{\psi}_{v})\,\psi^{\sigma}\,\partial_{v}h^{a}_{\sigma}-\partial(H^{a}_{b}\,\bar{\psi}_{v}\psi^{u})\,h^{b}_{a} + \bar{\psi}_{\mu}\,\partial(H^{a}_{b}\,x^{b}\,\psi^{u})\,\partial_{v}h^{\mu}_{a}\right]\,.
\end{multline}
Although we have succeeded in giving an explicit formula for the descended vertex operator on a plane wave background (something which, so far, has been impossible in ordinary string theories), the result is a rather unwieldy expression. Indeed, one might worry that \eqref{dgvo} is so complicated that it is impossible to actually obtain tractable formulae for worldsheet correlation functions -- even at 3-points. Thankfully, this is not the case: many of the terms appearing in \eqref{dgvo} do not contribute to the worldsheet correlator at 3-points, and one is left with a much more manageable operator to deal with.


\subsubsection{3-point function}

Using the vertex operators and the explicit representation for the graviton $h_{\mu\sigma}$, we want to compute the 3-point worldsheet correlation function:
\be\label{II3p1}
\left\la V_{1}(z_1)\,V_{2}(z_2)\,c(z_3)\bar{c}(z_3)\,U_{3}(z_3)\right\ra\,,
\ee
at genus zero, $\Sigma\cong\CP^1$. From \eqref{FixedVOBrinkmann} and the gauge condition $h_{v\mu}=0$, it follows that the fixed vertex operator $V_i$ does not contain any insertions of $\bar{\psi}_u$ or $\psi^{v}$. This means that any insertions of $\bar{\psi}_v$ or $\psi^u$ appearing in $U_3$ have no conjugate fields with which to Wick contract in the correlator \eqref{II3p1} due to normal ordering. Since $\bar{\psi}_{v}(z_3)$, $\psi^{u}(z_3)$ have no zero modes at genus zero, it follows that all terms in $U_3$ which are proportional to $\bar{\psi}_v$ or $\psi^u$ cannot contribute to the 3-point correlator.

This drastically reduces the number of terms which need to be considered in the descended vertex operator \eqref{dgvo}:
\begin{multline}\label{egU1}
U\rightarrow \Pi_{\mu}\,\Pi_{\sigma}\,h^{\mu\sigma}-\Pi_{\mu}\,\bar{\psi}_{\rho}\psi^{\sigma}\,\partial^{\rho}h^{\mu}_{\sigma} +\Pi_{\sigma}\,\bar{\psi}_{\mu}\psi^{\rho}\,\partial_{\rho}h^{\mu\sigma} +\bar{\psi}_{\rho}\bar{\psi}_{\mu}\psi^{\sigma}\psi^{\lambda}\, \partial^{\rho}\partial_{\lambda} h^{\mu}_{\sigma} \\
+\frac{\partial H}{2}\,\bar{\psi}_{a}\psi^{b}\,\partial^{2}_{v}h^{a}_{b}\,.
\end{multline}
In fact, the term in the second line, proportional to $\partial H$, can also be discarded. Since there are no $\Pi_{\mu}$ insertions in $V_1$ or $V_2$ which can contract with $\partial H$, it follows that $H(X)$ can be treated as a function of the zero-modes of the worldsheet field $X^{\mu}$. These zero-modes are constants on the worldsheet, and thus $\partial H=0$. So for the 3-point function \eqref{II3p1}, one is able to consider an `effective' descended vertex operator:
\be\label{egU2}
U^{\mathrm{eff}}=\Pi_{\mu}\,\Pi_{\sigma}\,h^{\mu\sigma}-\Pi_{\mu}\,\bar{\psi}_{\rho}\psi^{\sigma}\,\partial^{\rho}h^{\mu}_{\sigma} +\Pi_{\sigma}\,\bar{\psi}_{\mu}\psi^{\rho}\,\partial_{\rho}h^{\mu\sigma} +\bar{\psi}_{\rho}\bar{\psi}_{\mu}\psi^{\sigma}\psi^{\lambda}\, \partial^{\rho}\partial_{\lambda} h^{\mu}_{\sigma}\,,
\ee
with $h_{\mu\sigma}$ given by~\eqref{grsol2}.

The ghost sector of the correlation function decouples from the matter systems, so it is easy to see that the worldsheet correlator reduces to:
\begin{multline}\label{II3p2}
\left\la V_{1}(z_1)\,V_{2}(z_2)\,c(z_3)\bar{c}(z_3)\,U_{3}(z_3)\right\ra = 
\\
\frac{z_{23}^2\,z_{31}^2}{\d z_{1}\,\d z_{2}\,\d z_{3}^2}\,\left\la \bar{\psi}_{\mu}\psi^{\sigma} h_{1\,\sigma}^{\mu}(z_1)\,\bar{\psi}_{\rho}\psi^{\lambda} h_{2\,\lambda}^{\rho}(z_2)\,U_3^{\mathrm{eff}}(z_3)\right\ra_{\Pi X}^{\bar{\psi}\psi}\,,
\end{multline}
where $z_{ij}\equiv z_{i}-z_{j}$ and $\la \cdots \ra_{\Pi X}^{\bar{\psi}\psi}$ indicates a correlation function with respect to the $(\Pi,X)$ and $(\bar{\psi},\psi)$ worldsheet systems. To evaluate the remaining correlation function, it is useful to have explicit expressions for the (effective) vertex operators:
\be\label{IIfe}
\bar{\psi}_{\mu}\,\psi^{\sigma}\, h_{\sigma}^{\mu}=\left(\vepsilon^{\mu}\bar{\psi}_{\mu}\,\vepsilon_{\sigma}\psi^{\sigma}-\frac{\im}{k_0}\,\bar{\psi}_{v}\,\psi^{u}\,\epsilon^{a}\epsilon^{b}\sigma_{ab}\right)\Omega\,\e^{\im\phi_k}\,,
\ee
and 
\begin{multline}\label{IIde}
 U^{\mathrm{eff}}=\left[(\vepsilon^{\mu}\Pi_{\mu})^2 -\frac{\im}{k_0}\Pi_v^2\,\bar{\psi}_{v}\psi^{u}\,\epsilon^{a}\epsilon^{b}-\im\vepsilon^{\sigma}\Pi_{\sigma}\,(\bar{\psi}_{a}\epsilon^{a}\,K_{\rho}\psi^{\rho}+K^{\mu}\bar{\psi}_{\mu}\,\epsilon_{a}\psi^{a}) \right. \\
 -K^{\rho}\bar{\psi}_{\rho}\,\vepsilon^{\mu}\bar{\psi}_{\mu}\,\vepsilon_{\sigma}\psi^{\sigma}\,K_{\lambda}\psi^{\lambda} +\im k_{0}\,\bar{\psi}_{b}\psi^{a}\sigma_{a}^{b}\,\vepsilon^{\mu}\bar{\psi}_{\mu}\,\vepsilon_{\sigma}\psi^{\sigma} \\
 \left.+\Pi_{v}\,\epsilon^{b}\sigma_{ab}\,\vepsilon^{\mu}\bar{\psi}_{\mu}\,\psi^{a} -\Pi_{v}\,\epsilon^{b}\sigma_{b}^{a}\,\bar{\psi}_{a}\,\vepsilon_{\sigma}\psi^{\sigma}\right]\,\Omega\,\e^{\im\phi_k}\,,
\end{multline}
where the curved-space polarisation $\vepsilon_\mu$ and momentum $K_\mu$ are as given in~\eqref{polar} and~\eqref{momentum}, respectively.

Observe that in the remaining correlation function, the only $v$-dependence is in the exponentials $\e^{\im\phi_k}$ and takes the form $\exp(\im\sum_{r=1}^{3}k_{r\,0}v)$. This can be absorbed into the exponential of the action in the remaining path integral:
\be\label{react}
S_{(\Pi,X)}\rightarrow \frac{1}{2\pi}\int_{\Sigma}\Pi_{\mu}\,\dbar X^{\mu} + 2\pi\im \sum_{r=1}^{3} k_{r\,0}\,v(z)\,\delta^{2}(z-z_r)\,.
\ee
The path integral over $v$ can now be done explicitly; the zero mode integral results in a momentum conserving delta function $\delta(\sum_{r=1}^3 k_{r\,0})$, while the integral over non-zero-modes imposes an equation of motion on the conjugate field $\Pi_v$:
\be\label{pieom}
\dbar \Pi_{v}(z)=2\pi\im\sum_{r=1}^{3}k_{r\,0}\,\delta^{2}(z-z_r) \quad \Rightarrow \quad \Pi_{v}(z)=\d z\,\sum_{r=1}^{3}\frac{k_{r\,0}}{z-z_r}\,.
\ee
This allows us to replace every insertion of $\Pi_v$ in the correlator with the solution obtained from \eqref{pieom}. Similarly, there are no insertions of $\Pi_{u}$ anywhere in the remaining correlator; performing the path integral over the non-zero-modes of $u$ imposes $\dbar u=0$, reducing the worldsheet field $u(z)$ to its constant zero mode everywhere.

This leaves $(\Pi_{a}, x^a)$ as the only fields of the $(\Pi,X)$ system with non-zero-modes still in play. The only $\Pi_a$ insertions appear in $U_{3}^{\mathrm{eff}}$, while there is $x^a$-dependence lurking in the polarisation components $\vepsilon_{u}$ or $\vepsilon^{v}$ as well as in the exponential factors. In the latter case, conservation of the $k_0$ momentum components (and the assumption that all three states are incoming) reduces the exponential dependence on $x^a$ to:
\be\label{expx}
\exp\left(\im E^{i}_{a}\,x^{a}\,\sum_{r=1}^{3}k_{r\,i}\right)\,.
\ee
At this point, the path integral over $\Pi_{a}$ can be done explicitly by taking all possible Wick contractions. After also taking all possible contractions in the $(\bar{\psi},\psi)$ system, the result is the following rather unwieldy-looking expression (stripped of overall factors):
\begin{multline}\label{II3p3}
\left( \frac{\vepsilon_1 \cdot \vepsilon_2}{z_{12}}\right)^2 \left( \im \sum_{r=1}^2 \frac{\vepsilon^\mu_3  {K_r}_\mu}{z_{r3}} \right)^2
-\left(  \frac{\vepsilon_1 \cdot \vepsilon_2}{z_{12}}\right)^2 \left( \im\sigma_{ab} \sum_{r=1}^2 k_{r\,0}\frac{\epsilon^a_3 \epsilon^b_3}{z_{r3}^2} \right)
\\
+2\im\left[ \left(  \frac{\vepsilon_2 \cdot \vepsilon_3}{z_{23}}\right) \left(  \frac{\vepsilon_1 \cdot \vepsilon_2}{z_{12}}\right) \left(  \frac{\vepsilon_1\cdot {K_3}}{z_{13}}\right)  \left( \im \sum_{r=1}^2 \frac{\vepsilon_3\cdot  K_r}{z_{r3}} \right) + (1\leftrightarrow 2)\right]
\\
+2\im\left[ \left(k_{3\,0}  \frac{\vepsilon_2 \cdot \vepsilon_3}{z_{23}}\right) \left(  \frac{\vepsilon_1 \cdot \vepsilon_2}{z_{12}}\right)  \left(  \frac{\epsilon^b_{1} \sigma_{ab} \epsilon_3^a}{z_{13}^2}\right)   + (1\leftrightarrow 2)\right]
\\
+\left[ \left(  \frac{\vepsilon_2 \cdot \vepsilon_3}{z_{23}}\right)\left(  \frac{\vepsilon_2\cdot K_3}{z_{23}}\right) \left(  \frac{\vepsilon_1 \cdot K_3}{z_{13}}\right)  \left(  \frac{\vepsilon_1 \cdot \vepsilon_3}{z_{13}}\right) 
- \left(  \frac{\vepsilon_2 \cdot \vepsilon_3}{z_{23}}\right)^2 \left(  \frac{\vepsilon_1 \cdot K_3}{z_{13}}\right)^2  + (1\leftrightarrow 2)\right]
\\
- 2\left[\left( \im \sum_{r=1}^2 \frac{ {K_r}_v}{z_{r3}} \right)   \left(  \frac{\vepsilon_1 \cdot \vepsilon_2}{z_{12}}   \right)\left(  \frac{\vepsilon_2 \cdot \vepsilon_3}{z_{23}}   \right) \frac{\epsilon^b_1 \epsilon^a_3}{z_{13}}\, \sigma_{ab} + (1\leftrightarrow 2)\right]
\\
+ \im\,k_{3\,0}\left[ \left(  \frac{\vepsilon_2 \cdot \vepsilon_3}{z_{23}}   \right)^2 \frac{\epsilon_1^a \epsilon_1^b \sigma_{ab}}{z_{13}^2} -\left(  \frac{\vepsilon_2 \cdot \vepsilon_3}{z_{23}}   \right) \left(  \frac{\vepsilon_1 \cdot \vepsilon_3}{z_{13}} \right) \frac{\epsilon_1^a \epsilon_2^b \sigma_{ab}}{z_{13} z_{23}} 
 + (1\leftrightarrow 2)\right] 
\\
-\left(  \frac{\vepsilon_1 \cdot \vepsilon_2}{z_{12}}   \right)^2 \left( \frac{\epsilon^a_{3} \epsilon^b_{3}}{k_{3\,0}} \sigma_{ab}  \right) \left(\sum_{r=1}^2 \frac{ {K_r}_v}{z_{r3}} \right)^2+ \im\left[\frac{\epsilon^a_2 \epsilon^b_2}{k_{2\,0}} \sigma_{ab}\, \frac{{k_3}_0^2}{z_{23}^2}  \left(  \frac{\vepsilon_1 \cdot \vepsilon_3}{z_{13}}   \right)^2 + (1\leftrightarrow 2)\right] \,.
\end{multline}
This expression can be considerably simplified by performing the path integral over the remaining $x^a$ zero modes, which results in $d-2$ additional delta functions and a Jacobian factor:
\be\label{adelta}
\delta^{d-2}\!\left(\sum_{r=1}^3 k_{r\,i}\right)\,|E|\,,
\ee
where $|E|$ is the determinant of the vielbein $E_{i}^{a}$. 

On the support of these delta functions, and utilising the curved momentum and polarisation identities~\eqref{polrels} and~\eqref{polrels2}
\begin{eqnarray*}
K_{{r}}\cdot \varepsilon_{{s}}&=&\left\{\begin{array}{c c}
                                          0 & \mathrm{if}\;\; {r}={s} \\
                                          E^{i\,a}(k_{{r}\,0}\frac{k_{{s}\,i}}{k_{s\, 0}}\epsilon_{{s}\,a}-k_{{r}\,i}\epsilon_{{s}\,a})\quad & \mathrm{otherwise}
                                         \end{array}\right. \,,
\\
\vepsilon_{{r}}\cdot\varepsilon_{{s}}&=&\left\{\begin{array}{c c}
                \quad    \qquad  \qquad  0 \qquad\quad\qquad& \quad \mathrm{if}\;\; {r}={s} \\
-\epsilon_{r} \cdot \epsilon_{s} &
\qquad  \mathrm{otherwise}
                                         \end{array}\right.\,,
\end{eqnarray*}
the contribution \eqref{II3p3} can be massaged into a much more palatable form:
\be\label{II3p4}
 \frac{1}{z_{23}^{2}z_{31}^{2}}\left[\left(\vepsilon_1 \cdot \vepsilon_2\, \vepsilon_3\cdot K_2 +\vepsilon_2 \cdot \vepsilon_3\, \vepsilon_1 \cdot K_3+ \vepsilon_1 \cdot \vepsilon_3\, \vepsilon_2\,\cdot K_1 \right)^2 \\
 - \im  k_{1\,0}k_{2\,0}k_{3\,0}\,\sigma^{ab}\,\mathcal{C}_{a}\,\mathcal{C}_{b}\right]\,,
\ee
where 
\be\label{corr1}
\mathcal{C}_{a}:= \vepsilon_{2}\cdot\vepsilon_{3}\,\frac{\epsilon_{1\,a}}{k_{1\, 0}} + \vepsilon_{1}\cdot\vepsilon_{3}\, \frac{\epsilon_{2\,a}}{k_{2\, 0}}+\vepsilon_{1}\cdot\vepsilon_{2}\, \frac{\epsilon_{3\,a}}{k_{3\, 0}}\,,
\ee
encodes a `correction' to the tensor structure of the 3-point function in flat spacetime.

These manipulations leave us with:
\begin{multline}\label{II3p5}
 \left\la \bar{\psi}_{\mu}\psi^{\sigma} h_{1\,\sigma}^{\mu}(z_1)\,\bar{\psi}_{\rho}\psi^{\lambda} h_{2\,\lambda}^{\rho}(z_2)\,U_3^{\mathrm{eff}}(z_3)\right\ra_{\Pi X}^{\bar{\psi}\psi} = \delta^{d-1}\!\left(\sum_{r=1}^{3}k_r\right)\,\frac{\d z_{1}\,\d z_{2}\,\d z_{3}^2}{z_{23}^{2}z_{31}^{2}} \\
 \times \int \frac{\d u}{\sqrt{|E|}}\,\left[\left(\vepsilon_{1}\cdot\vepsilon_{3}\,K_{1}\cdot\vepsilon_{2}+ \mathrm{cyclic}\right)^{2} 
-\im\, k_{1\,0}k_{2\,0}k_{3\,0}\,\sigma^{ab}\mathcal{C}_{a}\mathcal{C}_{b}\right]\, \exp\left(\im F^{ij}\sum_{{s}=1}^{3}\frac{k_{{s}\,i}k_{{s}\,j}}{2k_{{s}\,0}}\right)\,,
\end{multline}
where 
\begin{equation*}
 \delta^{d-1}\!\left(\sum_{{r}=1}^{3}k_{{r}}\right):= \delta\!\left(\sum_{r=1}^{3} k_{r\,0}\right)\, \delta^{d-2}\!\left(\sum_{r=1}^{3} k_{r\,i}\right)
\end{equation*}
encodes all of the delta functions resulting from zero mode integrations. Taking into account the ghost contributions from \eqref{II3p2}, it follows that all dependence on the vertex operator locations $z_i$ drops out (as required for M\"obius invariance of the worldsheet correlator on $\Sigma\cong\CP^1$), leaving
\begin{multline}\label{II3p6}
 \left\la V_{1}(z_1)\,V_{2}(z_2)\,c(z_3)\bar{c}(z_3)\,U_{3}(z_3)\right\ra =
\\
\delta^{d-1}\!\left(\sum_{r=1}^{3}k_r\right)\,\int \frac{\d u}{\sqrt{|E|}}\,\bigg[\left(\vepsilon_{1}\cdot\vepsilon_{3}\,K_{1}\cdot\vepsilon_{2}+ \mathrm{cyclic}\right)^{2} 
\\
 -\im\, k_{1\,0}k_{2\,0}k_{3\,0}\,\sigma^{ab}\mathcal{C}_{a}\mathcal{C}_{b}\bigg]\, \exp\left(\im F^{ij}\sum_{{s}=1}^{3}\frac{k_{{s}\,i}k_{{s}\,j}}{2k_{{s}\,0}}\right)\,.
\end{multline}
The right-hand side of this expression is equal to the 3-point amplitude for (incoming) gravitons on a plane wave spacetime~\eqref{gr3p3}.

\medskip

The result of this calculation can be succinctly summarised as
\be\label{II3p}
\left\la V_{1}(z_1)\,V_{2}(z_2)\,c(z_3)\bar{c}(z_3)\,U_{3}(z_3)\right\ra=\cM_{3}^{\mathrm{pw}}(h_{1},h_{2},h_{3})\,,
\ee
where $\cM_{3}^{\mathrm{pw}}(h_{1},h_{2},h_{3})$ is the 3-point amplitude for graviton scattering on a sandwich plane wave metric. This demonstrates that the type II ambitwistor string encodes the correct interactions for gravity on non-trivial backgrounds in a practical way: the 3-point amplitude is obtained as a worldsheet correlation function with no reference whatsoever to a spacetime action or Lagrangian. Furthermore, the correct linearised external states are automatically encoded by the ambitwistor string's BRST cohomology.


\subsection{Heterotic model on a gauge field plane wave background}
\label{gtPW}

Plane wave gauge fields are the gauge theory analogues of plane wave metrics. The class of such background gauge fields we consider here are the highly symmetric solutions of the Yang-Mills equations valued in the Cartan of the gauge group, which we discussed in chapter~\ref{GTPW}. As such, they are consistent backgrounds for the heterotic ambitwistor string. Like gravitational plane waves, all of the higher curvature invariants (e.g., $F^2$, $F^3$) of a plane wave gauge field vanish, which makes them candidate solutions to the gauge sector of the equations of motion of heterotic or type I string theory, although (to our knowledge) this fact has not been explored in the literature. 

Unlike conventional string theory on a gauge field background, we saw in chapter~\ref{C3.Heterotic} that the heterotic ambitwistor string remains a free worldsheet CFT. To keep this chapter self-contained, we briefly consider the quantisation of the heterotic ambitwistor string on a plane wave gauge field background and derive the fixed gluon vertex operator. We then go on to calculate the descended picture vertex operator and compute the 3-point functions explicitly. Once more, we find that the 3-point correlation functions on the Riemann sphere reproduce the known results for 3-point gluon scattering on a plane wave gauge theory background.


\subsubsection{Worldsheet model and vertex operators}

On a plane wave gauge field background, the heterotic ambitwistor string is anomaly free, up to a conformal anomaly which can be eliminated with an appropriate choice of gauge group and is, in any case, irrelevant for our purposes at genus zero. The currents $\sG$ and $\sH$ of the heterotic ambitwistor string, given by~\eqref{Gcurr} and~\eqref{Hcurr}, take the form
\be\label{gBrG}
\sG=\Psi^{\mu}\,\Pi_{\mu}-\Psi^{u}\,x^{b}\,\dot{\sA}^{\sa}_{b}\,j^{\sa}\,,
\ee
\be\label{gBrH}
\sH=\eta^{\mu\sigma}\,\Pi_{\mu}\,\Pi_{\nu}-2\Pi_{v}\,x^{b}\,\dot{\sA}^{\sa}_{b}\,j^{\sa}+2\Psi^{b}\,\Psi^{u}\,\dot{\sA}^{\sa}_{b}\,j^{\sa}\,.
\ee
Note that there are no terms proportional to worldsheet derivatives in $\sH$, since the background gauge field obeys $\partial^{\mu} \sA_{\mu}=0$ in Brinkmann gauge. 

Equipped with the explicit BRST operator $Q$ and the free worldsheet OPEs, we can investigate the vertex operators in the BRST cohomology. We restrict our attention to vertex operators in the NS sector which should correspond to small perturbations of the background gauge field (i.e., gluons). Such vertex operators can appear with picture number $-1$ or zero. We know the fixed vertex operator from equation~\eqref{CurvedGluonVO} to be
\begin{align}\label{gluonfvo}
V=c\,\bar{c}\,\delta(\gamma)\, \Psi^\mu\, a_\mu^{\sa} \, j^{\sa}\,,
\end{align}
where we chose $a_\mu^{\sa}$ to obey $a_{v}=0$ as in chapter~\ref{RaisingGluon}. Note that although the background gauge field is valued in the Cartan subalgebra, this vertex operator carries generic colour charge with respect to the gauge group. 

As shown in~\ref{glVO}, the $Q$-closure of~\eqref{gluonfvo} enforces the Lorenz gauge condition~\eqref{gQV} and the linearised Yang-Mills equations~\eqref{linYM}. In plane wave backgrounds, the relevant OPEs simplify and take the explicit form
\be\label{gG2pol}
\sG(z)\,\Psi^{\mu} a_{\mu}^{\sa} j^{\sa}(w)\sim -\frac{\partial^{\mu}a^{\sa}_{\mu}\,j^{\sa}}{(z-w)^2} + \frac{1}{z-w}\left(\cdots \right)
\ee
and
\begin{multline}\label{gH2pol}
\sH(z)\,\Psi^{\mu} a_{\mu}^{\sa} j^{\sa}(w)\sim \frac{\Psi^{\mu}\,j^{\sa}}{(z-w)^2}\left[\partial_{\sigma}\partial^{\sigma} a_{\mu}^{\sa}+2 f^{\sa\mathsf{b}\mathsf{c}}\,x^{b}\,\dot{\sA}^{\mathsf{b}}_{b}\,\partial_{v}a^{\mathsf{c}}_{\mu} \right. \\
\left. +2 f^{\sa\mathsf{b}\mathsf{c}}\,\delta^{u}_{\mu}\,\dot{\sA}^{\mathsf{c}\,b}\,a^{\mathsf{b}}_{b}\right] + \frac{1}{z-w} \left(\cdots\right)\,.
\end{multline}
The resulting vanishing conditions agree with the corresponding equations found in chapter~\ref{RaisingGluon}, in particular~\eqref{cgfeom}.
\medskip

One can now ask for an explicit representative of $a_{\mu}$ analogous to a momentum eigenstate in flat space. Such a wavefunction can be obtained from the gauge-covariant spin raising process acting on solutions to the charged wave equation on the plane wave background as described in chapter~\ref{FF}. Recall that the key ingredient is the function~\eqref{cssol2}
\begin{equation*}
\tilde{\phi}_{k}=k_{0}\,v +\left(k_{a}+e\sA_{a}\right)\,x^{a} +\frac{f(u)}{2\,k_0}\,,
\end{equation*}
where $(k_{0},k_a)$ are $d-1$ constants parametrising the momenta, $e$ is the charge of the gluon with respect to the background gauge field under the Cartan of the gauge group, and
\begin{equation*}
f(u):=\int^{u}\d s\,(k_{a}+e \sA_{a}(s))\,(k^{a}+e\sA^{a}(s))\,,
\end{equation*}
as defined in~\eqref{ffunc}. Note that $\sA_{a}$ is not gauge-invariant: the addition of a constant preserves the field strength. We take an in-state representation for which $\sA_{a}=0$ in the in-region of the sandwich plane wave (i.e., $u<u_1$) but $\sA_{a}\neq0$ as $u\rightarrow +\infty$ even though the field strength vanishes in the out region.

In chapter~\ref{RaisingGluon} the gluon $a_{\mu}$ is then constructed from $\tilde{\phi}_k$ as:
\be\label{gluon}
a^{\sa}_{\mu}=\mathsf{T}^{\sa}\,\tilde{\vepsilon}_{\mu}\,\e^{\im\,\tilde{\phi}_k}\,,
\ee
with the generator $\mathsf{T}^{\sa}$ of the gauge group encoding the colour charge, and the polarisation given by equation~\eqref{gfpol}:
\begin{equation*}
\tilde{\vepsilon}_\mu   \d X^\mu=\tilde{\epsilon}_{a} \left(\d x^a  +
\frac{1}{k_0}\,(k^{a}+e\sA^{a})\d u\right)
\end{equation*}
Here, $\tilde{\epsilon}_{a}$ is a constant vector in $d-2$ dimensions, encoding the polarisation information. In equation~\eqref{csmom1} the natural local null momentum associated with the gluon is given as
\begin{align}\label{glumom}
{\sK}_{\mu}\,\d X^\mu & :=-\im\e^{-\im\tilde \phi_k}\,D_{\mu}\,\e^{\im\tilde{\phi}_{k}}\,\d X^\mu \nonumber \\
& = k_0\, \d v+ \frac{1}{2\,k_{0}}(k_{a}+e\sA_{a})(k^{a}+e\sA^{a})\d u +(k_{a}+e\sA_{a})\d x^a\,.
\end{align}
It is straightforward to verify that $\eta^{\mu\sigma} \sK_{\mu}\tilde{\vepsilon}_{\sigma}=\eta^{\mu\sigma}\sK_{\mu}\sK_{\sigma}=0$.

The descended gluon vertex operator (i.e. the picture number zero version of $V$) can be obtained by isolating contributions to the simple pole between $\sG$ and $V$ or equivalently by linearising the constraint $\sH$. The resulting vertex operator is given by~\eqref{Dgluon}, which takes the following explicit form in this case:
\be\label{dgluvo}
 c\,\bar{c}\,U=c\,\bar{c}\,j^{\sa}\left[\Pi_{\sigma}\,a^{\sa\,\sigma}-\Psi^{\sigma}\,\Psi^{\mu}\,\partial_{\sigma} a_{\mu}^{\sa}-f^{\sa\mathsf{b}\mathsf{c}}\,x^{a}\,\dot{\sA}^{\mathsf{b}}_{a}\,\Psi^{u}\,\Psi^{\mu}\,a_{\mu}^{\mathsf{c}}\right]
\ee
Unlike the descended graviton vertex operator on a plane wave metric \eqref{dgvo}, this gluon vertex operator contains only one additional term relative to its flat space counterpart. The third term, proportional to $\dot{\sA}_{a}$, ensures that the resulting operator is covariant with respect to the background gauge field.


\subsubsection{3-point function}

The fixed and descended vertex operators can now be used to compute the 3-point correlation function on the Riemann sphere,
\be\label{H3p1}
\left\la V_{1}(z_1)\,V_{2}(z_2)\,c(z_3)\bar{c}(z_3)\,U_{3}(z_3)\right\ra\,,
\ee
using \eqref{gluon} for an explicit representation of the incoming gluon. In order for the colour structure to produce a non-vanishing result, the sum of charges for the vertex operators under the background gauge field must vanish: $e_1+e_2+e_3=0$. 

The ghost and current algebra portions of the correlator are easily evaluated, leaving an effective correlator:
\begin{multline}\label{H3p2}
 \left\la V_{1}(z_1)\,V_{2}(z_2)\,c(z_3)\bar{c}(z_3)\,U_{3}(z_3)\right\ra = \tr\left(\mathsf{T}_{1}\mathsf{T}_{2}\mathsf{T}_{3}\right) \frac{z_{23}\,z_{31}}{\sqrt{\d z_{1}\,\d z_{2}}\,\d z_3} \\
 \times \left\la \Psi\cdot\tilde{\vepsilon}_{1}(z_1)\,\Psi\cdot\tilde{\vepsilon}_{2}(z_2)\left(\Pi\cdot\tilde{\vepsilon}_3-\Psi\cdot\sK_{3}\,\Psi\cdot\tilde{\vepsilon}_3\right)(z_3)\,\e^{\im(\tilde{\phi}_1+\tilde{\phi}_2+\tilde{\phi}_3)}\right\ra^{\Psi\Psi}_{\Pi X}\,.
\end{multline}
The remaining correlator does not contain any insertions of $\Pi_{u}$ (since $\tilde{\vepsilon}^{u}=0$) so all $u$-dependence is immediately reduced to zero modes. This means that there are no Wick contractions into the $u$-dependent components of momenta $\sK$ or polarisations $\tilde{\varepsilon}$, or into the $u$-dependent terms appearing in the exponential through $\tilde{\phi}_k$. Since the $u$-dependence is totally relegated to zero mode integrations, the remaining fermion correlator can be seen to have the exact same structure as the 3-point function in flat space (cf. \cite{Mason:2013sva}). 

With this in mind, it is easy to see that the remaining correlation function is reduced to:
\be\label{H3p3}
2\im\,\tr\left(\mathsf{T}_{1}\mathsf{T}_{2}\mathsf{T}_{3}\right)\,\delta^{d-1}\!\left(\sum_{r=1}^3 k_r\right) \int \d u\,\left[\tilde{\vepsilon}_{1}\cdot\tilde{\vepsilon}_3\,\sK_{1}\cdot\tilde{\vepsilon}_2+\mathrm{cyclic}\right]\,\exp\left(\im\sum_{r=1}^3\frac{f_r(u)}{2\,k_{r\,0}}\right)\,,
\ee
with the $d-1$ delta functions emerging after performing the zero mode integrals over $v$ and $x^a$. As expected, this is precisely the 3-point amplitude for gluon scattering on a gauge field plane wave background~\eqref{cgf3p4}:
\be\label{H3p4}
\left\la V_{1}(z_1)\,V_{2}(z_2)\,c(z_3)\bar{c}(z_3)\,U_{3}(z_3)\right\ra=\cA_{3}^{\mathrm{pw}}(a_1,a_2,a_3)\,.
\ee
So the heterotic ambitwistor string correctly encodes the interactions of gauge theory on a curved background, with the appropriate linear perturbations (i.e., gluons) emerging from the worldsheet BRST cohomology.


\subsection{Discussion}
The work of~\cite{Adamo:2014wea} showed that ambitwistor strings can be consistently defined on a type II supergravity background.  This suggested that it might be possible to calculate amplitudes on such backgrounds following an extension of the flat space strategy.  In the first part of this chapter, we have seen that this does indeed turn out to be the case on a plane wave background at three points.  In chapter~\ref{C3.Heterotic}, we analogously saw that the heterotic ambitwistor string is consistent on Yang-Mills backgrounds.  In the second part of this chapter, we have found the corresponding result for amplitudes: Non-abelian gluon scattering can correctly be computed from this model on plane wave gauge field backgrounds.  The results are all checked against the three point amplitudes on plane wave backgrounds as computed in chapter~\ref{C4.Spacetime} and~\cite{Adamo:2017nia}.  

There are many further directions to explore. The first perhaps is to go to higher numbers of points. The computations of \cite{Adamo:2017nia} were limited to three points because the propagator would be needed at four points, and that was not available in simple enough form.  In terms of the ambitwistor string in flat spacetime, the new phenomenon at four points is the appearance of integrated vertex operators. These incorporate the scattering equations.  To take the calculations in this chapter to four points we will therefore need to introduce some analogue of scattering equations on a curved background.  If this is successful, they will effectively encode the propagators.

Another natural direction to consider is  other backgrounds, such as (anti-) de Sitter, or black-hole or brane spacetimes.  These offer different challenges, with more sophisticated global issues to be addressed already in the spacetime version of the calculations. In the (anti-) de Sitter case, the background is not actually a vacuum solution with respect to the equations of motion arising in the RNS-like formulation of the ambitwistor string used here. Ostensibly, AdS backgrounds would require a manifestly supersymmetric worldsheet model, such as the pure spinor formalism, where the scalar curvature of the background (times a compact space) is supported by Ramond-Ramond flux. While there has been some progress in describing the pure spinor ambitwistor string on such backgrounds (cf. \cite{Chandia:2015sfa,Chandia:2015xfa,Azevedo:2016zod}), there is currently no formulation which is quantum mechanically consistent as a worldsheet theory. If these issues could be resolved, then it would enable computations akin to the ones performed in this chapter on (A)dS background geometries. 

A further direction is to take more seriously the fact that ambitwistor strings have target ambitwistor space.  We should therefore  construct the corresponding ambitwistor spaces more explicitly for the curved backgrounds under consideration.   We must then learn how to quantise ambitwistor strings for amplitude calculations in such backgrounds.  

A key theme of \cite{Adamo:2017nia} was the extent to which the double copy relationship between gravity and Yang-Mills amplitudes, as expressed for example in colour kinematics duality \cite{Bern:2010ue}, survives in curved space.  The answer was that this is indeed the case with suitable modifications.  However, the curved space formulation of ambitwistor strings in \cite{Adamo:2014wea} was not expressed in such a way that the double copy is apparent. Finding a version of the ambitwistor string which manifests the double copy relationship on a curved background would provide further evidence that colour kinematics duality persists in a useful way on non-trivial backgrounds.

%% file: conclusions.tex
\chapter{Conclusions}\label{conclusions}

In this thesis we have demonstrated how some of the modern techniques of the amplitudes program can be carried over into curved backgrounds.  

Ambitwistor strings share many of the nice properties of more conventional string theories, however without the additional complication of $\alpha'$ corrections. On curved backgrounds this difference makes itself felt in a particularly nice way: the \emph{exact} background equations of motion are imposed by quantum consistency conditions for the heterotic and type II ambitwistor string. In even more stark contrast to conventional string theory, heterotic and type II ambitwistor strings remain free CFTs on curved backgrounds (see chapter~\ref{C3.Heterotic} and~\cite{Adamo:2014wea}). It would be intriguing to verify analogous results for the other ambitwistor string models present in the literature~\cite{Ohmori:2015sha,Casali:2015vta,Azevedo:2017lkz}. 

The only case in which vertex operators for conventional strings on a curved background have been obtained is for AdS$_3$~\cite{Maldacena:2000hw,Maldacena:2000kv,Maldacena:2001km}. The fact that ambitwistor strings remain free theories on curved backgrounds has made it possible to write down fixed vertex operators for \emph{arbitrary} gauge theory and supergravity backgrounds. These vertex operators are BRST closed if and only if they correspond to linear on-shell perturbations of the background fields as shown in chapter~\ref{C3.VertexOps}. A clear open question is to determine the corresponding integrated vertex operators. From a practical point of view, they are simply necessary for calculating correlators for more than three points. However more importantly, integrated ambitwistor string vertex operators encode the scattering equations~\cite{Mason:2013sva}. It is completely unclear if and how they manifest themselves in curved backgrounds.

In view of the previous paragraph, it is maybe not too surprising that there are no string theory calculations of curved background three point functions yet\footnote{Several exceptions and their limitations have been pointed out in footnote~\ref{FN3Ptcorr} on page~\pageref{FN3Ptcorr}.}. Such calculations for three gluon vertex operators on a sandwich plane wave gauge background in the heterotic ambitwistor string and three graviton vertex operators on a sandwich plane wave Einstein gravity background in the type II model have been performed in chapter~\ref{C4.Worldsheet}. In keeping with the spirit of the flat space ambitwistor string, these correlators are shown to agree with the corresponding amplitudes\footnote{Here the use of the word \emph{amplitude} is appropriate due to the special nature of the sandwich plane wave background. See chapter~\ref{FF} for details.} obtained from QFT calculations in these curved backgrounds. Several natural questions arise at this point: Can these correlators be calculated for a higher number of points (this ties in with the problem in the previous paragraph) and if yes, do they still yield tree level amplitudes (this ties in with the open question in the next paragraph). There is a considerable amount of work on ambitwistor strings at loop level~\cite{Adamo:2013tsa,Casali:2014hfa,Adamo:2015hoa,Geyer:2015bja,Geyer:2015jch,Geyer:2016wjx,Geyer:2017ela,Geyer:2018xwu}, can these results be useful for the type II model in this context? Another interesting problem is to consider different backgrounds. Despite the exceptional simplicity of plane waves, the explicit calculations in this chapter are by no means short. So other backgrounds will have to be sufficiently simple to admit explicit solutions for on-shell external states and render calculations feasible. Given that there are no general amplitudes formulae on curved backgrounds, proving that this model yields amplitudes on arbitrary backgrounds seems like very high hanging fruit at this point.
\medskip

Due to their simplicity, (sandwich) plane waves have been a popular playground for quantum field theory in curved backgrounds and it was known that scattering of scalars is well defined~\cite{Gibbons:1975jb,Garriga:1990dp}. In chapter~\ref{C4.Spacetime}, we generalised this result to gluons and gravitons making use of a twistor theory inspired spin raising formula. This then allowed us to compute the corresponding three point amplitudes on these backgrounds. It is subsequently shown that a suitably adapted version of the double copy still works for three points on sandwich plane wave backgrounds. One can optimistically conjecture that this prescription will also hold at a higher number of points. To test this conjecture, one needs to calculate the corresponding $n$-point amplitudes on a plane wave background, which requires explicit and manageable expressions for the gluon and graviton propagators. A Feynman gauge version of the gluon propagator along with a preliminary investigation of the colour kinematics duality on plane waves will appear shortly~\cite{Adamo:2018mpq}\footnote{A Feynman gauge graviton propagator is known to the authors of said paper as well.}. Another obvious direction for future research is the generalisation of the double copy prescription to other curved backgrounds, with the same caveat about calculational feasibility as in the previous paragraph.
\bigskip

The work presented here is a small piece of the puzzle leading to a better understanding of perturbative quantum field theory. It provides a first pillar for the bridge between the flat space QFT~\cite{Kawai:1985xq,Bern:2008qj,Bern:2010ue,Bern:2010yg} and classical incarnation~\cite{Anastasiou:2013hba,Monteiro:2014cda,Luna:2015paa,Ridgway:2015fdl,Borsten:2015pla,Luna:2016due,Goldberger:2016iau,Cardoso:2016amd,Luna:2016hge,Goldberger:2017frp} of the double copy. Both yield very surprising results and their relation and true origin is not fully understood. It also strengthens the role of ambitwistor strings as a simplified testing ground for string theory, something that has previously been explored in~\cite{Ohmori:2015sha,Reid-Edwards:2015stz,Reid-Edwards:2017goq,Roehrig:2017gbt}. It might pave the way to further insight into the nature and origin of the scattering equations and CHY formulae~\cite{Cachazo:2013gna,Cachazo:2013hca,Cachazo:2013iea} by leading towards an understanding on generic backgrounds. 

These and the many other results of the amplitudes community, as well as all the non-perturbative QFT results of the recent years, suggest that it still is a long path to a thorough understanding of quantum field theories. I am confident that scattering amplitudes will keep providing many fascinating and important insights along this way.

%% file: appendix1.tex
\chapter{Usefull current algebra identities}\label{CA}

Recall that for a level $k$ current algebra, the OPE between two currents is simply given by equation~\eqref{calg}
\begin{align*}
j^\sa(z) j^\sb(w) \sim \frac{k\delta^{ab}}{(z-w)^2}+\frac{f^{\sa\sb\sc} j^\sc(w)}{z-w} \,.
\end{align*}
To obtain the anomalies~\eqref{anom2}, we need to know the OPE between a current and a normal ordered product of two currents. From the current algebra OPE, one finds
\begin{align}
j^\sa(z) : j^\sb j^\sc:(w) \sim &\frac{kf^{\sa\sb\sc}}{(z-w)^3}+\frac{k\delta^{\sa\sb}j^{\sc}(w)}{(z-w)^2}+\frac{k\delta^{\sa\sc}j^{\sb}(w)}{(z-w)^2}\nonumber\\
+&\frac{f^{\sa\sb\sd} f^{\sd\sc\se} j^\se(w)}{(z-w)^2} + \frac{f^{\sa\sc\sd} :j^\sb j^\sd:(w)}{z-w} + \frac{f^{\sa\sb\sd} :j^\sd j^\sc:(w)}{z-w}\,,
\end{align}
which can be derived following steps similar to those of the Sugawara construction as described in~\cite{Blumenhagen:2013fgp}. 

Identity~\eqref{nocurr} also plays a crucial role in determining the anolmalies~\eqref{anom2}, however it is difficult to obtain directly form the current algebra OPE. To derive
\begin{align*}
: j^\sa j^\sb:(z) - : j^\sb j^\sa:(z) = f^{\sa\sb\sc} \partial j^\sc(z)\,,
\end{align*}
we use the notation from chatper 11 of~\cite{Blumenhagen:2013fgp}: Given the usual mode expansion 
\begin{align}
j^\sa(z) = \sum_n z^{-n-1}j^\sa_n\,, 
\end{align}
the derivative of the current has modes
\begin{align}
\partial j^\sa_m &= -(m+1)  j^\sa_m \,.
\end{align}
Furthermore, the current algebra OPE\footnote{Here we set $k=0$, it is easy to convince yourself that this does not affect the resulting identity using a symmetry argument.} is equivalent to
\begin{align}
\left[ j^\sa_m,  j^\sb_n \right] &= f^{\sa\sb\sc} j^\sc_{m+n}
\end{align}
and normal ordered products can be expanded as $ : j^\sa j^\sb:(z) = \sum_{m,k} z^{-m-k-2} :j^\sa_m  j^\sb_k:$ with
\begin{align}
: j^\sa_n  j^\sb_m : &= j^\sa_n  j^\sb_m -\theta(n)  f^{\sa\sb\sc} j^\sc_{m+n}
\\
\theta(n) &= \begin{cases} 1 \qquad n \ge 0 \\ 0 \qquad n < 0 \end{cases}\,.
\end{align}
Then we can define $\Delta^{\sa\sb} = : j^\sa j^\sb:(z) - : j^\sb j^\sa:(z)$ and expanding this into modes, taking into account that we can only shift summation variables in normal ordered expressions, yields
\begin{align}
\Delta^{\sa\sb}_n &= f^{\sa\sb\sc} j^\sc_{n} \sum_m \left(1 - \theta(m) - \theta(n-m) \right)
\\
&= f^{\sa\sb\sc} \partial j^\sc_{n}
\end{align}
as expected.

%% file: appendix2.tex
\chapter{Linearised supergravity equations}\label{LinSugra}

In this appendix, we provide some details of the linearisation of the supergravity equations~\eqref{SugraEOM}
\begin{align*}
 &0=R+4\nabla_\mu\nabla^\mu\Phi-4\nabla_\mu\Phi\nabla^\mu\Phi-\frac{1}{12}H^2\nonumber\\
 &0=R_{\mu\nu}+2\nabla_\mu\nabla_\nu\Phi-\frac{1}{4}H_{\mu\rho\sigma}H_\nu{}^{\rho\sigma}\\
 &0=\nabla_\kappa H^\kappa{}_{\mu\nu}-2H^\kappa{}_{\mu\nu}\nabla_\kappa\Phi\,.\nonumber
\end{align*}
The linearised supergravity equations are obtained by perturbing these equations with respect to all three background fields and combining the corresponding linear terms in the scalar, symmetric and antisymmetric equation respectively. We will derive the individual contributions from all three sectors of the spectrum here. It is then easy to see that their sums agree with equations~\eqref{NSsym} -- \eqref{NSscalar}.


\section{Metric perturbation}\label{LinSugraG}
Recall that we are considering a trace free graviton perturbation $h_{\mu\nu}$ of the background metric in modified de Donder gauge $\nabla_\mu h^\mu{}_\nu=2h^\mu{}_\nu\partial_\mu\Phi$. This yields the following linear perturbations of the curvature
\begin{align}
\delta R_{\mu\nu} &= -\frac{1}{2} \left(\Box h_{\mu \nu} - 2 R_{\mu \alpha \beta \nu} h^{\alpha \beta} -R^\lambda_{\; \mu} h_{\lambda \nu}  -R^\lambda_{\; \nu} h_{ \mu \lambda} \right) +2 \nabla_{(\nu} \left(h_{\mu)}^{\sigma} \partial_\sigma \Phi \right)
\\
\delta R &= - h^{\mu\nu}R_{\mu\nu} + 4 h^{\rho\sigma} \partial_\rho \Phi \; \partial_\sigma \Phi + 2h_\mu^\sigma \nabla^\mu \partial_\sigma \Phi
\end{align}
as well as the perturbation
\begin{align}
\delta \left( \nabla_\kappa H^\kappa _{\;\; \mu\nu}\right) &= - h^{\rho \kappa} \nabla_{\rho} H_{\kappa \mu \nu} - 2 \left( \nabla_\rho h_{\sigma [\mu} \right)H_{\nu ]}^{\;\; \rho \sigma} - 2 H_{\alpha\mu\nu} h^{\alpha\lambda} \partial_\lambda \Phi \,.
\end{align}
Using these equations, it is straight forward to obtain the perturbations of~\eqref{SugraEOM}. We find the following first order expressions
\begin{align}
\begin{split}
\label{SLinEOM}
\delta \left( R_{\mu\nu} -\frac{1}{4} H_\mu^{\;\;\rho\sigma} H_{\nu\rho\sigma}  +2 \nabla_\mu \partial_\nu \Phi \right)
=-\frac{1}{2} \left(\Box h_{\mu \nu} - 2 R_{\mu \alpha \beta \nu} h^{\alpha \beta} -R^\lambda_{\; \mu} h_{\lambda \nu}  -R^\lambda_{\; \nu} h_{ \mu \lambda} \right)
\\
+\frac{1}{2} H_{\mu\kappa\lambda} H_{\nu\rho}{}^{\lambda} h^{\kappa \rho} + h_{\mu\sigma} \nabla_\nu \partial^\sigma \Phi + h_{\nu\sigma} \nabla_\mu \partial^\sigma \Phi + \nabla_\sigma h_{\mu\nu} \partial^\sigma \Phi
\end{split} 
\end{align}
\begin{align}
\begin{split}
\label{ASLinEOM}
\delta \left( \nabla^\kappa H_{\kappa\mu\nu } -2H_{\kappa\mu\nu } \partial^\kappa \Phi \right) = - h^{\rho \kappa} \nabla_{\rho} H_{\kappa \mu \nu} - 2 \left( \nabla_\rho h_{\sigma [\mu} \right)H_{\nu ]}^{\;\; \rho \sigma}
\end{split} 
\end{align}
for the tensorial equations and their scalar counterpart
\begin{align}
\begin{split}
\label{ScalLinEOM}
\delta \left( R + 4 \Box \Phi - 4 (\partial \Phi)^2 - \frac{1}{12} H^2 \right) = -h^{\mu\nu} \left( R_{\mu\nu} +2 \nabla_\mu \partial_\nu \Phi -\frac{1}{4} H_\mu^{\;\;\rho\sigma} H_{\nu\rho\sigma} \right) \,,
\end{split} 
\end{align} 
which is proportional to the symmetric term in~\eqref{SugraEOM} and therefore vanishes identically on supergravity backgrounds.


\section{B-field perturbation}\label{LinSugraB}
The usual convention for the field strength of the B-field is 
\be
H_{\mu\alpha\sigma}=\partial_\mu B_{\alpha\sigma}+\partial_\alpha B_{\sigma\mu}+\partial_\sigma B_{\mu\alpha}
\ee
and we denoted the linearised version $(\d b)_{\mu\alpha\sigma}=\partial_\mu b_{\alpha\sigma}+\partial_\alpha b_{\sigma\mu}+\partial_\sigma b_{\mu\alpha}$. Then the linearisation of~\eqref{SugraEOM} with respect to the B-field in the gauge $\nabla^\mu b_{\mu\nu} = 2 b_{\mu\nu} \partial^\mu \Phi$ is
\begin{align}
\begin{split}
\label{BScalLinEOM}
\delta \left( R + 4 \Box \Phi - 4 (\partial \Phi)^2 - \frac{1}{12} H^2 \right) =& -\frac{1}{6} H \cdot \d b
\end{split} 
\end{align} 
\begin{align}
\begin{split}
\label{BSLinEOM}
\delta \left( R_{\mu\nu} -\frac{1}{4} H_\mu^{\;\;\rho\sigma} H_{\nu\rho\sigma}  +2 \nabla_\mu \partial_\nu \Phi \right) =& -\frac{1}{2} H^{\rho \sigma}{}_{(\mu}  \d b_{\nu) \rho \sigma}
\end{split} 
\end{align} 
\begin{align}
\begin{split}
\label{BASLinEOM}
\delta \left( \nabla^\kappa H_{\kappa\mu\nu } -2H_{\kappa\mu\nu } \partial^\kappa \Phi \right) =& \Box b_{\mu\nu} -2 R^\lambda{}_{\mu \nu}{}^{\rho} b_{\lambda \rho}+2R^\sigma{}_{[\mu} b_{\nu] \sigma}\\ &-2 \partial^\kappa \Phi \nabla_\kappa b_{\mu\nu} + 4 b_{\alpha [\mu} \nabla_{\nu]} \partial^\alpha \Phi \,.
\end{split} 
\end{align}


\section{Dilaton perturbation}\label{LinSugraPhi}
The linearisation of~\eqref{SugraEOM} with respect to the dilaton field is straight forward and yields
\begin{align}
\begin{split}
\label{DilatonScalLinEOM}
\delta \left( R + 4 \Box \Phi - 4 (\partial \Phi)^2 - \frac{1}{12} H^2 \right) = 4 \Box  \varphi - 8 \partial_\mu \Phi \, \partial^\mu \varphi
\end{split} 
\end{align} 
\begin{align}
\begin{split}
\label{DilatonSLinEOM}
\delta \left( R_{\mu\nu} -\frac{1}{4} H_\mu^{\;\;\rho\sigma} H_{\nu\rho\sigma}  +2 \nabla_\mu \partial_\nu \Phi \right) = 2 \nabla_\mu \partial_\nu \varphi
\end{split} 
\end{align} 
\begin{align}
\begin{split}
\label{DilatonASLinEOM}
\delta \left( \nabla^\kappa H_{\kappa\mu\nu } -2H_{\kappa\mu\nu } \partial^\kappa \Phi \right) = -2H_{\kappa\mu\nu } \partial^\kappa \varphi \,.
\end{split} 
\end{align}

%% file: appendix3.tex
\chapter{Classical S-matrix \& Tree level integrands}\label{S-matrix-integrands}

This appendix reviews the notion of classical S-matrix which is used throughout chapter~\ref{C4.Spacetime}, as well as providing a precise definition for the \emph{tree level integrand}. On a sandwich plane wave background (for either gauge theory or gravity), the tree level S-matrix for a theory encodes the evolution of asymptotic free states from the in-region of the spacetime (i.e., $u<u_1$) through the non-trivial, or radiation region ($u_{1}\leq u\leq u_{2}$), to the out-region ($u>u_2$) as governed by the classical theory. 


\section{Classical S-matrix}

Rather than work out the curved space Feynman rules, we use a definition of the classical S-matrix in which tree level amplitudes are given by extracting certain multi-linear pieces of the classical action evaluated on a perturbative solution to the non-linear equations~\cite{Arefeva:1974jv,Jevicki:1987ax,Rosly:1996vr}.  In general this has the interpretation of the field-theoretic Hamilton-Jacobi generating function for the evolution and gives the tree level contribution to the S-matrix. For the 3-point calculations in chapter~\ref{C4.Spacetime}, there is no need to iterate the perturbative solution, but here we present the general framework.   

Let $S[\Phi]$ be the classical action, a functional of some fields $\Phi$ which is defined on the sandwich plane wave background (gravitational or gauge theoretic -- at this stage it makes no difference). We assume that this action takes the generic form:
\be\label{smat1}
S[\Phi]=\int \d^{d}X \left(\cL_{\mathrm{kin}}+\cL_{\mathrm{int}}\right)\,,
\ee
where $\cL_{\mathrm{kin}}$ is the kinetic portion of the action, which is quadratic in $\Phi$ and governs the free theory, and $\cL_{\mathrm{int}}$ contains all higher-point interactions.

Define the following object:
\be\label{smat2}
\Phi^{[n]}(X):=\sum_{i=1}^{n}\epsilon_{i}\,\varphi_{i}(X)+\int \d^{d}Y\,\Delta(X,Y)\,\left.\frac{\delta\cL_{\mathrm{int}}}{\delta\Phi}\right|_{\Phi=\sum_{j=1}^{n}\epsilon_{j}\varphi_{j}(Y)}.
\ee
This is essentially an integral form of the full non-linear equations from the action $S$ with data given by the first term on the right hand side. Here, the $\{\epsilon_{i}\}$ are $n$ parameters that will eventually  be thought of as infinitesimal; $\{\varphi_{i}\}$ are $n$ solutions to the free equations of motion of $\cL_{\mathrm{kin}}$ with specified asymptotic behaviour; and $\Delta(X,Y)$ is a Green's function defined by $\cL_{\mathrm{kin}}$. There are precise formulae for various useful definitions of this $\Delta(X,Y)$ (e.g., advanced, retarded, Feynman) in scalar, gauge, and gravitational theories on plane wave backgrounds~\cite{Gibbons:1975jb,Harte:2012uw}, though we will not make explicit use of them here. Specifying the asymptotic behaviour of the free solution $\varphi_{i}$ boils down to saying whether it looks like an `in' or `out' state.

Both the in- and out-regions are flat, so asymptotically free states $\varphi_{i}$ should look like free states in Minkowski space in at least one of these regions. In a momentum space representation, such free states in Minkowski space are modelled on massless plane wave momentum eigenstates, $\e^{\im\,k\cdot X}$ for $k^2=0$. Unlike Minkowski space, in the sandwich plane wave a state which looks like $\e^{\im\,k\cdot X}$ in the in-region will \emph{not} look like $\e^{\im\,k\cdot X}$ in the out-region. This is a consequence of the `memory' relations \eqref{newvb}, \eqref{gmem}. Hence, the specification of asymptotic behaviour for $\varphi_{i}$ boils down to stating whether it is an \emph{incoming} or \emph{outgoing} state, denoted respectively as $\varphi_{i}^{-}$ or $\varphi_{i}^{+}$. An incoming state is one which looks like a free solution in Minkowski space the in-region; an out state looks like a free solution in Minkowski space in the out-region. More precisely,
\be\label{smat3} 
\varphi^{-}_{i}|_{\mathrm{in}}\sim\e^{\im\,k\cdot X}\sim\varphi^{+}_{i}|_{\mathrm{out}}\,,
\ee
for both the gravitational and gauge theory backgrounds.

The $n$-point tree level scattering amplitude for the states $\{\varphi_{i}\}$ -- with their given asymptotic configuration of in and out states -- is then a multi-linear piece of the classical action:
\be\label{smat4}
M^{(0)}_{n}(\varphi_{1},\ldots,\varphi_{n})=\left.\frac{\partial^{n}S[\Phi^{[n]}]}{\partial\epsilon_{1}\cdots\partial\epsilon_{n}}\right|_{\epsilon_{1}=\cdots=\epsilon_{n}=0}\,.
\ee
For flat backgrounds, this agrees with the usual definition of the S-matrix and would also correspond with a Feynman diagram definition for sandwich plane waves.


\section{Tree level integrands}

For the purposes of investigating the double copy, a notion of \emph{tree level integrand} closely related to the tree level amplitude is useful. Indeed, it is actually this tree level integrand that appears in the KLT relations of the standard double copy. From the definition \eqref{smat4} it is straightforward to see that the tree level scattering amplitude will always take the form:
\be\label{tint1}
M^{(0)}_{n} = \int \d^{d}X\,\cM_{n}(X)\,\prod_{i=1}^{n} f_{i}(X)\,,
\ee
where each of the $f_{i}(X)$ is a solution to the free scalar wave equation on the plane wave background. The object $\cM_{n}$ is defined to be the tree level integrand; generically, it will be formed of polarisations, momenta and propagators and depends on the background geometry. It captures everything that is encoded by the kinematic numerators and denominators which would result from a conventional Feynman diagram approach. In more heuristic terms, the tree level integrand is what remains after removing the final integral that forms the action functional in \eqref{smat4}, along with `universal' spin-independent functions.  

In Minkowski space, it is easy to see that
\begin{equation*}
\prod_{i=1}^{n} f_{i}(X)=\e^{\im(k_{1}+\cdots+k_{n})\cdot X}\,,
\end{equation*}
so the effect of isolating $\cM_{n}$ is to strip off an overall momentum conserving delta function. On non-trivial backgrounds such as the sandwich plane wave, the result of the final $\d^{d}X$ integrals is more complicated, but the principle is the same: $\cM_{n}$ contains all of the information which one could expect to be `squared' in taking the double copy. Another interesting property of the integrand is that it is functionally independent of the asymptotic conditions of the states being scattered. This enables the investigation of the double copy by considering the computationally simplest configuration of incoming and outgoing states.

Clearly, there is a sense in which the tree level integrand is not a gauge-invariant object, just as one can add boundary terms to an action. This lack of gauge invariance is analogous to the statement that individual Feynman diagrams -- or individual terms contributing to \eqref{smat4} -- are not gauge invariant. However, once a gauge for performing perturbative calculations has been fixed (i.e., specific linearised solutions $\{\varphi_i\}$ and a Green's function $\Delta(X,Y)$ have been consistently chosen), the object $\cM_{n}$ is well-defined. In our calculations, we always work in a Lorenz or de Donder gauge, so the resulting expressions for the integrand should be viewed as expressions in these particular gauges. Their \emph{integrals}, however, do not depend on the gauge choice.

Throughout chapter~\ref{chapter4}, the tree level integrand for theories on the gravitational plane wave background is denoted by $\cM_{n}$, and the tree level integrand for theories on the gauge theory plane wave background by $\cA_{n}$. After performing the integration, we denote the actual amplitudes on these backgrounds by $\cM^{\text{pw}}_{n}$ and $\cA^{\text{pw}}_{n}$ respectively.

%% file: appendix4.tex
\chapter{The impulsive plane wave}\label{Impulse}

For both gauge theory and gravitational sandwich plane waves, the computation of 3-point amplitudes (rather than integrands) boils down to performing integrations that depend on the particulars of the background geometry. In this appendix, we consider the simplest concrete example of a sandwich plane wave: the \emph{impulsive plane wave}~\cite{Aichelburg:1970dh,Penrose:1972xrn,Dray:1984ha,Klimcik:1988az,Ferrari:1988cc}. Impulsive plane waves correspond to gluing two flat regions together along an infinitesimal burst of radiation; in other words, the radiation region of the sandwich plane wave has delta function support. In the case of the impulsive gauge theory background, the scalar and gluon 3-point amplitudes can be computed in closed form. For the impulsive gravitational background, the 3-point amplitudes can be written in terms of integrals which are suitable to numerical approximation.


\section{Gauge theory background}

For an impulsive gauge theory plane wave, we have
\be\label{gimp1}
\dot{\sA}_{a}(u)=\delta(u)\,\sa_{a}\,,
\ee
for $\sa_{a}$ a set of $d-2$ constants which characterise the impulsive wave. Using the asymptotic conditions \eqref{gfbc}, it follows that
\be\label{gimp2}
\sA^{-}_{a}(u)=\Theta(u)\,\sa_{a}\,, \qquad \sA^{+}_{a}(u)=-\Theta(-u)\,\sa_{a}\,,
\ee
where $\Theta(u)$ is the Heaviside step function. Proceeding from \eqref{cs3p1} it is a straightforward calculation to obtain the 3-point amplitudes of charged scalars on this background. The results for the two independent configurations -- all incoming or two incoming and one outgoing -- are given by:
\begin{multline}\label{gimps1}
 M_{3}(\Phi^{-}_{1},\Phi^{-}_{2},\Phi^{-}_{3})=\frac{\lambda}{6}\,\delta^{d-1}\!\left(\sum_{{r}=1}^3 k_{r}\right)\left[\left(\sum_{{s}=1}^{3}\frac{\mathbf{k}_{{s}}^{2}}{2\,k_{{s}\,0}}\right)^{-1} \right. \\
 \left.-\left(\sum_{{s}=1}^{3}\frac{\mathbf{k}_{{s}}^{2}+2 e_{s} k_{{s}}^{a}\sa_{a}+e_{s}^{2}\sa^{2}}{2\,k_{{s}\,0}}\right)^{-1}\right]\,,
\end{multline}
and
\begin{multline}\label{gimps2}
 M_{3}(\Phi^{-}_{1},\Phi^{-}_{2},\Phi^{+}_{3})=\frac{\lambda}{6}\,\delta^{d-1}\!\left(\sum_{{r}=1}^3 k_{r}\right)\,\left[\left(\frac{\mathbf{k}_{3}^{2}-2e_{3} k_{3}^{a}\sa_{a}+e_{3}^{2}\sa^{2}}{2\,k_{3\,0}}+\sum_{{s}=1,2}\frac{\mathbf{k}_{{s}}^{2}}{2\,k_{{s}\,0}}\right)^{-1} \right. \\
 \left.-\left(\frac{\mathbf{k}_{3}^2}{2\,k_{0\,3}}+\sum_{{s}=1,2}\frac{\mathbf{k}_{{s}}^{2}+2e_{s} k_{{s}}^{a}\sa_{a}+e_{s}^{2}\sa^{2}}{2\,k_{{s}\,0}}\right)^{-1}\right]\,,
\end{multline}
where $\mathbf{k}_{{s}}^{2}:=k_{{s}\,a}k^{a}_{{s}}$ for any ${s}=1,2,3$.

The 3-point amplitudes for gluons on the impulsive gauge theory background follow similarly from \eqref{cgf3p4}. A calculation leads to:
\begin{multline}\label{gimpg1}
 M_{3}(a^{-}_{1},a^{-}_{2},a^{-}_{3})=2\,g\,\delta^{d-1}\!\left(\sum_{{r}=1}^3 k_{r}\right)\left[\left(\sum_{{s}=1}^{3}\frac{\mathbf{k}_{{s}}^{2}}{2\,k_{{s}\,0}}\right)^{-1}\,F(\{k_{t}, \tepsilon_{t}\}) \right. \\
 -\left(\sum_{{s}=1}^{3}\frac{\mathbf{k}_{{s}}^{2}+2e_{s} k_{{s}}^{a}\sa_{a}+e_{s}^{2}\sa^{2}}{2\,k_{{s}\,0}}\right)^{-1} \left(F(\{k_{t},\tepsilon_{t}\}) -\sa^{a} \left(\frac{\tepsilon_{1}\cdot \tepsilon_{3}}{k_{2\,0}}\,\tepsilon_{2\,a}(e_{2}k_{1\,0}-e_{1}k_{2\,0}) \right.\right. \\
 +\left.\left.\left.\frac{\tepsilon_{1}\cdot \tepsilon_{2}}{k_{3\,0}}\,\tepsilon_{3\,a}(e_{3}k_{2\,0}-e_{2}k_{3\,0}) +\frac{\tepsilon_{2}\cdot \tepsilon_{3}}{k_{1\,0}}\,\tepsilon_{1\,a} (e_{1}k_{3\,0}-e_{3}k_{1\,0})\right)\right)\right]\,,
\end{multline}
and
\begin{multline}\label{gimpg2}
 M_{3}(a^{-}_{1},a^{-}_{2},a^{+}_{3})=2\,g\,\delta^{d-1}\!\left(\sum_{{r}=1}^3 k_{r}\right)\,\left[\left(\frac{\mathbf{k}_{3}^{2}-2e_{3} k_{3}^{a}\sa_{a}+e_{3}^{2}\sa^{2}}{2\,k_{3\,0}}+\sum_{{s}=1,2}\frac{\mathbf{k}_{{s}}^{2}}{2\,k_{{s}\,0}}\right)^{-1} \right. 
\\
 \times\left(F(\{k_{t},\tepsilon_{t}\})+e_{3}\,\sa^{a}\left(\frac{k_{2\,0}}{k_{3\,0}}\,\tepsilon_{1}\cdot \tepsilon_{2}\,\tepsilon_{3\,a}-\tepsilon_{2}\cdot \tepsilon_{3}\,\tepsilon_{1\,a}\right)\right) 
\\
 -\left(\frac{\mathbf{k}_{3}^2}{2\,k_{0\,3}}+\sum_{{s}=1,2}\frac{\mathbf{k}_{{s}}^{2}+2e_{s} k_{{s}}^{a}\sa_{a}+e_{s}^{2}\sa^{2}}{2\,k_{{s}\,0}}\right)^{-1}\,\bigg(F(\{k_{t},\tepsilon_{t}\}) 
\\
-\sa^{a} \left(\frac{\tepsilon_{1}\cdot \tepsilon_{3}}{k_{2\,0}}\,\tepsilon_{2\,a}(e_{2}k_{1\,0}-e_{1}k_{2\,0})\right.
 \left.-e_{2}\,\tepsilon_{1}\cdot \tepsilon_{2}\,\tepsilon_{3\,a}+e_{1}\,\frac{k_{3\,0}}{k_{1\,0}}\,\tepsilon_{2}\cdot \tepsilon_{3}\,\tepsilon_{1\,a}\right)\bigg)\Bigg]\,,
\end{multline}
where the function $F$ of the kinematic data is defined by \eqref{flatamp}.

In each of these expressions a Hartle-Hawking contour deformation is used to dampen rapidly oscillating contributions to the $u$-integrations near $u=\pm\infty$. This is the same as the prescription on Minkowski space, and corresponds to selecting the physical vacuum.


\section{Gravitational background}

For an impulsive gravitational wave, the non-trivial metric component $H(u,\mathbf{x})$ in Brinkmann coordinates has delta function support:
\be\label{imp1}
H(u,\mathbf{x})=\delta(u)\,H_{ab}\,x^{a}\,x^{b}\,,
\ee
with $H_{ab}$ a trace-free and constant $(d-2)\times(d-2)$ matrix. Assuming that $H_{ab}$ is corank zero with distinct eigenvalues, it can be diagonalised using rotations in the $x^{a}$-plane. So without loss of generality, we take
\be\label{imp2}
H_{ab}=\lambda_{(a)}\,\delta_{ab}\,, \qquad \sum_{a=1}^{d-2}\lambda_{(a)}=0\,.
\ee
The vielbein $E^{a}_{i}$ must solve the equation
\be\label{imp3}
\ddot{E}_{a\,i}=\lambda_{(a)}\,\delta_{ab}\,\delta(u)\,E^{b}_{i}\,,
\ee
subject to incoming or outgoing boundary conditions \eqref{vbbc}. In each case, one finds
\be\label{imp4}
E^{-}_{a\,i}=\delta_{ai}\left(1+u\,\lambda_{(a)}\,\Theta(u)\right)\,, \qquad E^{+}_{a\,i}=\delta_{ai}\left(1-u\,\lambda_{(a)}\,\Theta(-u)\right)\,,
\ee
so the transverse metric $\gamma_{ij}(u)$ is given in incoming or outgoing coordinates by:
\be\label{impmet}
\gamma_{ij}^{-}(u)=\delta_{ij}\left(1+u\,\lambda_{(i)}\,\Theta(u)\right)^{2}\,, \qquad \gamma_{ij}^{+}(u)=\delta_{ij}\left(1-u\,\lambda_{(i)}\,\Theta(-u)\right)^{2}\,,
\ee
where $\lambda_{(i)}$ is identified with $\lambda_{(a)}$ using $\delta_{a}^{i}$. This demonstrates that the impulsive gravitational wave is two copies of Minkowski space glued together along a single pulse of gravitational radiation. While the metrics \eqref{impmet} are continuous across the pulse at $u=0$, they have discontinuous first derivatives.

To compute 3-point amplitudes, it is also important to have the inverse vielbeins:
\be\label{imp5}
E^{i\,-}_{a}=\delta^{i}_{a}\left(1+u\,\lambda_{(a)}\,\Theta(u)\right)^{-1}\,, \qquad E^{i\,+}_{a}=\delta^{i}_{a}\left(1-u\,\lambda_{(a)}\,\Theta(-u)\right)^{-1}\,,
\ee
leading to expressions for $F^{ij}_{\pm}(u)$:
\begin{subequations}\label{imp6}
 \be
 F^{ij}_{-}(u)=\frac{u\,\delta^{ij}}{1+u\,\lambda_{(i)}\,\Theta(u)},
 \ee
 \be
 F^{ij}_{+}(u)=\frac{u\,\delta^{ij}}{1-u\,\lambda_{(i)}\,\Theta(-u)}\,.
 \ee
\end{subequations}
So in both cases $F^{ij}(u)$ gets an infinite series of $O(u^2)$ corrections upon crossing the pulse at $u=0$.

Even at the level of scalar amplitudes, the situation on the gravitational background is more complicated than on the gauge theory background. Unlike \eqref{gimps1}--\eqref{gimps2}, on the impulsive gravitational wave (relatively) compact expressions for the $u$-integrations are not available. Instead, we find explicit expressions which could be evaluated (numerically or possibly analytically) when the momenta and eigenvalues $\{\lambda_{(a)}\}$ are specified. For instance, one finds:
\begin{multline}\label{imps1}
 M_{3}(\Phi^{-}_{1},\Phi_{2}^{-},\Phi^{-}_{3})=\frac{\lambda\,\im}{6}\,\delta^{d-1}\!\left(\sum_{{r}=1}^{3}k_{r}\right)\,\left[\left(\sum_{{s}=1}^{3}\frac{\mathbf{k}_{{s}}^{2}}{2\,k_{{s}\,0}}\right)^{-1} \right. \\
 \left. +\im\int\limits_{0}^{\infty+i\epsilon}\d u\,\prod_{a=1}^{d-2}(1+u\,\lambda_{(a)})^{-\frac{1}{2}}\,\exp\left(\im\,u \sum_{{s}=1}^{3}\sum_{i=1}^{d-2}\frac{k_{s\,i}^{2}}{2k_{s\,0}\,(1+u\lambda_{(i)})}\right)\right]\,,
\end{multline}
for the all-incoming configuration.

The expression for the two-incoming, one-outgoing configuration is similarly given in terms of $u$-integrals over the in- and out-regions:
\begin{multline}\label{imps2}
 M_{3}(\Phi^{-}_{1},\Phi_{2}^{-},\Phi^{+}_{3})=-\frac{\lambda}{6}\,\sqrt{\frac{(2\,\pi\im)^{d-2}}{k_{3\,0}^{d-2}}}\;\delta\!\left(\sum_{{r}=1}^{3}k_{{r}\,0}\right) \\
 \times\left[\int\limits_{-\infty-\im\epsilon}^{0}\frac{\d u}{\prod_{a=1}^{d-2}\sqrt{\lambda_{(a)}}}\,\exp\left(-\frac{\im}{2\,k_{3\,0}}J_{a}J_{b} (A^{-1})^{ab}+\im\sum_{{s}=1}^{3}\frac{k_{{s}\,i}k_{{s}\,j}}{2\,k_{{s}\,0}} F^{ij}_{{s}}\right)\right. \\
 \left.+\int\limits^{\infty+\im\epsilon}_{0} \d u\,\prod_{a=1}^{d-2}(\lambda_{(a)}+u\,\lambda^{2}_{(a)})^{-\frac{1}{2}}\,\exp\left(-\frac{\im}{2\,k_{3\,0}}J_{a}J_{b} (A^{-1})^{ab}+\im\sum_{{s}=1}^{3}\frac{k_{{s}\,i}k_{{s}\,j}}{2\,k_{{s}\,0}} F^{ij}_{{s}}\right)\right]\,.
\end{multline}
Here, the $F^{ij}_{{s}}(u)$ are given by \eqref{imp6}, while
\be\label{imp7}
A_{ab}(u)=\frac{-\lambda_{(a)}\,\delta_{ab}}{1+|u|\,\lambda_{(a)}}\,,
\ee
and
\be\label{imp8}
J_{a}(u)=\frac{k_{1\,a}+k_{2\,a}+k_{3\,a}+u\,\lambda_{(a)}(k_{3\,a}\,\Theta(u)-(k_{1\,a}+k_{2\,a})\,\Theta(-u))}{1+|u|\,\lambda_{(a)}}\,.
\ee